\documentclass[fleqn,usenatbib]{mnras}

\usepackage{newtxtext,newtxmath}

\usepackage[T1]{fontenc}

\DeclareRobustCommand{\VAN}[3]{##2}
\let\VANthebibliography\thebibliography
\def\thebibliography{\DeclareRobustCommand{\VAN}[3]{##3}\VANthebibliography}

\usepackage{graphicx}	
\usepackage{amsmath}	
\usepackage{subcaption}
\usepackage{cleveref}
\usepackage{comment}

\title[TNG50 morphological properties]{The building up of \emph{observed} stellar scaling relations of massive galaxies and the connection to black hole growth in the TNG50 simulation}

\author[S. Varma et al.]{
S. Varma$^{1}$, 
M. Huertas-Company$^{1,2,8}$\thanks{E-mail: mhuertas@iac.es}, A. Pillepich$^{3}$, D. Nelson$^{4}$,
V. Rodriguez-Gomez$^{5}$,
\newauthor A. Dekel$^{6,8}$, S.M. Faber$^{7}$, P. Iglesias-Navarro$^{1}$, D.C. Koo$^{7}$, J. Primack$^{8}$ \\
$^{1}$Instituto de Astrofısica de Canarias (IAC); Departamento de Astro´ısica, Universidad de La Laguna (ULL), E-38200, La Laguna, Spain\\
$^{2}$ LERMA, Observatoire de Paris, CNRS, PSL, Universit\'e de Paris, France \\
$^{3}$Max-Planck-Institut f{\"u}r Astronomie, K{\"o}nigstuhl 17, 69117 Heidelberg, Germany\\
$^{4}$Universit\"{a}t Heidelberg, Zentrum f\"{u}r Astronomie, Institut f\"{u}r theoretische Astrophysik, Albert-Ueberle-Str. 2, 69120 Heidelberg, Germany\\
$^{5}$Instituto de Radioastronom\'ia y Astrof\'isica, Universidad Nacional Aut\'onoma de M\'exico, Apdo. Postal 72-3, 58089 Morelia, Mexico\\
$^{6}$Center for Astrophysics and Planetary Science, Racah Institute ofPhysics, The Hebrew University, Jerusalem, Israel \\
$^{7}$UCO/Lick Observatory, Department of Astronomy and Astrophysics,
University of California, Santa Cruz, CA, USA \\
$^{8}$SCIPP, University of California, Santa Cruz, CA, USA\\}

\date{Accepted XXX. Received YYY; in original form ZZZ}

\pubyear{2015}

\begin{document}
\label{firstpage}
\pagerange{\pageref{firstpage}--\pageref{lastpage}}

\maketitle

\begin{abstract}
\newgeometry{verbose, tmargin = 0cm, bmargin = 2cm, lmargin = 3cm, rmargin = 3cm}
 \vskip -20pt
We study how \emph{mock-observed} stellar morphological and structural properties of  massive galaxies are built up between $z=0.5$ and $z=3$ in the TNG50 cosmological simulation. We generate mock images with the properties of the CANDELS survey and derive Sersic parameters and optical rest-frame morphologies as usually done in the observations. Overall, the simulation reproduces the observed evolution of the abundances of different galaxy morphological types of star-forming and quiescent galaxies. The $\log{M_*}-\log R_e$ and $\log{M_*}-\log\Sigma_1$ relations of the simulated star-forming and quenched galaxies also match the observed slopes and zeropoints to within 1-$\sigma$. 
In the simulation, galaxies increase their observed central stellar mass density ($\Sigma_1$) and transform in morphology from irregular/clumpy systems to normal Hubble-type systems in the Star Formation Main Sequence at a characteristic stellar mass of $\sim 10^{10.5}~M_\odot$ which is reflected in an increase of the central stellar mass density ($\Sigma_1$). 
This morphological transformation is connected to the activity of the central Super Massive Black Holes (SMBHs). At low stellar masses ($10^9$ < $M_*/M_\odot$ < $10^{10}$) SMBHs grow rapidly, while at higher mass SMBHs switch into the kinetic feedback mode and grow more slowly. During this low-accretion phase, SMBH feedback leads to the quenching of star-formation, along with a simultaneous growth in $\Sigma_1$, partly due to the fading of stellar populations.
More compact massive galaxies grow their SMBHs faster than extended ones of the same mass and end up quenching earlier. In the TNG50 simulation, SMBHs predominantly grow via gas accretion before galaxies quench, and $\Sigma_1$ increases substantially after SMBH growth slows down. The simulation predicts therefore that quiescent galaxies have higher $\Sigma_1$ values than star-forming galaxies for the same SMBH mass, which disagrees with alternative models, and may potentially be in tension with some observations. 
\end{abstract}




\section{Introduction}

Understanding the origins of the morphological diversity of galaxies is one of the main challenges in the field of galaxy evolution. How morphological transformations happen and how they are related to the regulation and quenching of star formation remain open questions today.

Thanks to a variety of photometric and spectroscopic surveys, we have reached a reasonably good description of the properties of star-forming and quenched galaxies over the last $\sim9$ Gyrs. It is now well established that the morphologies of galaxies have significantly evolved over cosmic time, and that quenched and star-forming galaxies present different structural properties at all times (e.g.\citealp{2013MNRAS.428.1460B,2015ApJS..221....8H}).  As a matter of fact, quiescent and star-forming galaxies have been shown to follow distinct linear relations between $\log M_*$ and $\log R_e$ (e.g. \citealp{vanderwel2014ApJ...788...28V,2015ApJ...813...23V}), the log surface density within the effective radius $R_e$, and the $\log$
surface density within 1 kpc ($\log \Sigma_1$) \citep{barro}. Quenched galaxies have always been smaller and denser than star-forming galaxies at fixed stellar mass. Additionally, quiescent galaxies typically present spheroid like shapes at all redshifts, with Sersic indices $n>2$, while star-forming galaxies transition from irregular/clumpy systems at $z>1$ to more regular disks at later epochs (e.g.~\citealp{2015ApJS..221....8H,2015ApJ...800...39G}). Their surface brightness distributions are nevertheless well described by exponential profiles at all redshifts~\citep{vanderwel2014ApJ...788...28V}.

While the observational relations are well established and measured, the evolutionary tracks that build them remain highly debated. Some works argue that quenching is tightly linked to morphological transformations. Major merging remains a classical channel to produce spheroidal morphologies while quenching star formation by rapidly consuming gas within the galaxy (e.g.\citealp{1977egsp.conf..401T,2021NatAs.tmp....5P}). Other internal physical process such as sudden gas inflows (the so-called compaction events) can also build the central densities of stars by provoking a rapid consumption of gas (e.g.~\citealp{2015MNRAS.450.2327Z,2019arXiv190408431D}), which in turn can trigger the growth of a Super Massive Black Hole (SMBH) in the centers of galaxies~\citep{2020arXiv201209186L}. The gas inflows can not only be provoked by mergers, but also by other mechanisms such as fly-by's, or counter rotating streams. The Black Hole growth in that view is a consequence of the inflow and therefore helps to keep the galaxy quenched, but is not strictly required for quenching. A similar view has been put forward in~\cite{2019MNRAS.484.4413H} using the IllustrisTNG simulations (300 and 100 runs). In a recent work~\citep{2020ApJ...897..102C}, the authors develop a semi-empirical model - hereafter \emph{Black Hole versus Halo} (BHvH) model- to explain the building up of galaxy scaling laws. In this model, the SMBH is the main cause for quenching star formation in galaxies. The key assumption of the model is that galaxies quench when the total emitted SMBH radiation is a multiple of the halo gas binding energy. This threshold is achieved at lower stellar masses for small galaxies than for large ones, which explains the tilt of the quenched ridgeline in the mass-size plane. The BHvH Model also clearly predicts that the SMBH rapidly grows when the galaxy starts quenching and crossing the so-called Green Valley. Some authors have also pointed out that observational biases, such as the progenitor bias effect, can also contribute to explain the different scaling laws of quiescent and star-forming galaxies. The passive population we observe at a given time indeed comes from a parent population of star-forming galaxies which were more compact than star-forming galaxies at the time of observation (e.g.\citealp{2016ApJ...833....1L,2019MNRAS.487.5416T}). This might imply that the morphological bimodality we observe between star-forming and quenched galaxies does not require strong morphological transformations.

Numerical simulations offer an attractive way to consistently explore how morphology and quenching are related since they allow to follow individual objects over cosmic time on a volume limited sample and can also be forward modeled to properly reproduce the observing conditions~(e.g. \citealp{2015MNRAS.451.4290S,2018ApJ...858..114H}). Cosmological hydrodynamic simulations therefore represent the ideal tool in order to have a more statistically meaningful sample of mock images that can be compared against observations. However, for many years, cosmological simulations have struggled to reproduce the diversity and the structural properties of galaxies we observe in the local universe, making them impossible to use to learn about morphological evolution. This is changing quickly with the latest generation of simulations, which have made significant progress. Several works have shown that recent simulations are able to reproduce the abundances and sizes of different morphological types at $z\sim0$ with fairly good accuracy. In particular, the TNG suite has been shown to present significant improvements regarding the structure of $z=0$ galaxies compared to its predecessor, the Illustris simulation (e.g.~\citealp{2018MNRAS.474.3976G, 2019MNRAS.483.4140R,marc2019MNRAS.489.1859H,2020MNRAS.tmp.3638Z,2021MNRAS.tmp..841W}). A logical step forward consists in exploring the morphological evolution of galaxies in the simulations and compare to observations at higher redshifts in a consistent way. This is particularly useful because it (i) provides additional constraints to the models which are typically {\it tuned} to match observables at $z=0$ but not beyond that, and (ii) allows us to track down morphological transformations without progenitor biases.

However, this remains challenging with state-of-the art large volume hydrodynamic simulations. The usual $\sim1$ kpc spatial resolution is typically not enough to accurately infer the morphology of high redshift galaxies. Zoom-in cosmological simulations obviously provide a higher spatial resolution at the expense of statistics. In that context, the latest run of the TNG simulation suite, TNG50~\citep{2019MNRAS.490.3234N,2019MNRAS.490.3196P}, offers an excellent trade off between volume and resolution, thus filling the gap between very high resolution zoom-in simulations and large volume low resolution ones. It is therefore the perfect simulated dataset to explore the morphological properties of galaxies at $z>0$.

This is precisely the main goal of this work. We compare the \emph{mock-observed} morphological and structural properties of a statistical sample of simulated galaxies in a cosmological context in the redshift range $0.5<z<3$ with available Hubble Space Telescope observations in the same redshift range. Following a similar approach as in previous works (e.g.~\citealp{marc2019MNRAS.489.1859H}), we \emph{observe} simulations to ensure that all observational biases are properly taken into account. We generate to that purpose realistic mock observations of the simulated galaxies and apply the same methodologies as in the observational samples. We then use the simulations to trace back in time the evolution of the stellar morphology of galaxies and its connection to quenching. 

The paper proceeds as follows. We first describe the datasets used and the methods to estimate morphologies and sizes in sections \ref{sec:data} to~\ref{sec:qfrac}. Section \ref{sec:evol} and \ref{sec:scaling} show the main results, which are then discussed in terms of evolutionary tracks in section \ref{sec:discuss}. We use a Chabrier IMF~\citep{2003PASP..115..763C} and a Planck 2013 cosmology~\citep{2014A&A...571A..16P}.

\section{Data}
\label{sec:data}
\subsection{Simulations: TNG50}

The IllustrisTNG Project \citep{2018MNRAS.475..624N, 2018MNRAS.480.5113M, 2018MNRAS.477.1206N, 2018MNRAS.475..676S, 2018MNRAS.475..648P}  is a suite of magneto-hydrodynamic cosmological simulations performed with the moving-mesh code AREPO~\citep{2010ARA&A..48..391S, 2011MNRAS.418.1392P, 2016MNRAS.455.1134P}. The model is an updated version of the original Illustris project~\citep{2014Natur.509..177V, 2014MNRAS.444.1518V, 2014MNRAS.445..175G, 2015MNRAS.452..575S}. We refer the reader to~\cite{marc2019MNRAS.489.1859H} and references therein for a description of the main differences which essentially affect to AGN feedback model, the galactic winds and the presence of magnetic fields . 

In this project, we make use of the highest resolution version of TNG, called TNG50~\citep{2019MNRAS.490.3234N,2019MNRAS.490.3196P} - a new class of cosmological simulation, designed to overcome the traditional trade-off between simulation volume and spatial resolution. The TNG50 simulation realizes a simulation cube with a volume of 51.7$^{3}$ Mpc, following the evolution of 2160$^{3}$ resolution elements. For this work, we consider 6 different snapshots at \textit{z} = 0.5, 1.0, 1.5, 2.0, 2.4 and 3.0 (snapshots 27, 30, 33, 40, 50, and 67 respectively), selecting galaxies with $Log(M_*/M_\odot)$ >9.0. This results in a sample of 11,048 galaxies. Galaxies were identified with the SUBFIND algorithm \citep{2001MNRAS.328..726S, 2009MNRAS.399..497D}, while the merger trees were constructed with SubLink \citep{2015MNRAS.449...49R}.

\begin{figure*}
    \centering
    \vspace{-10 pt}
    \subcaptionbox{TNG50 F160W}{\includegraphics[width=0.45\textwidth]{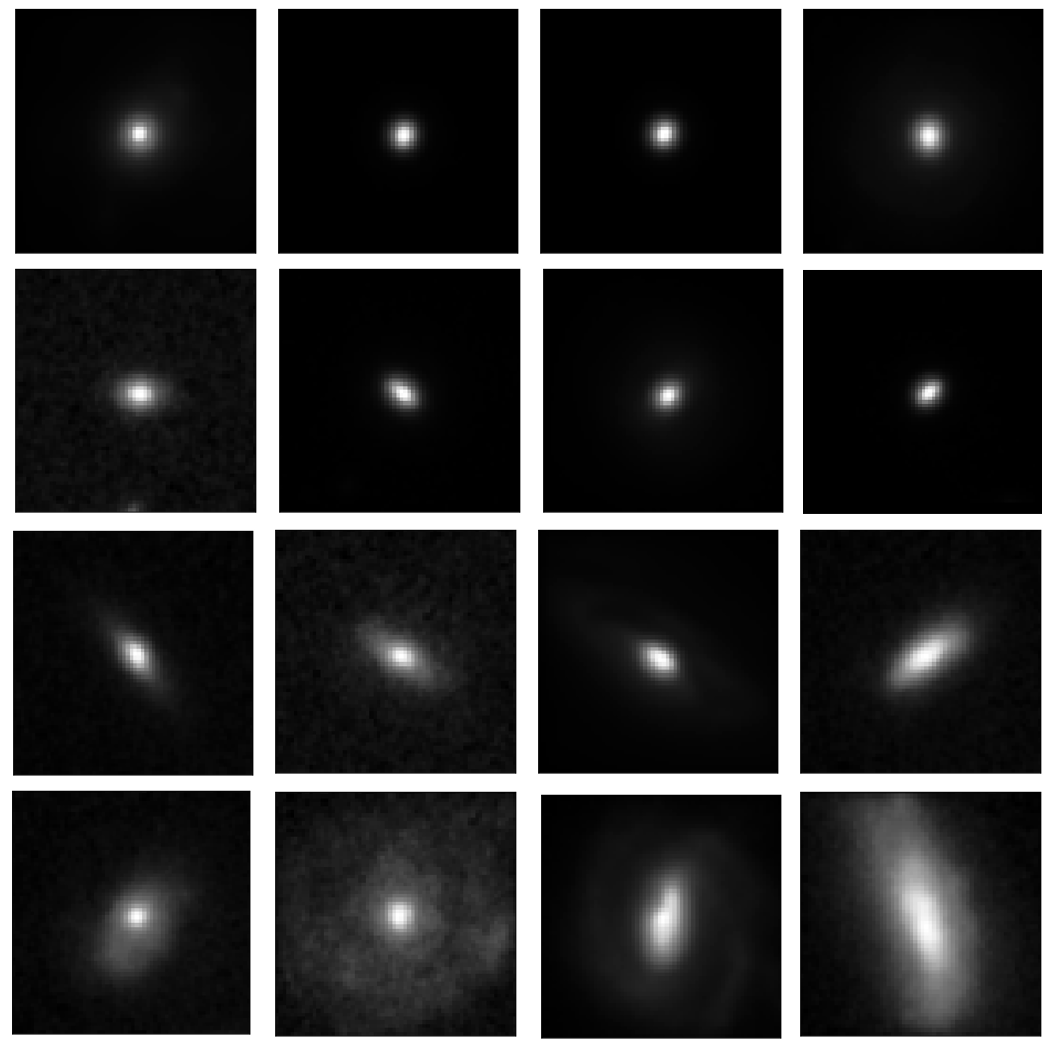}}%
     \qquad
      \subcaptionbox{CANDELS F160W}{\includegraphics[width=0.45\textwidth]{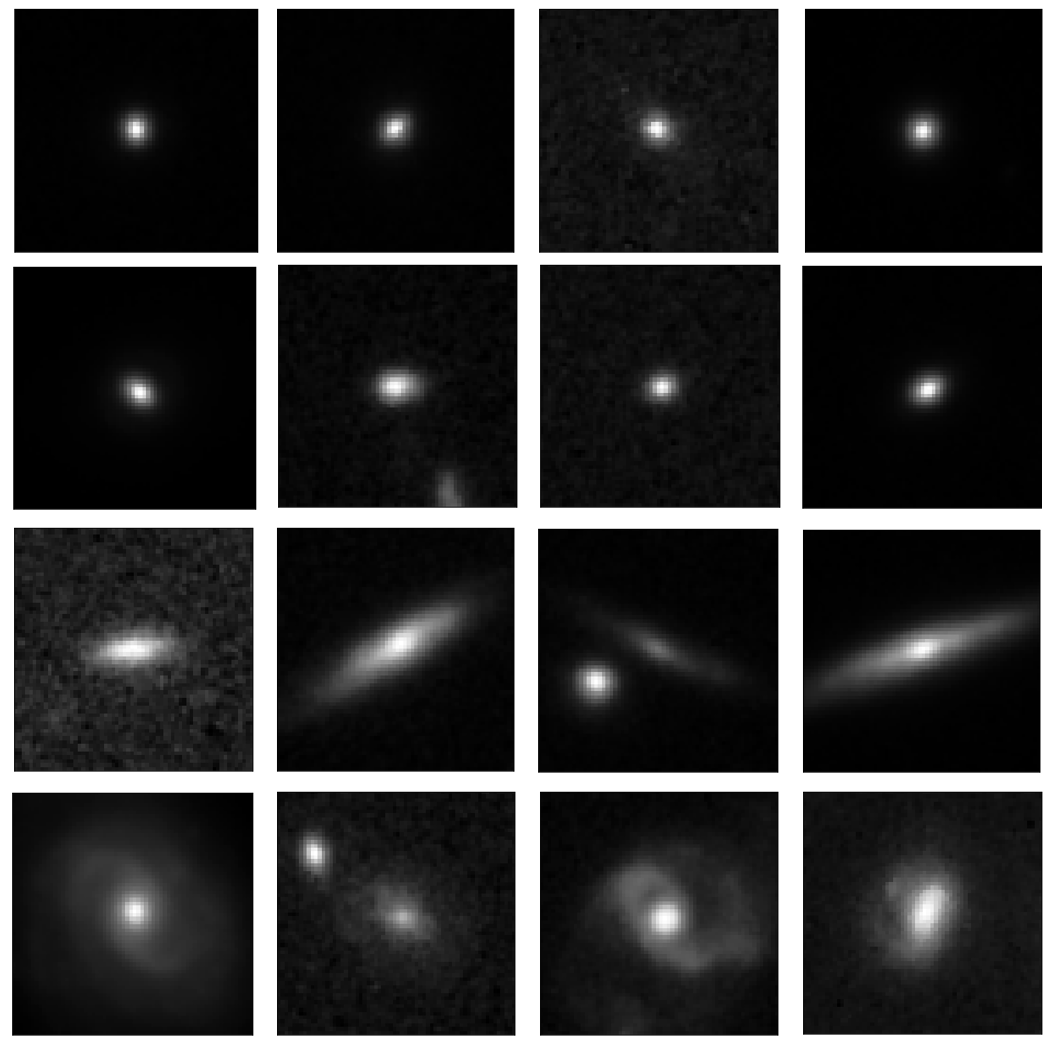}}%
     \qquad
    \subcaptionbox{TNG50 F775W}{\includegraphics[width=0.45\textwidth]{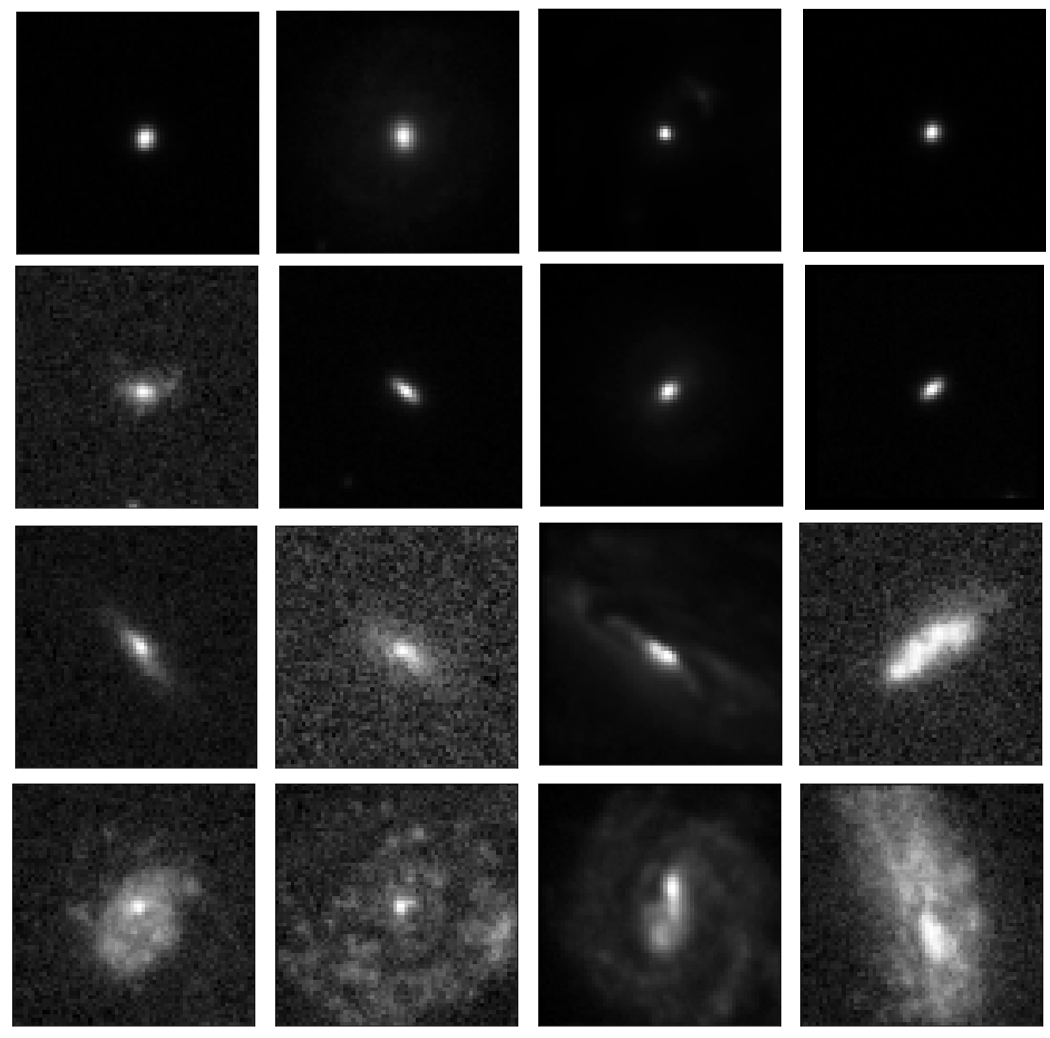}}
    \qquad
    \subcaptionbox{CANDELS F775W}{\includegraphics[width=0.45\textwidth]{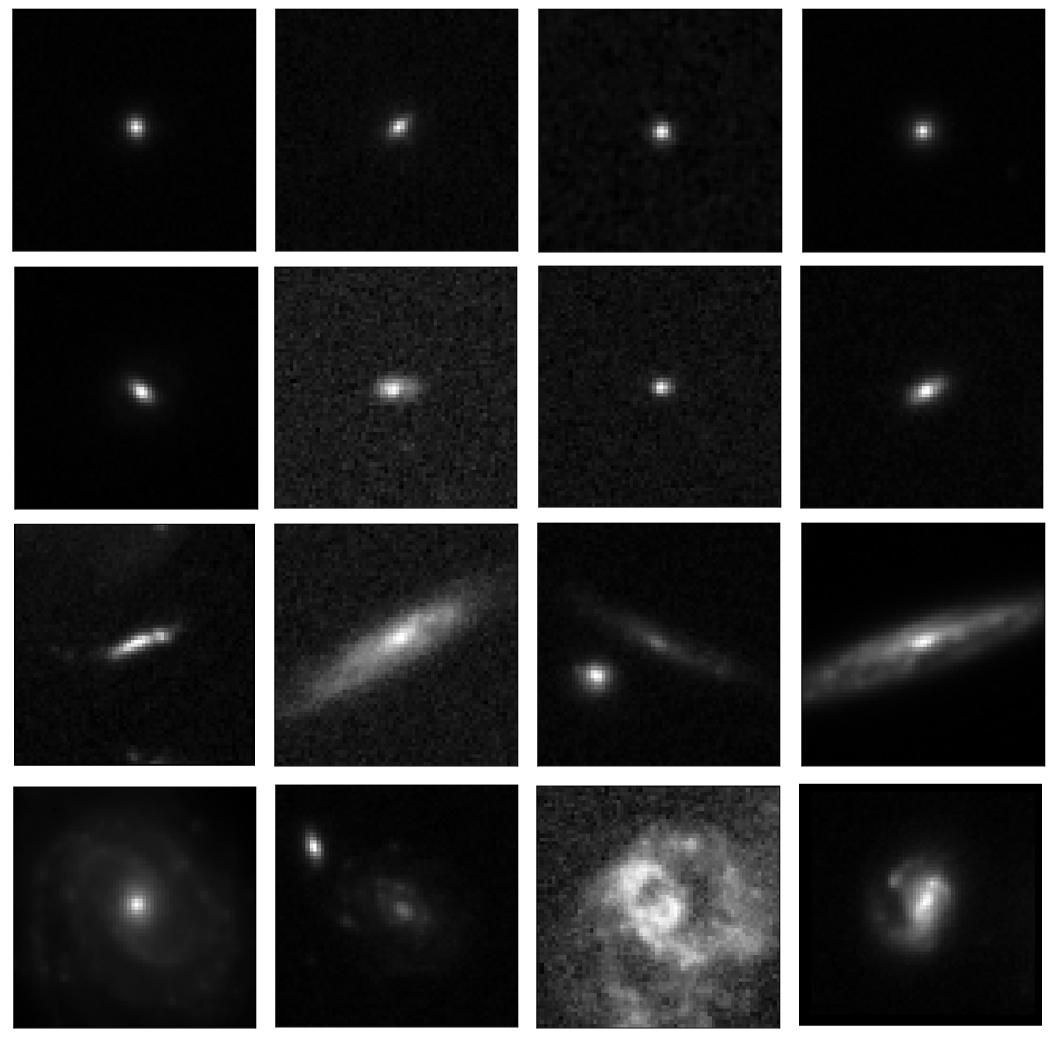}}%

    \caption{Example stamps of galaxies of different morphological types between $z=0.5$ and $z=1$. (a) and (c): TNG50 galaxies observed in the NIR (F160W) and optical (F775W) respectively. (b) and (d): same for observed CANDELS galaxies. Each row indicates a different morphological type. From top to bottom: spheroids, bulge+disk, disks, clumpy/irregular. The grey scale is arbitrary. TNG50 presents a comparable morphological diversity to observations.}%
    \label{fig:classifiedstamps}%
\end{figure*}

Forward modeling of the simulated galaxies. From the selected sample, we generate mock images in four different HST filters ($F435W$, $F606W$, $F775W$, and  $F160W$) to mimic the properties of the observational sample. We use to that purpose the radiative transfer code SKIRT~\citep{2011ApJS..196...22B}\footnote{http://www.skirt.ugent.be/root/index.html} and follow the procedure outlined in ~\cite{2019MNRAS.483.4140R} and in~\cite{marc2019MNRAS.489.1859H}. For computational reasons, dust is taken into account only if the fraction of star-forming gas is above $1\%$ of the total baryonic mass.  In order to make the mock images a more accurate representation of the observed CANDELS ones, we convolve each mock image by the point spread functions in the different filters and cameras. In addition, we add real CANDELS background noise seen in empty CANDELS regions as in the different filters. We note that the mock images contain only one subhalo, i.e. the projected galaxies in the line of sight are not rendered. In order to include background and foreground objects as in the observations, the regions in CANDELS selected to drop the simulated galaxies contain off centered objects.

\subsection{Observations: CANDELS}

Our observational sample is based on the CANDELS survey~\citep{2011ApJS..197...35G, 2011ApJS..197...36K}. We use official data products from the CANDELS collaboration: namely stellar masses, photometric redshifts as well as structural properties and central mass densities~\citep{2013ApJS..206...10G,2013ApJS..207...24G,2017ApJS..228....7N,2017ApJS..229...32S, 2019ApJS..243...22B, 2015ApJ...801...97S}. We refer the reader to the mentioned works for details on how these parameters are derived. Additionally we make use of the visual morphology catalog by \cite{Kartaltepe_2015} to train a Convolutional Neural Network as explained in the following. Our final sample is mass selected above $10^9$ solar masses in the redshift range $0.5<z<3.0$, to match the simulated sample.

\section{Estimation of galaxy morphology and structural properties of TNG50 galaxies}
\label{sec:morphs}
\subsection{Deep Learning visual like  morphologies}
We first derive global \emph{visual-like} morphologies for our TNG50 sample. To that purpose, we train a simple vanilla Convolutional Neural Network (CNN) model on the observed CANDELS images, with labels for each image being provided by the \cite{Kartaltepe_2015} catalog (hereafter K15 catalog) as already done in previous works~(e.g. \citealp{2015ApJS..221....8H,2020ApJS..248...20H}). The K15 catalog associates each galaxy with a numerical value between 0 and 1 based on the fraction of votes of the several classifiers.  In order to convert the task into a binary classification problem, we define three main morphological classes (spheroid, disk, irregular) by imposing a threshold of classification - following recommendations in K15: spheroid if $f_{sph}$ > 0.66, irregular if $f_{irr}$ > 0.66 and disk if $f_{disk}$ > 0.66.

We then train three identical vanilla CNNs to identify spheroids ($CNN_1$), disks ($CNN_2$) and irregulars ($CNN_3$) on the CANDELS observational data. We employ a similar architecture as the one used in \cite{marc2019MNRAS.489.1859H}, with 5 convolutional layers, with the filter sizes in each layer being 6, 32, 64, 128, and 128, with the dimensions of each filter being (2x2), (2x2), (2x2), (2x2) and (3x3) respectively. We add max-pooling filters of dimensions (2x2) after the second, third, and fourth convolutional layers, to reduce the spatial dimensions of the feature map created by the convolutional layers before it. Furthermore, we add batch normalization layers after each convolutional layer, before flattening the final feature map outputted by the last convolutional layer. This flattened, 1-D feature map is then connected to a fully-connected neural network with 64 neurons in its first layer, a dropout layer removing 20\% of neurons, and an output layer with a sigmoid activation.


We evaluate the model on an independent test set from CANDELS. We measure that the model classifying galaxies into being spheroidal or not has the highest Area Under the Curve (AUC) value, with a 90\% accuracy in spheroidal galaxy classification. This is then followed by the model classifying disks, and then irregulars with an AUC of 86\% and 84\% respectively. We emphasize that the main goal of this work is not to achieve a perfect classification, but to classify simulated and observed galaxies with the same methodology.

Once trained and tested on the CANDELS images, we use all three models to classify the mock TNG50 images. There is therefore no training done on simulations. The weights of the network are fixed and simply applied to the simulated dataset. Based on the outputs of the three models, we then define four composite morphological classes. The spheroid class ($output_{CNN1}>0.5$ and $output_{CNN2}<0.5$) contains then galaxies fully dominated by a bulge component, with little to no disk component to them. The disk-spheroid class ($output_{CNN1}>0.5$ and $output_{CNN2}>0.5$ and $output_{CNN3}<0.5$), contains galaxies with two clear components but are considered as bulge dominated as shown in~\cite{2015ApJS..221....8H}. The Sersic index distribution of the population peaks indeed at $n\sim2.5$. These could be therefore associated with lenticular galaxies.  Therefore, throughout the paper, we will refer to bulge dominated systems as the combination of galaxies classified as spheroids or disk+spheroids. The disk ($output_{CNN2}>0.5$, $output_{CNN1}<0.5$ and $output_{CNN3}<0.5$) class contains galaxies with a dominating disk component.  We define the irregular  class ($output_{CNN3}>0.5$) as galaxies with clumpy or disturbed surface brightness profiles. We acknowledge that this class is made of objects with diverse properties. Irregular galaxies can be for example low mass gas-rich systems, interacting galaxies or more massive clumpy galaxies. With the above definitions, galaxies for which all CNN outputs are lower than 0.5 are not included in any class and are therefore considered as unclassified. This fraction remains below $10\%$ at all redshifts and are removed from the sample when computing morphological fractions.

Figure \ref{fig:classifiedstamps} shows example stamps of CANDELS and TNG50 galaxies of different morphological types classified by the CNNs side-by-side. We see that simulated and observed galaxies of the same morphological class present similar features. It confirms that the CNN trained on the CANDELS galaxies found similar enough features in the simulated galaxies to classify them into the aforementioned morphologies. It also shows that the TNG50 simulation presents a reasonable morphological diversity in the redshift range considered here, confirming the results already reported at $z\sim0$~\citep{marc2019MNRAS.489.1859H,2019MNRAS.483.4140R}. We will quantify this further in the following sections.

\subsection{Parametric morphologies} 
\label{sec:statmorph}

In addition to visual like morphologies, we also estimate structural properties of galaxies. We employ \verb|statmorph|, an open-source Python package based on IDL code described in \cite{2004AJ....128..163L} to fit the F160W surface brightness profiles of the mock TNG50 galaxies. We refer the reader to \cite{2019MNRAS.483.4140R} for a full description of how \verb|statmorph| works. We employ the default settings recommended by \cite{2019MNRAS.483.4140R} and fit to the surface brightness distribution of every simulated galaxy a single Sersic model as done in CANDELS. This provides us with a measurement of the {\emph semi-major} half-light radius ($R_e$), the Sersic index ($n$) and the axis ratio ($b/a$) from the best Sersic model. In Appendix~\ref{app:sigma1}, we show how the Sersic light-weighted effective radius compares to the 3D half-mass size.

The parameters of the best fit model are also used to estimate the stellar mass density in the central kpc ($\Sigma_1$) following a similar approach as in observations. We compute the total luminosity in the central kpc ($L_{1}$) by integrating the best Sersic model. We then estimate a mass-to-light within 2 effective radii ($M_{2Re}/L_{2Re}$) by integrating the Sersic profile up to a radius of 2 times the effective radius combined with the 2D projected stellar mass computed as the sum of stellar particles that in projection fall within 2 times the effective radius with a $0.2$\textit{dex} Gaussian scatter (see next section for more details). The mass in the central kpc is finally obtained as: $M_{1}=M_{2Re}/L_{2Re}\times L_{1}$ which assumes a constant $M/L$ across the galaxy. In Appendix~\ref{app:sigma1}, we show how these observation-like measurements of $\Sigma_1$ (labelled throughout as $\Sigma_1$ - mocked ) differ from those obtained by directly summing up the simulated stellar mass within the central kpc ($\Sigma_1$ - simulation ).

\begin{figure*}
    \centering
    \vspace{-10 pt}
    \subcaptionbox{}{\includegraphics[width=0.3\textwidth]{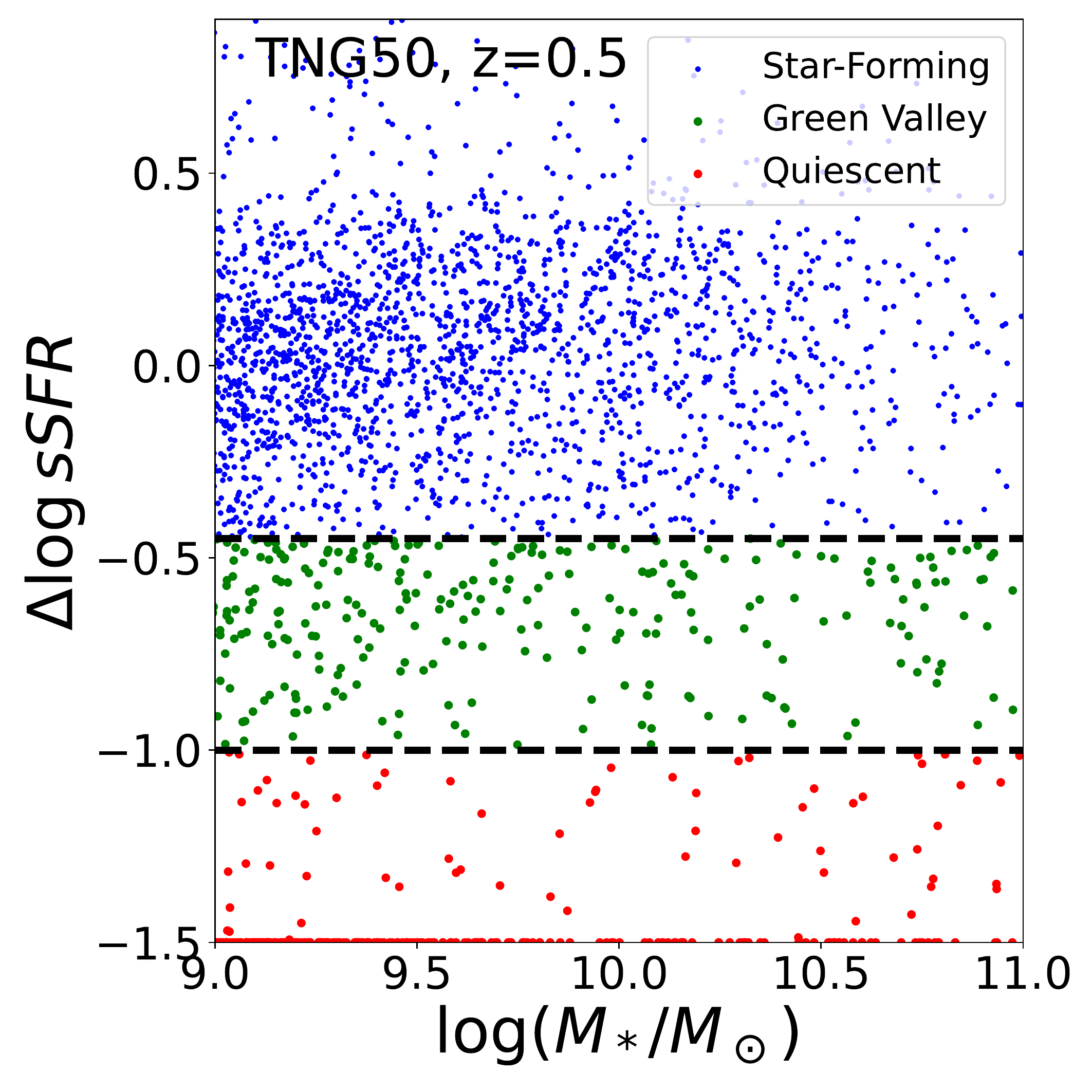}}%
     \qquad
      \subcaptionbox{}{\includegraphics[width=0.3\textwidth]{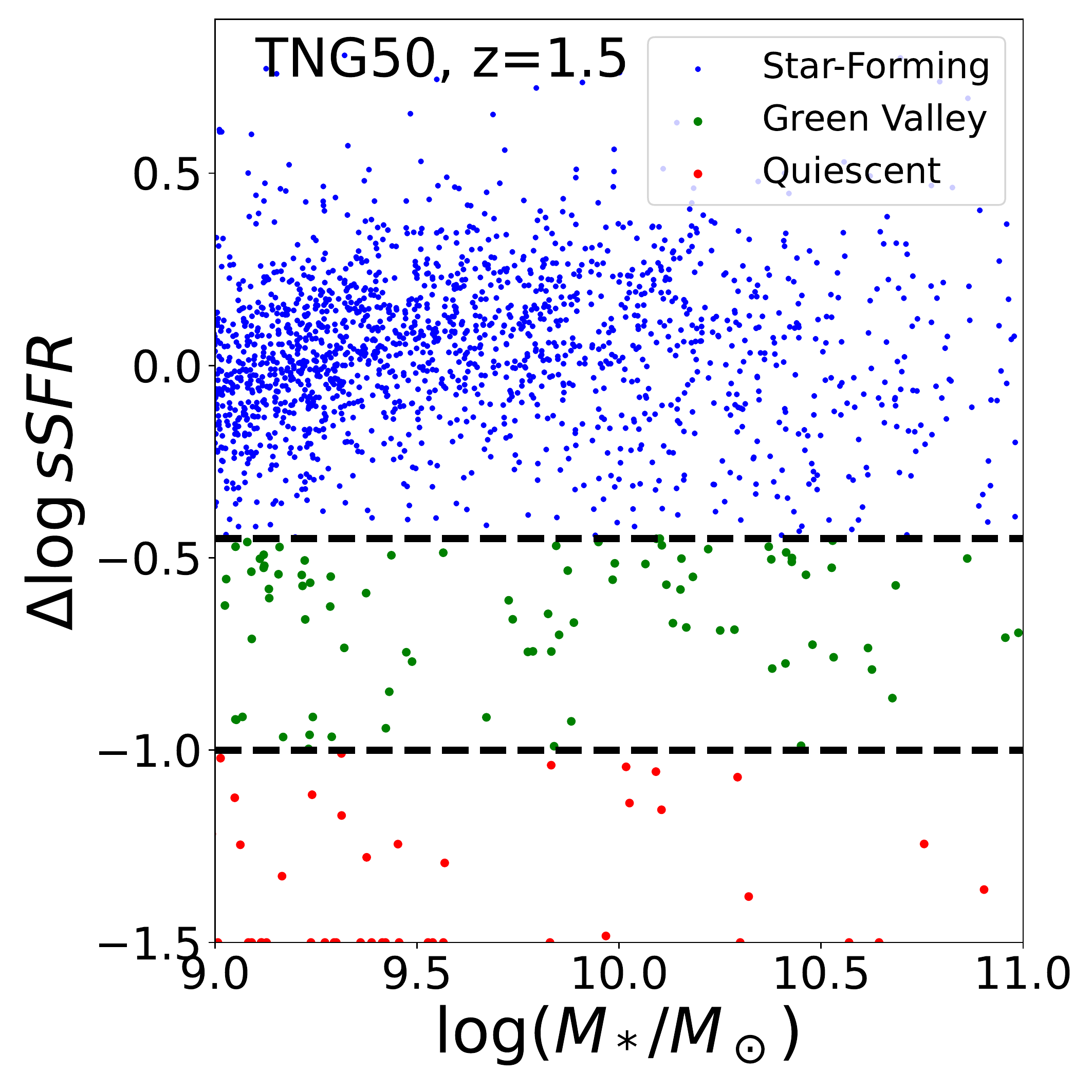}}%
     \qquad
    \subcaptionbox{}{\includegraphics[width=0.3\textwidth]{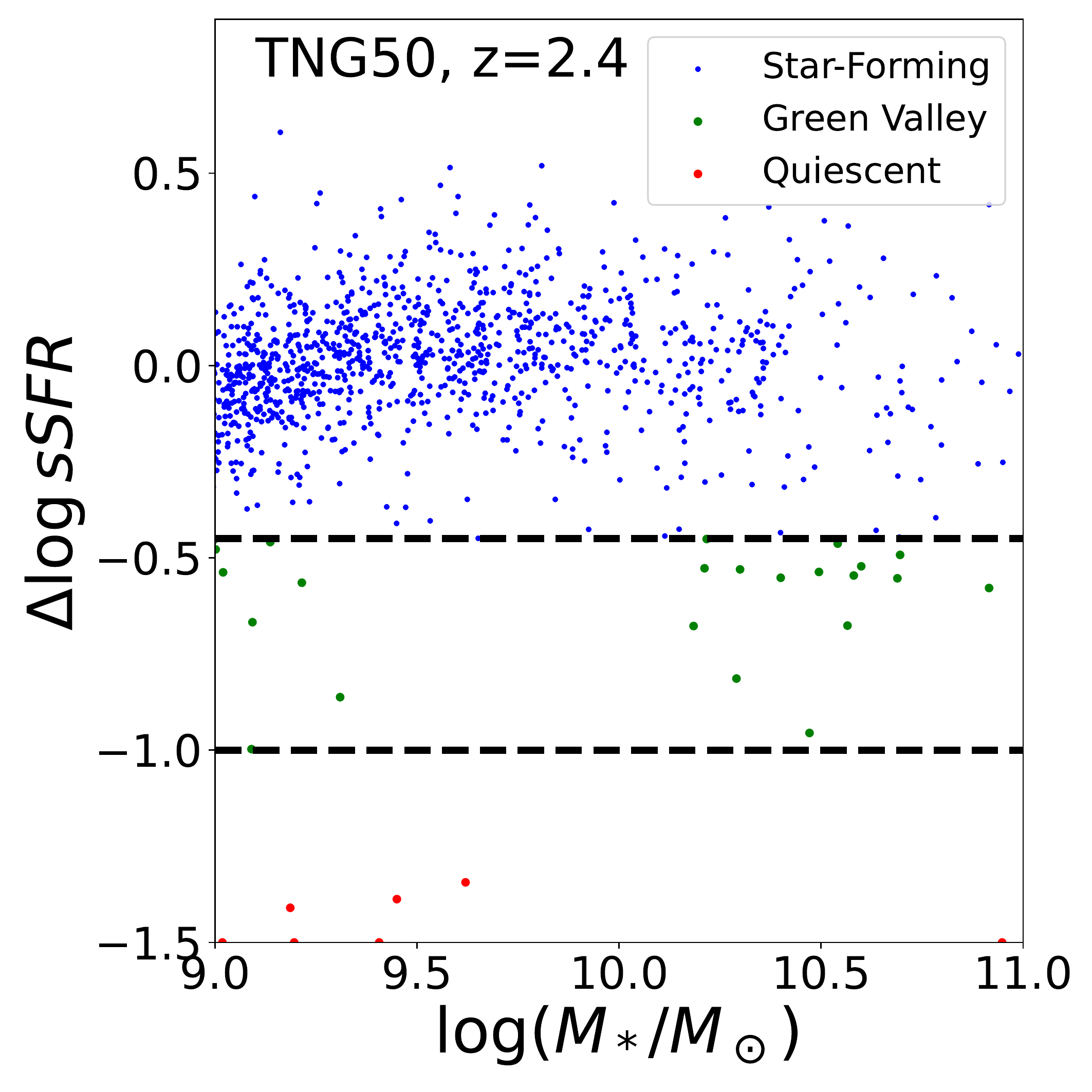}}
    \qquad
    \subcaptionbox{}{\includegraphics[width=0.3\textwidth]{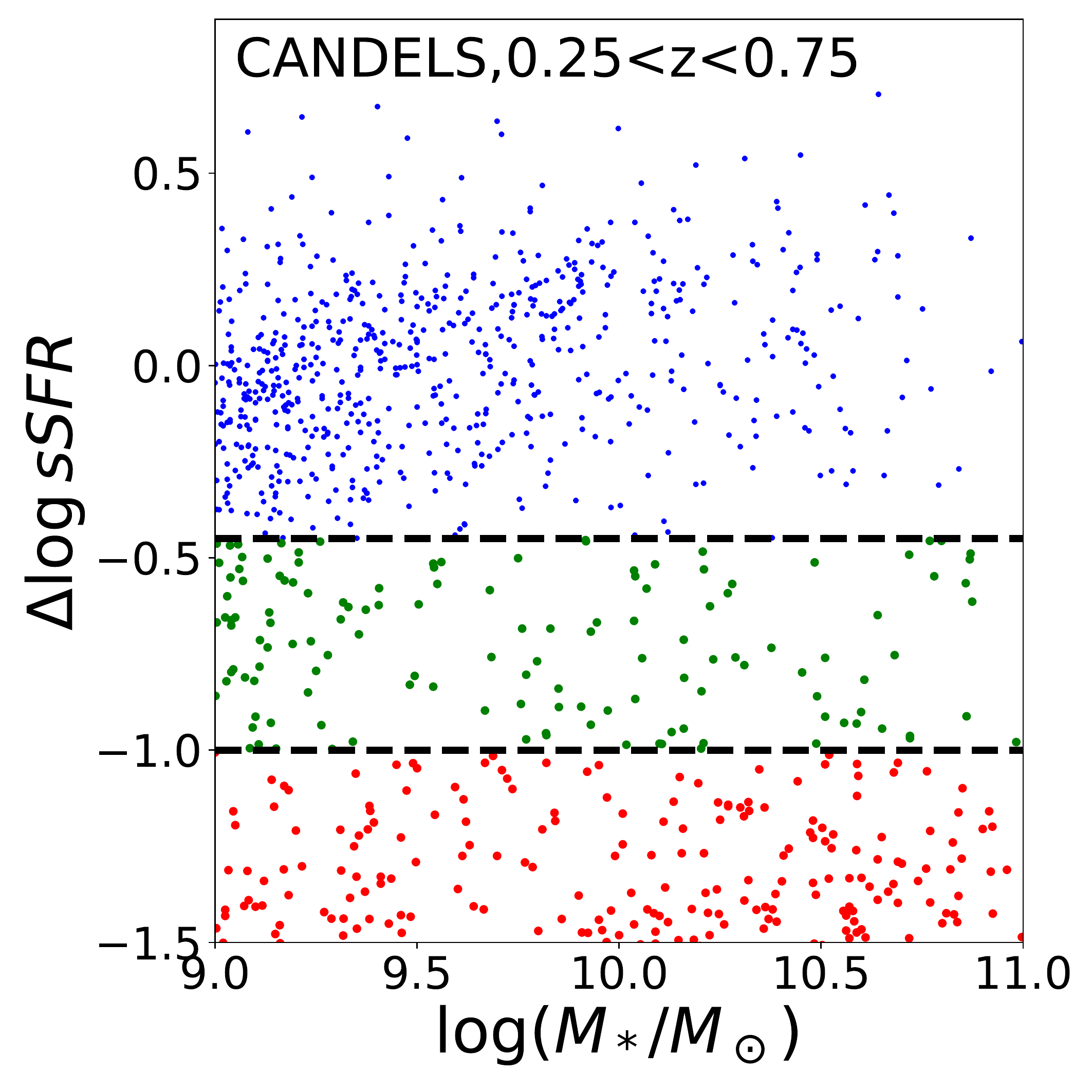}}%
    \qquad
    \subcaptionbox{}{\includegraphics[width=0.3\textwidth]{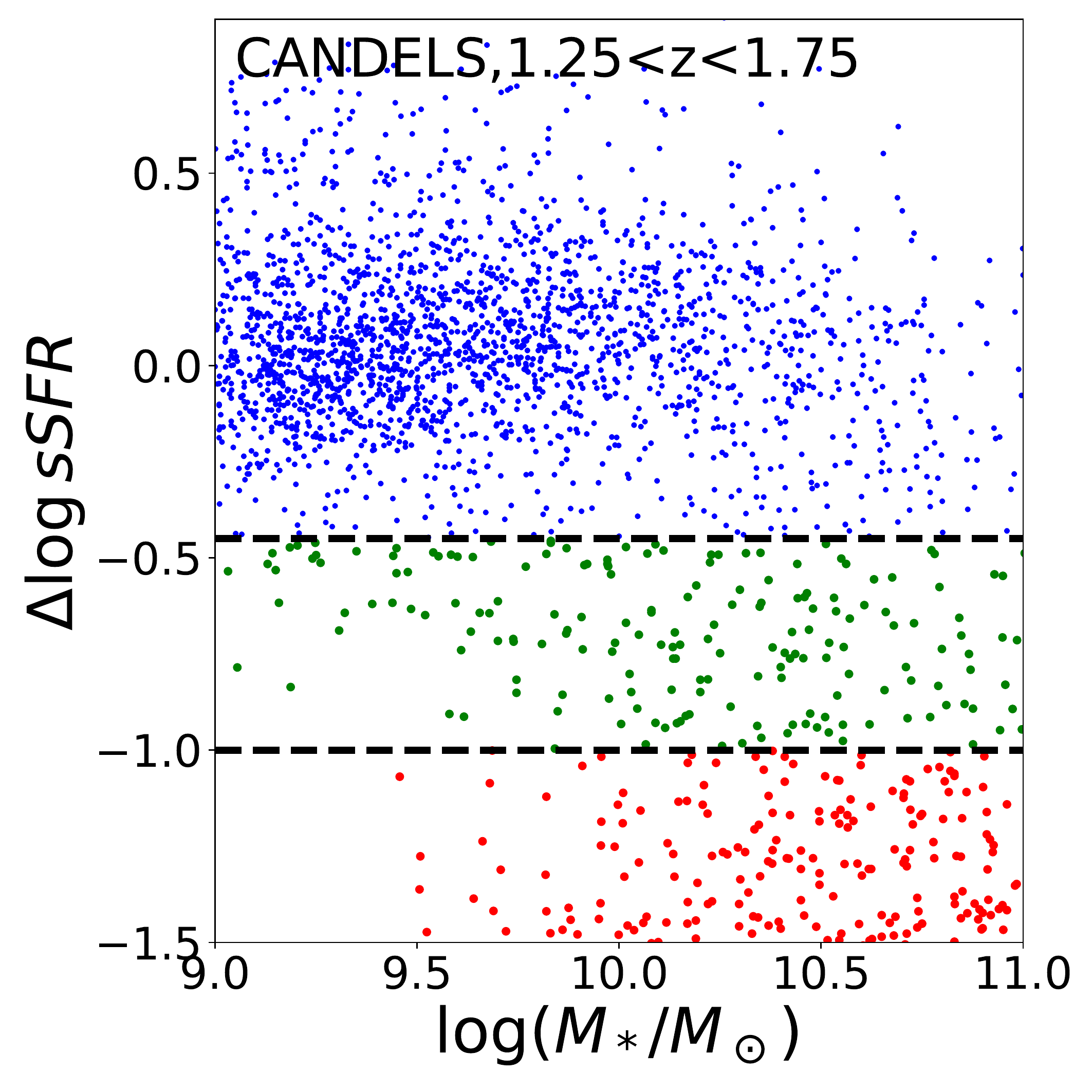}}%
    \qquad
    \subcaptionbox{}{\includegraphics[width=0.3\textwidth]{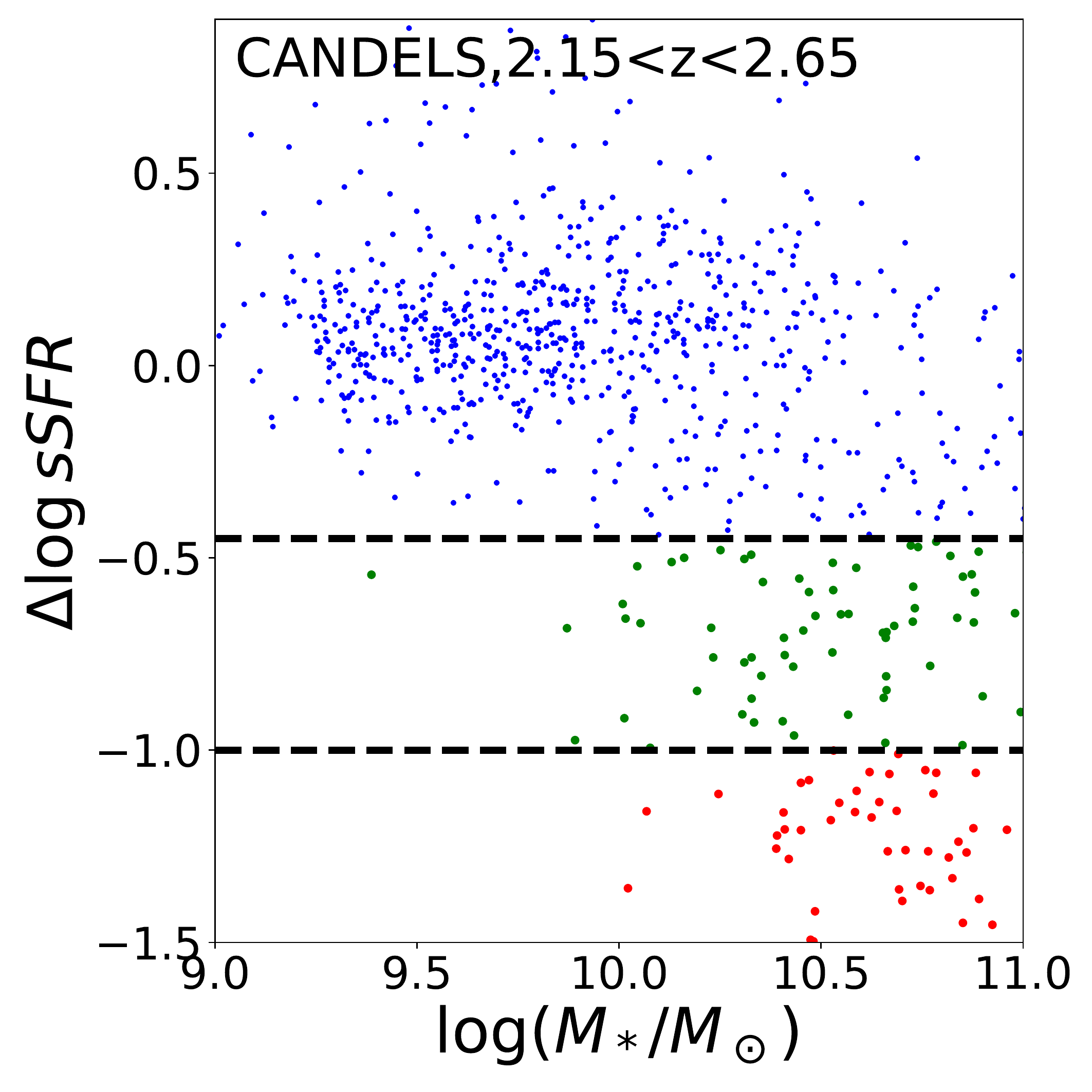}}%

      \vspace{-5 pt}
    \caption{$\Delta\log \rm{sSFR}$ as a function of stellar mass in TNG50 (top row) and CANDELS (bottom row). Each panel shows a different redshift as labelled. Blue, green and red dots show star-forming, green-valley and quiescent galaxies respectively. The horizontal dashes lines indicate the thresholds used to classify galaxies into these three classes. The TNG50 simulations tends to under predict the abundance of massive quiescent galaxies which produces a different distribution in the plane. }%
    \label{fig:SSFR-MSTAR}%
\end{figure*}

\section{Definition of Star-Forming and Quenched Galaxies}
\label{sec:qfrac}

One of the main objectives of this work is to explore the links between star formation and morphology in the TNG50 simulation as compared to observations. Therefore we first need to split galaxies between star-forming and quiescent in a consistent manner.

As already shown in previous works~\citep{2019MNRAS.485.4817D,Donnari2021,2021MNRAS.500.4004D},  a direct comparison between the SFRs derived from the simulation and the measured values in the observations is challenging given the different timescales involved. Different observational proxies are sensitive to different timescales, making an outright comparison of observed and simulated SFRs difficult. A possible solution is to apply the same SED fitting techniques used in observations to the simulated dataset. This however requires a proper simulation of the SEDs from UV to FIR. We refer the reader to~\cite{2021MNRAS.tmp.2068N} for a detailed comparison between 3D-HST and TNG50. Their main conclusion is that the observed and simulated Main Sequences agree within 0.1-0.2 dex. Since the main goal of this work is to analyze the structural properties, we have decided to follow a simpler approach.

Following~\cite{Donnari2021}, we use the SFR value from the simulation averaged over the last 200 Myrs, measured within an aperture of 2 effective radii, and add a 0.3 \textit{dex} Gaussian scatter to account for observational errors. The Specific Star Formation Rate (sSFR) is then derived by dividing the obtained value by the stellar mass of the galaxy with a 0.2 \textit{dex} scatter as well. We then fit a power law to the $\log \rm{sSFR}$-$\log M_*$ relation at every redshift slice and use the best fit to derive a distance to the Star Formation Main Sequence (SFMS) for every simulated galaxy ($\Delta\log \rm{sSFR}$). For consistency with previously published observational works, we  define a simulated galaxy as being star forming if its $\Delta \log \rm{sSFR}$ is above $-0.45$, and quenched if its $\Delta\log \rm{sSFR}$ is below $-1$. In between we consider galaxies to be in the Green Valley (GV). The same definition is applied to CANDELS galaxies using the observationaly derived sSFRs. Figure~\ref{fig:SSFR-MSTAR} compares the distribution of galaxies in the sSFR-$M_*$ plane of simulated and observed galaxies in three redshift slices. Both datasets present comparable behaviors, which confirms that our procedure is a reasonable approximation. Some differences can be appreciated though. The simulations tend to present a tighter main sequence at high redshift. Also the relative number of quiescent galaxies seems to be smaller in TNG50 at high redshift and high stellar masses especially. Massive quiescent galaxies are in fact rare at high redshift and the volume probed by the simulation is small. This, as well as different definitions can contribute to the effect ~\citep{Donnari2021} .

\begin{figure*}
    \subcaptionbox{}{\includegraphics[width=0.4\textwidth]{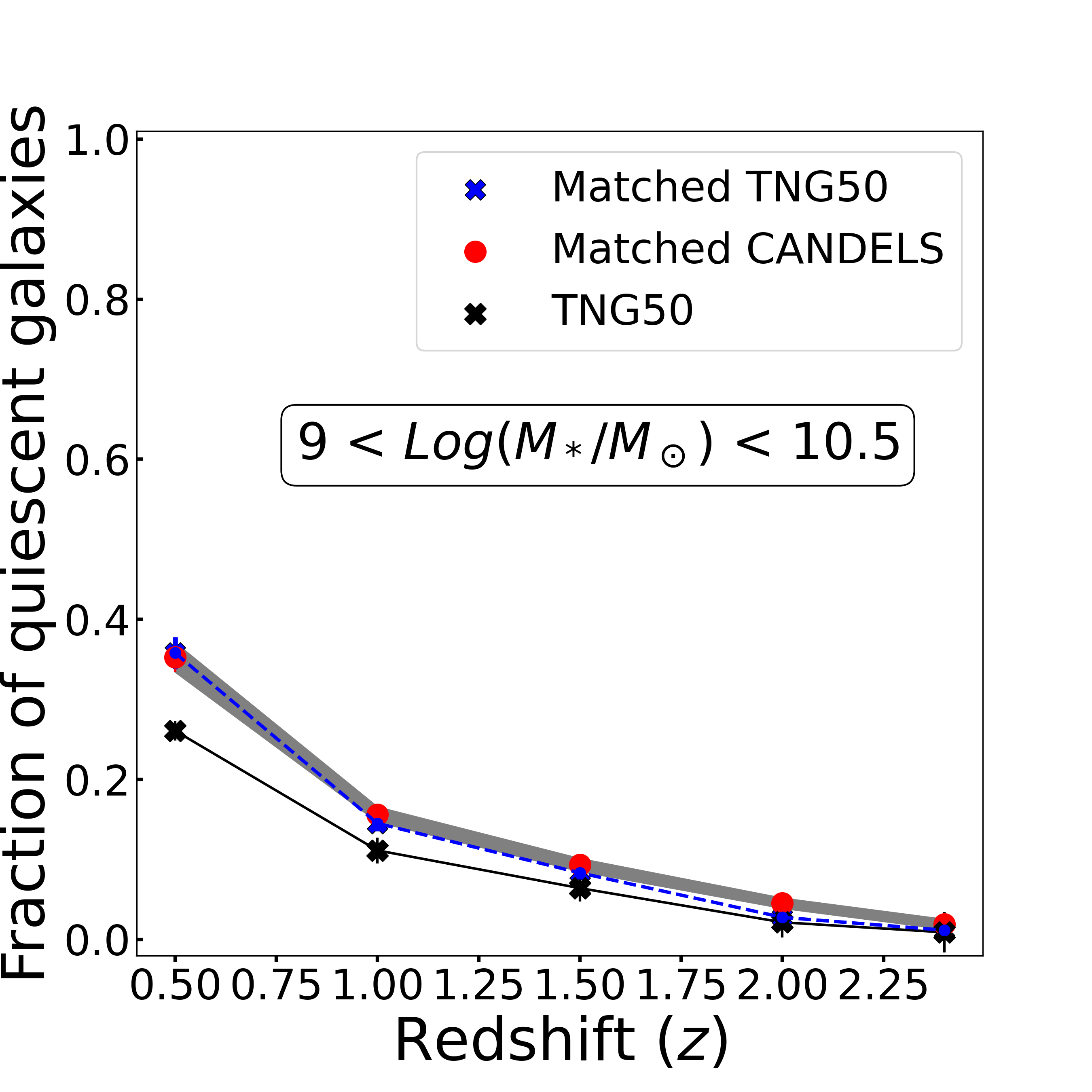}}%
\qquad
    \subcaptionbox{}{\includegraphics[width=0.4\textwidth]{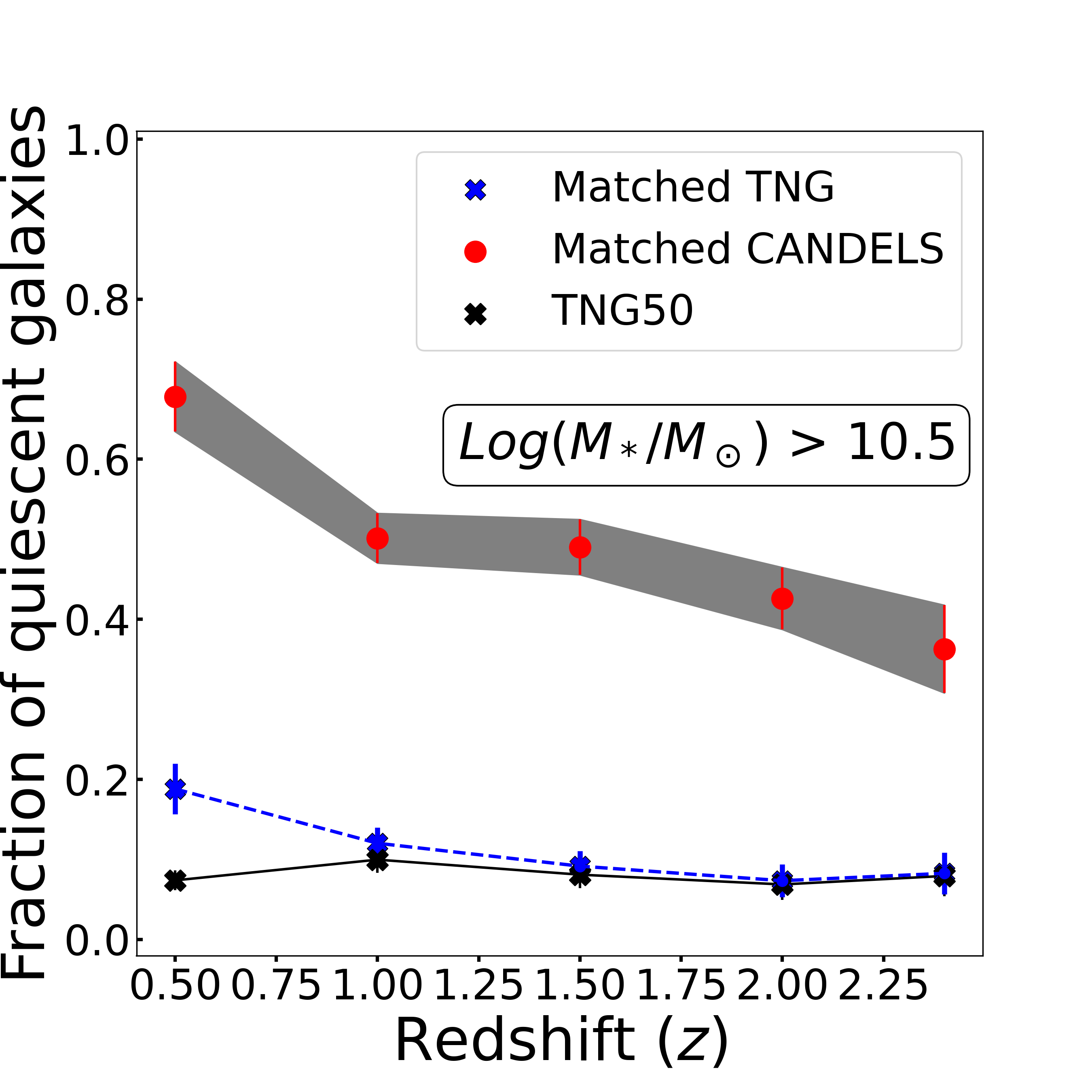}}%
\qquad

    \caption{Quiescent fraction (including GV galaxies) in the matched CANDELS (red markers), matched TNG50 (blue dashed line), and original TNG50 catalogue (black line) as a function of redshift for low mass ($9<\log M_*/M_\odot<10$, left panel) and massive galaxies ($\log M_*/M_\odot>10$, right panel.) The gray shaded area shows the variation of 10 random realizations. The error bars indicate Poisson errors. Both datasets show a similar decreasing trends with redshift. The TNG50 simulation tends to under estimate the abundance of massive quiescent galaxies at all redshifts, although the observed trends are similar.}
    \label{fig:fracquenched}
\end{figure*}

In order to further quantify the differences, we show in Figure~\ref{fig:fracquenched} the quiescent fractions in CANDELS and TNG50 in two different stellar mass bins( $9 < \log M_*/M_\odot  <10 $ and $\log M_*/M_\odot > 10.5$) using the method described in the previous paragraph. For the figure we also include Green Valley galaxies in the quiescent population. We therefore consider a galaxy to be passive if its $\Delta\log \rm{sSFR}<-0.45$. To ensure a fair comparison and quantify the impact of volume effects, we match the stellar mass and redshift distributions of the observed and simulated samples. For every galaxy in the TNG sample we draw a random galaxy in CANDELS with a similar stellar mass ($\pm 0.05 dex$) and redshift ($\pm 0.2$). In case there is no match, the TNG galaxy is not considered.This is why in figure~\ref{fig:fracquenched} the CANDELS and TNG samples are labeled as \emph{matched}. This procedure is repeated 10 times randomly. The figure shows similar trends in both datasets, suggesting that the global evolution of the abundances of passive galaxies is well captured by the simulation. However, at high stellar mass, observations tend to present a larger fraction of passive galaxies, by up to 20 percentage points than in TNG50, which confirms the trend seen in Figure~\ref{fig:SSFR-MSTAR}. The discrepancy persists even after matching the galaxy properties, so it does not seem to be directly related to the different volumes or mass distributions probed by the samples. The purpose of this work is to compare the structural and morphological properties of observed and simulated quiescent and star-forming galaxies. Therefore, the fact that the numbers do not match exactly is not a significant issue for this specific work and we refer the reader to the works by~~\cite{Donnari2021,2021MNRAS.500.4004D} for a detailed discussion on the topic: there it is shown that numerical resolution affects the outcome of TNG50, so that its quenched fraction can be up to $10-20$ percentage points lower than in TNG100 and TNG300.

\section{Evolution of galaxy morphological fractions}
\label{sec:evol}

We first investigate the evolution of visual morphological fractions in TNG50 as compared to CANDELS. We use the four morphological classes defined in Section \ref{sec:morphs} and study the evolution of their relative abundances as a function of redshift and stellar mass.\\

Figures~\ref{fig:mabundance}, ~\ref{fig:mabundance_SF} and~\ref{fig:mabundance_Q} show the fractions of the four morphological classes as a function of stellar mass in three snapshots ($z=0.5$, $z=1.5$ and $z=2.4$) for observed and simulated galaxies and divided into quiescent and star forming. We do not plot all snapshots to improve readability. The trends are however the same. In order to increase statistics, we include the Green Valley galaxies in the quiescent population in these plots as well.

\subsection{Global morphological diversity}

\begin{figure*}
\centering
\vspace{-10 pt}
\subcaptionbox{}{\includegraphics[width=0.3\textwidth]{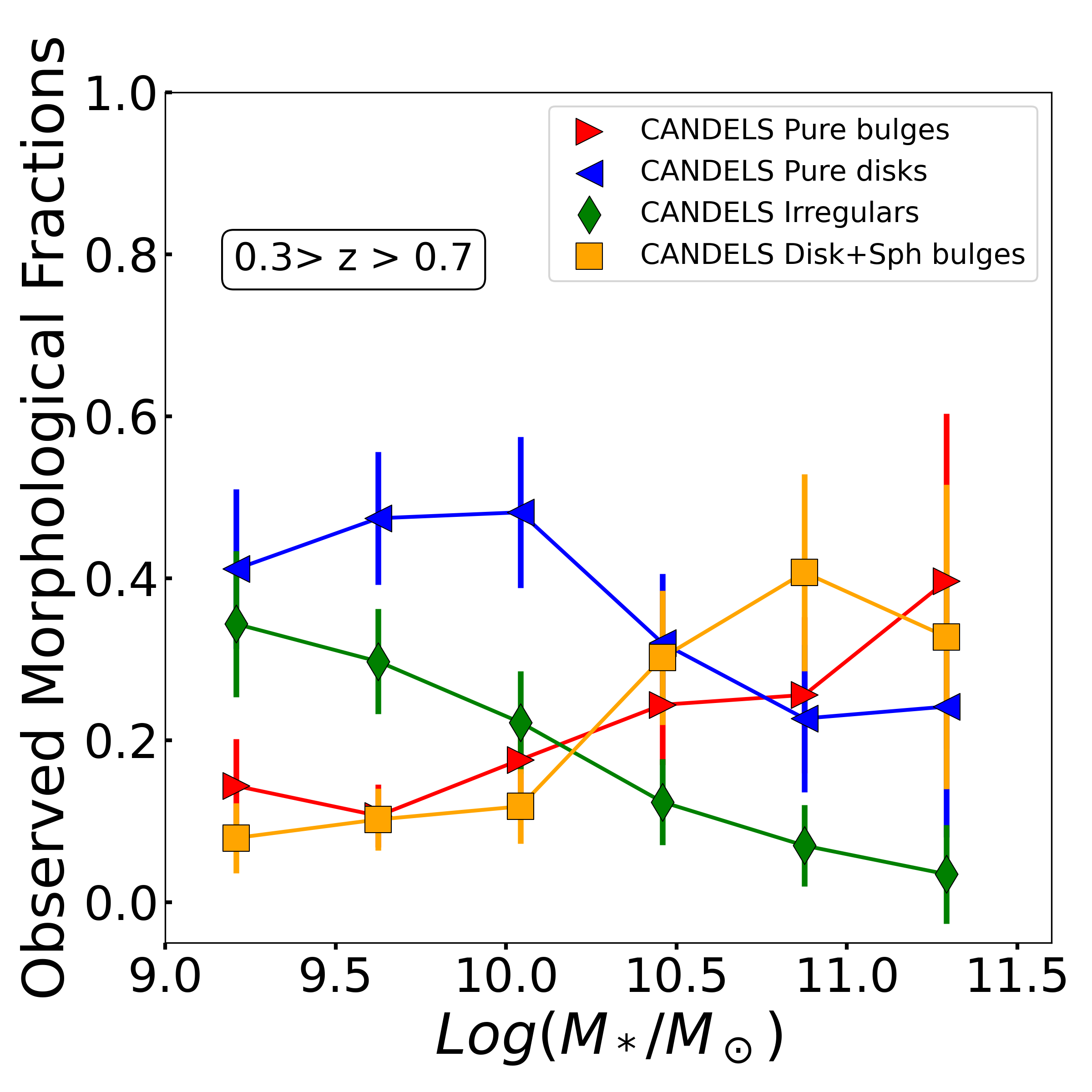}}%
\qquad
\subcaptionbox{}{\includegraphics[width=0.3\textwidth]{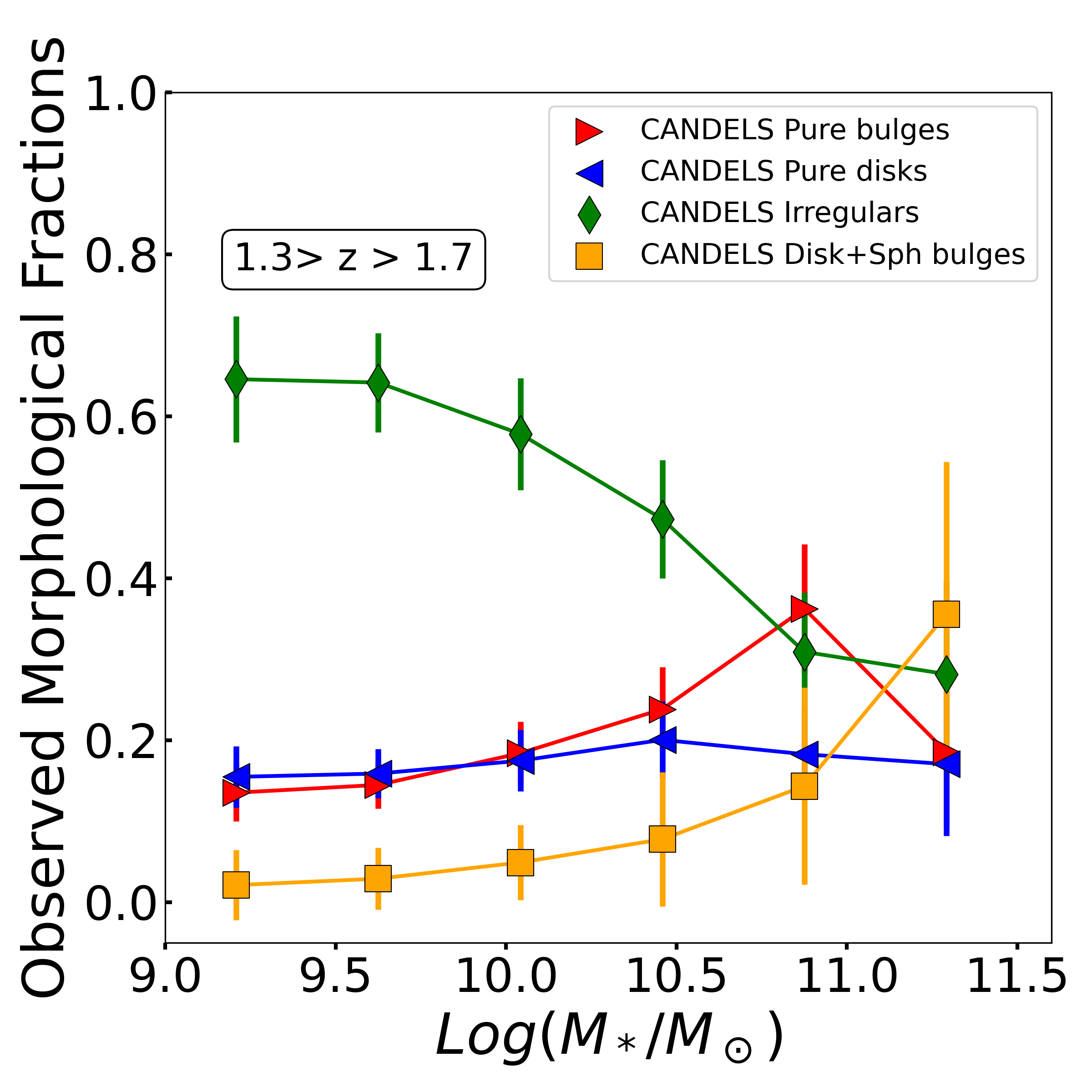}}%
\qquad
\subcaptionbox{}{\includegraphics[width=0.3\textwidth]{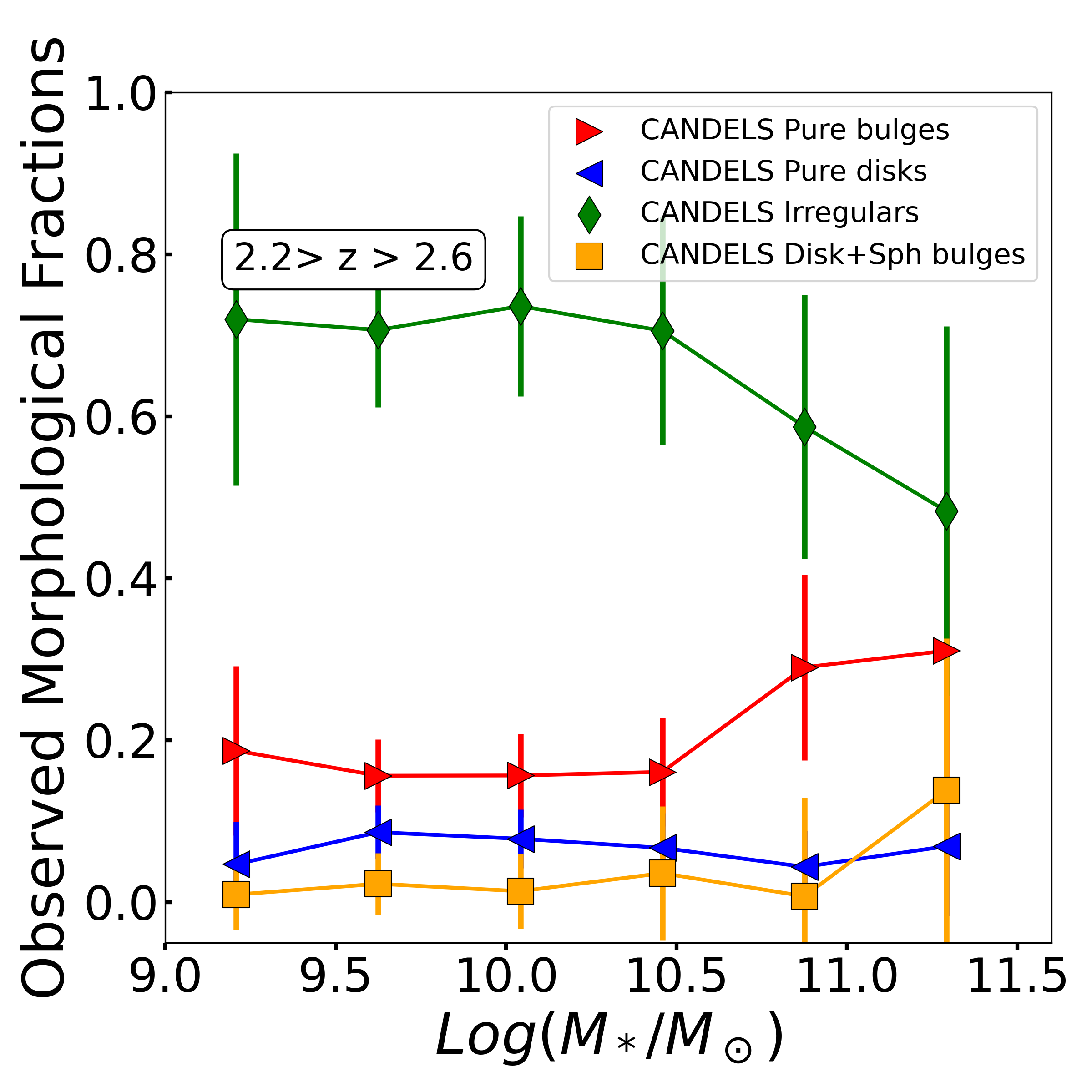}}%
\qquad
\subcaptionbox{}{\includegraphics[width=0.3\textwidth]{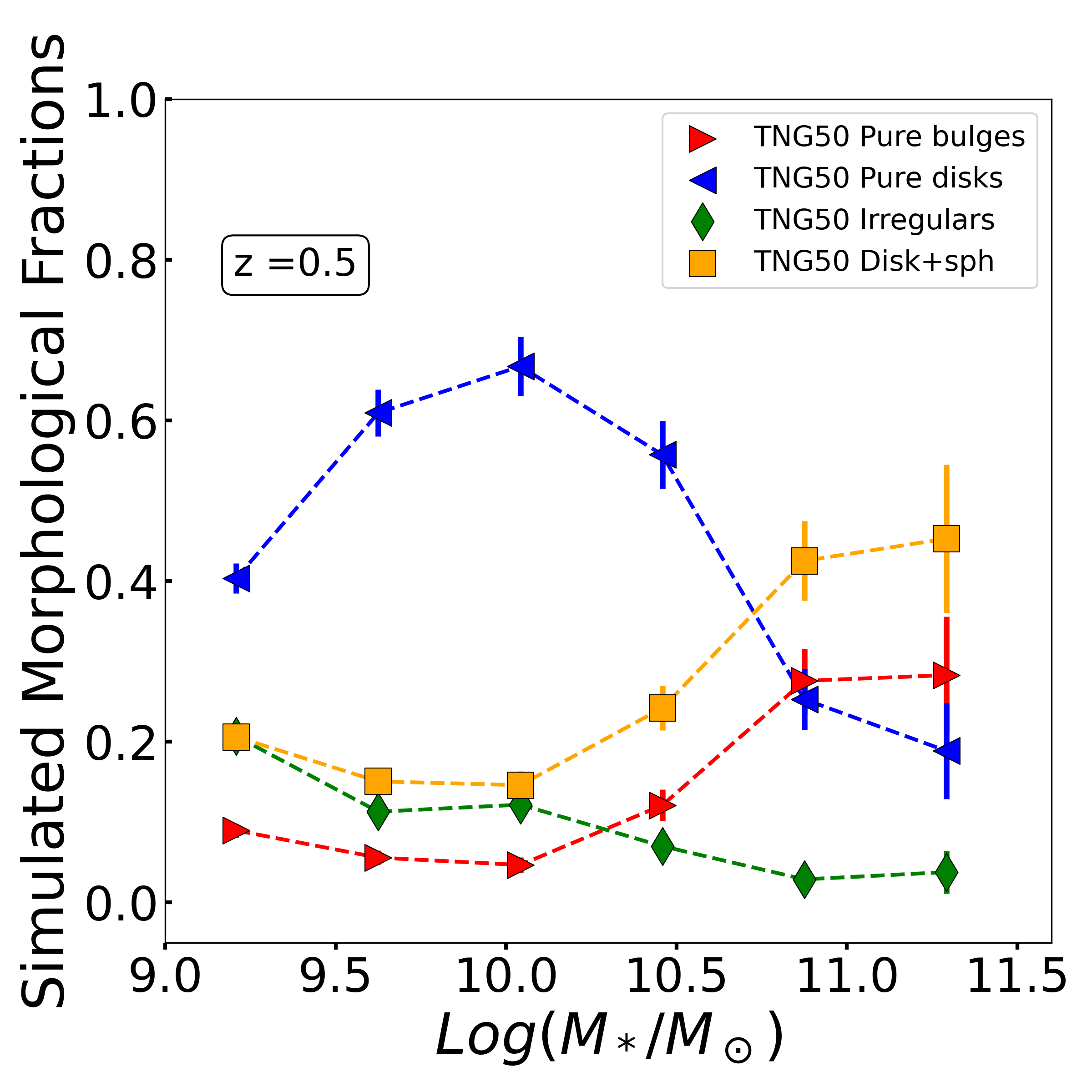}}%
\qquad
\subcaptionbox{}{\includegraphics[width=0.3\textwidth]{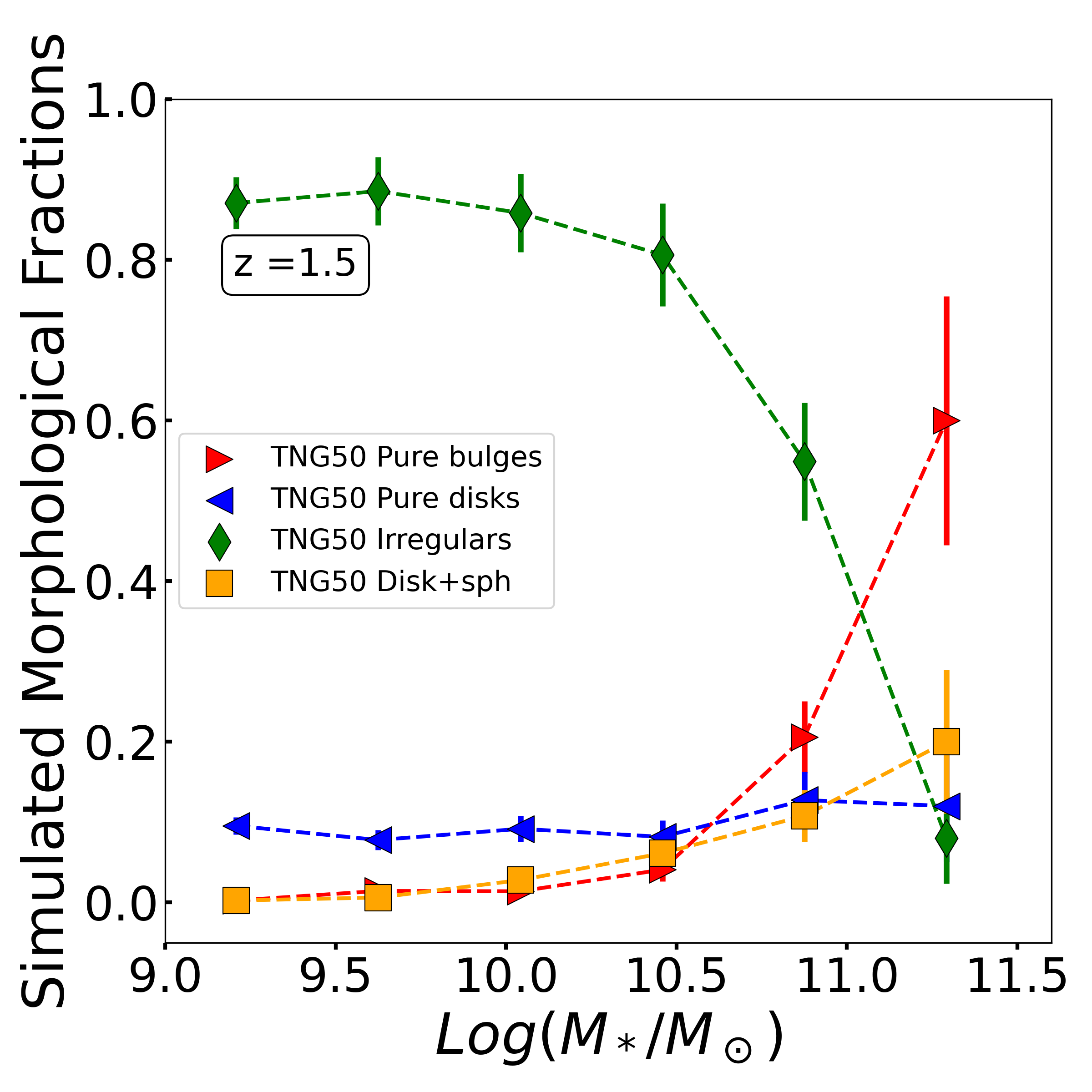}}
\qquad
\subcaptionbox{}{\includegraphics[width=0.3\textwidth]{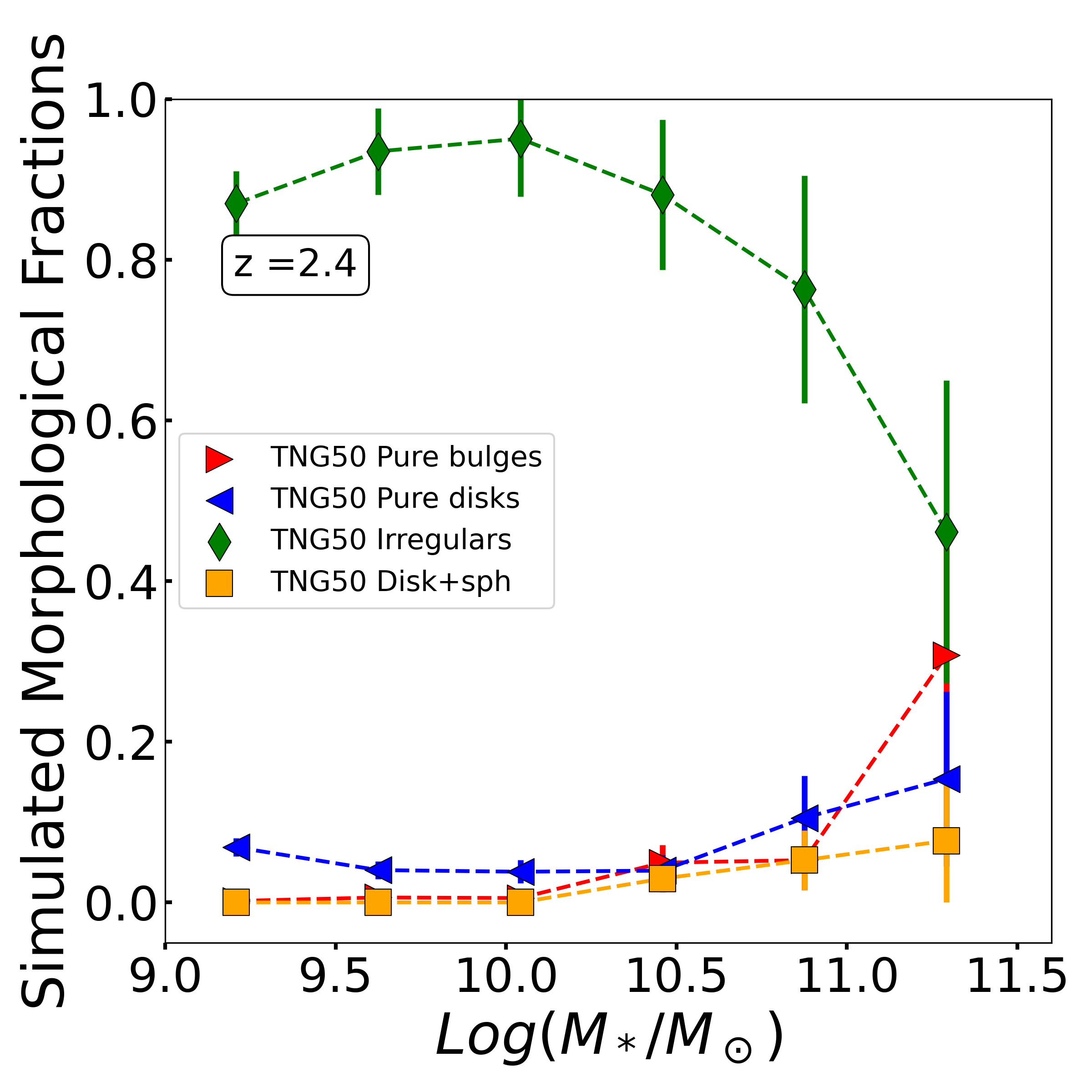}}

\vspace{-5 pt}
\caption{Morphological fractions as a function of stellar mass in three snapshots: from left to right: $z=0.5$, $z=1.5$, $z=2.4$. The top row shows observations from CANDELS. The bottom row shows the simulated sample from TNG50. The different lines indicate different morphological types, as labelled. The general observed trends of morphological abundances as a function of mass and redshift are well reproduced by the TNG50 simulation.}%
\label{fig:mabundance}%
\end{figure*}

As it can be appreciated in Figure~\ref{fig:mabundance}, trends previously reported in other observational samples can also be seen here. Namely that the galaxy population at $z\sim2$ is dominated by irregular/clumpy systems, which are gradually replaced by disks~(e.g.~\citealp{2015ApJ...800...39G,2016MNRAS.462.4495H,2017ApJ...843...46S}) at $z<1$. We also find a well known dependence of morphology with stellar mass. Bulge dominated systems tend to be more abundant at the high mass end. They represent $\sim60\%$ of massive galaxies at $z\sim0.5$. Interestingly, we find that TNG50 overall reproduces the main observational trends. This is a remarkable achievement of the simulation since it was not calibrated in any way to reproduce the galaxy morphological diversity, not at low nor at high redshift.

However, some differences in the absolute numbers between CANDELS and TNG50 are in place. The fraction of simulated galaxies defined as a pure bulge is consistently lower than its observed counterparts especially at low masses. The simulation tends therefore to over predict the fraction of irregular galaxies at high redshifts and the fraction of pure disks at later epochs. The physical mechanisms that form spheroids at low and high stellar masses might be different. For example, environment is thought to play a more dominant role at the low mass end (e.g.~\citealp{2010ApJ...721..193P}). The difference we see might therefore be a consequence of volume effects (as there are only a couple of cluster-like dense environments in TNG50) or that the formation channel of low mass spheroids is not well captured in the simulation.

\subsection{Morphologies of quiescent and star-forming galaxies}

\begin{figure*}
    \centering
    \vspace{-12 pt}
    \subcaptionbox{}{\includegraphics[width=0.3\textwidth]{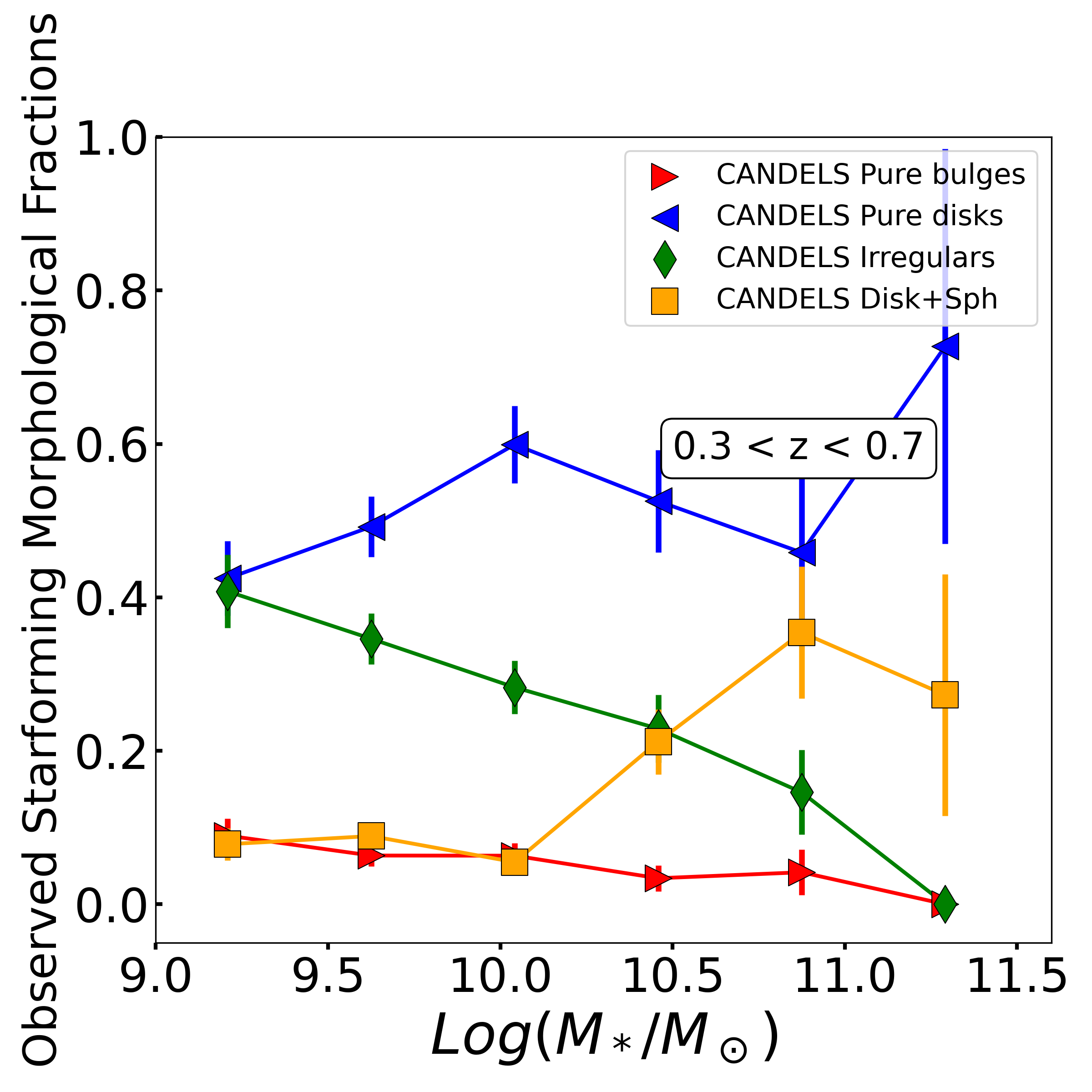}}%
     \qquad
      \subcaptionbox{}{\includegraphics[width=0.3\textwidth]{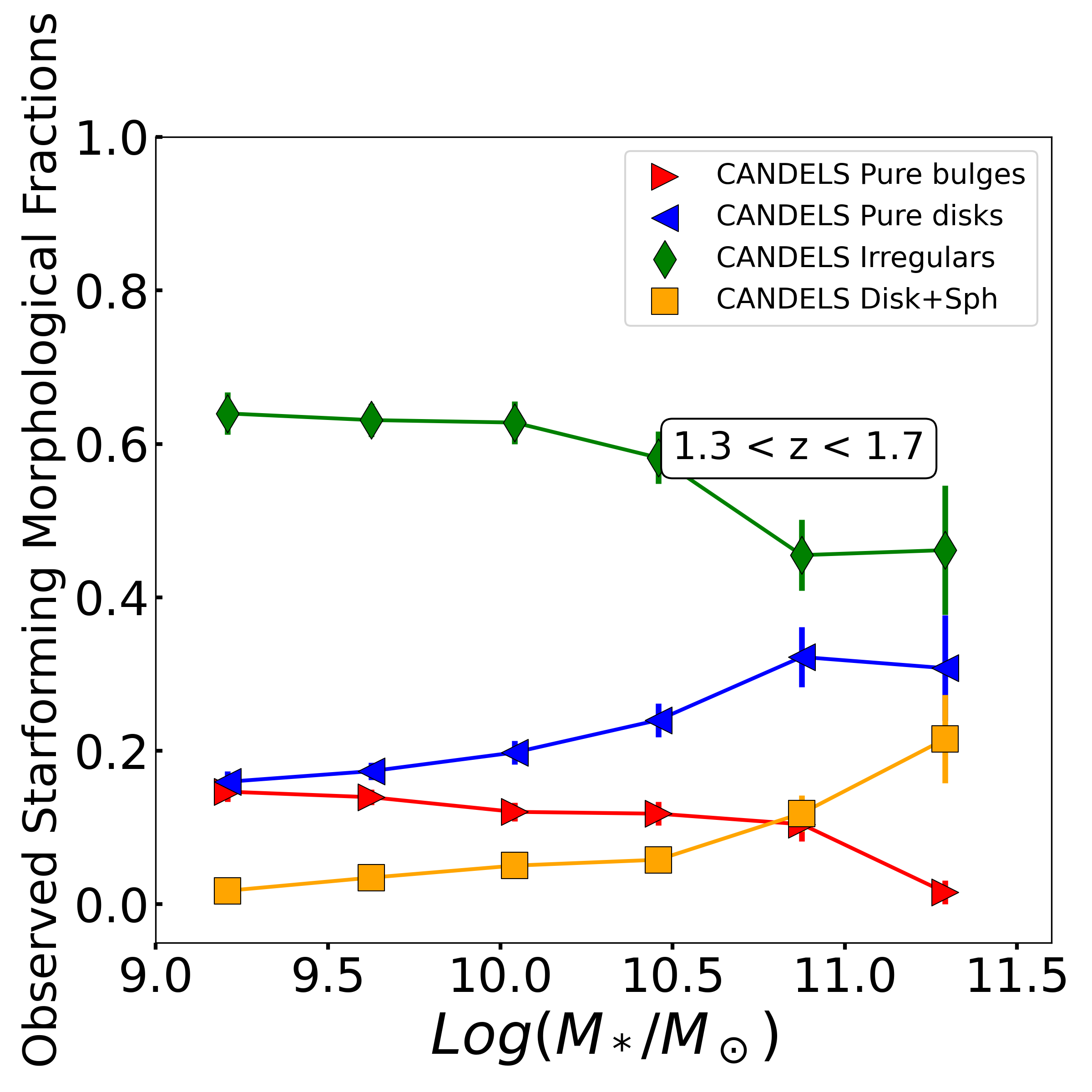}}%
     \qquad
    \subcaptionbox{}{\includegraphics[width=0.3\textwidth]{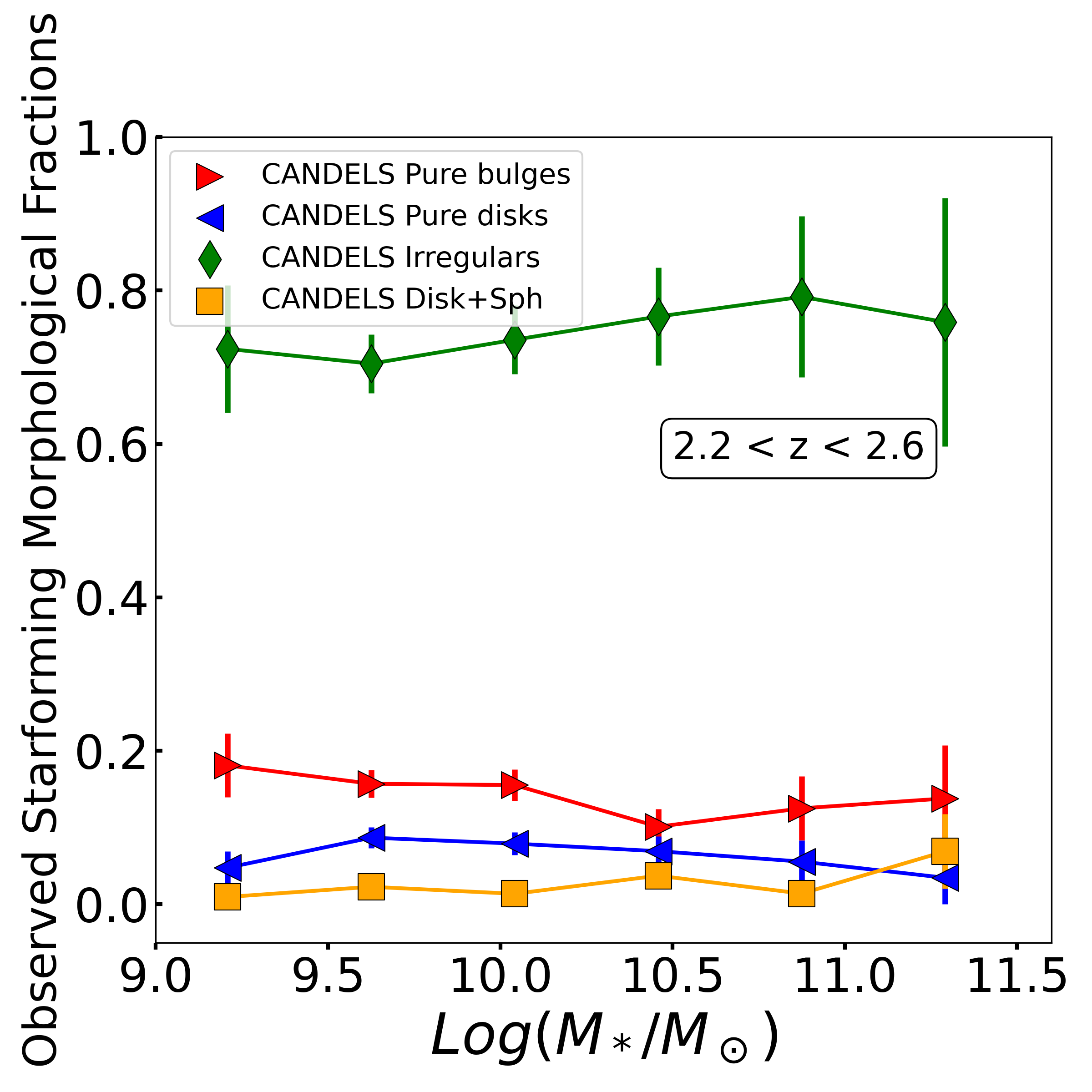}}%
    \qquad
    \subcaptionbox{}{\includegraphics[width=0.3\textwidth]{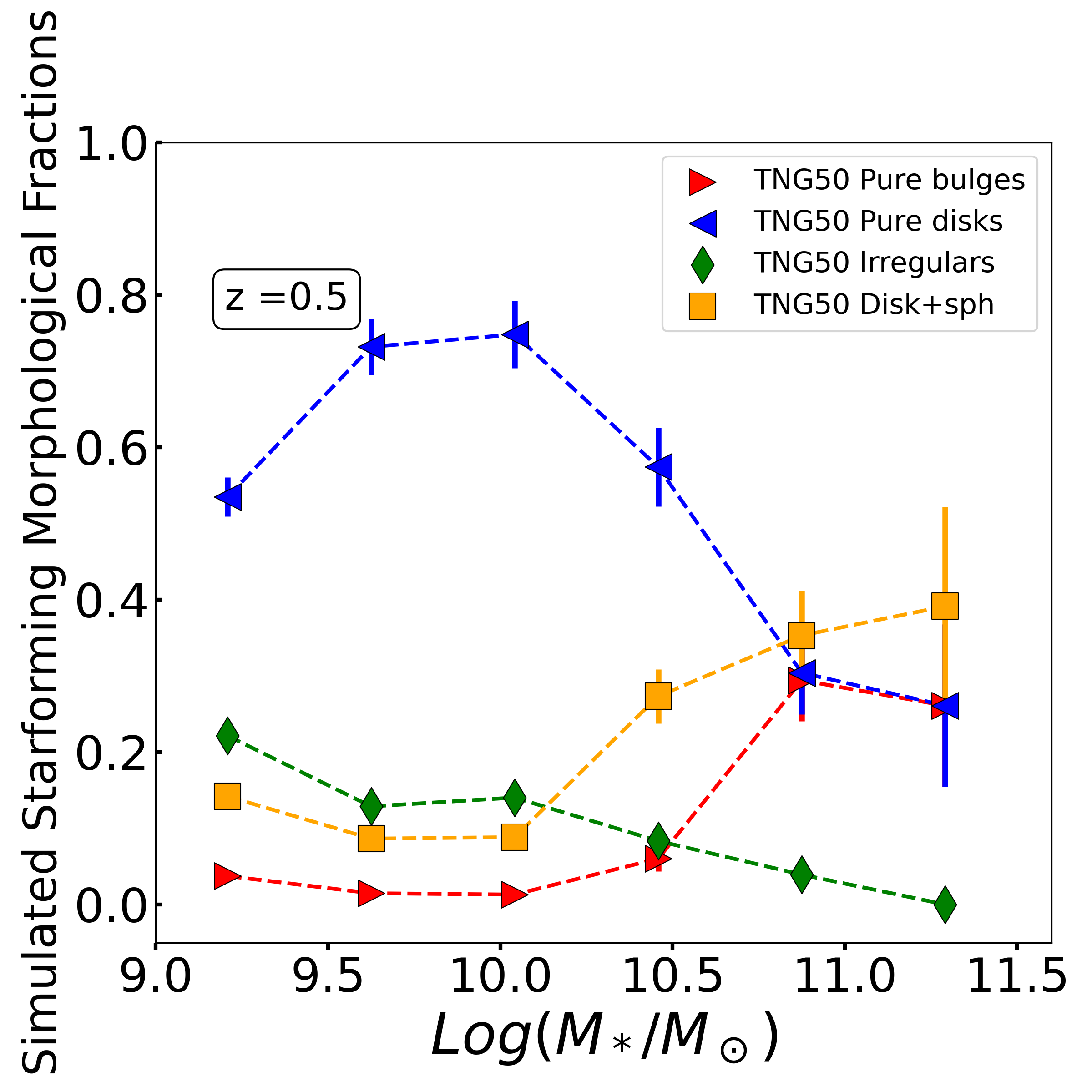}}%
    \qquad
    \subcaptionbox{}{\includegraphics[width=0.3\textwidth]{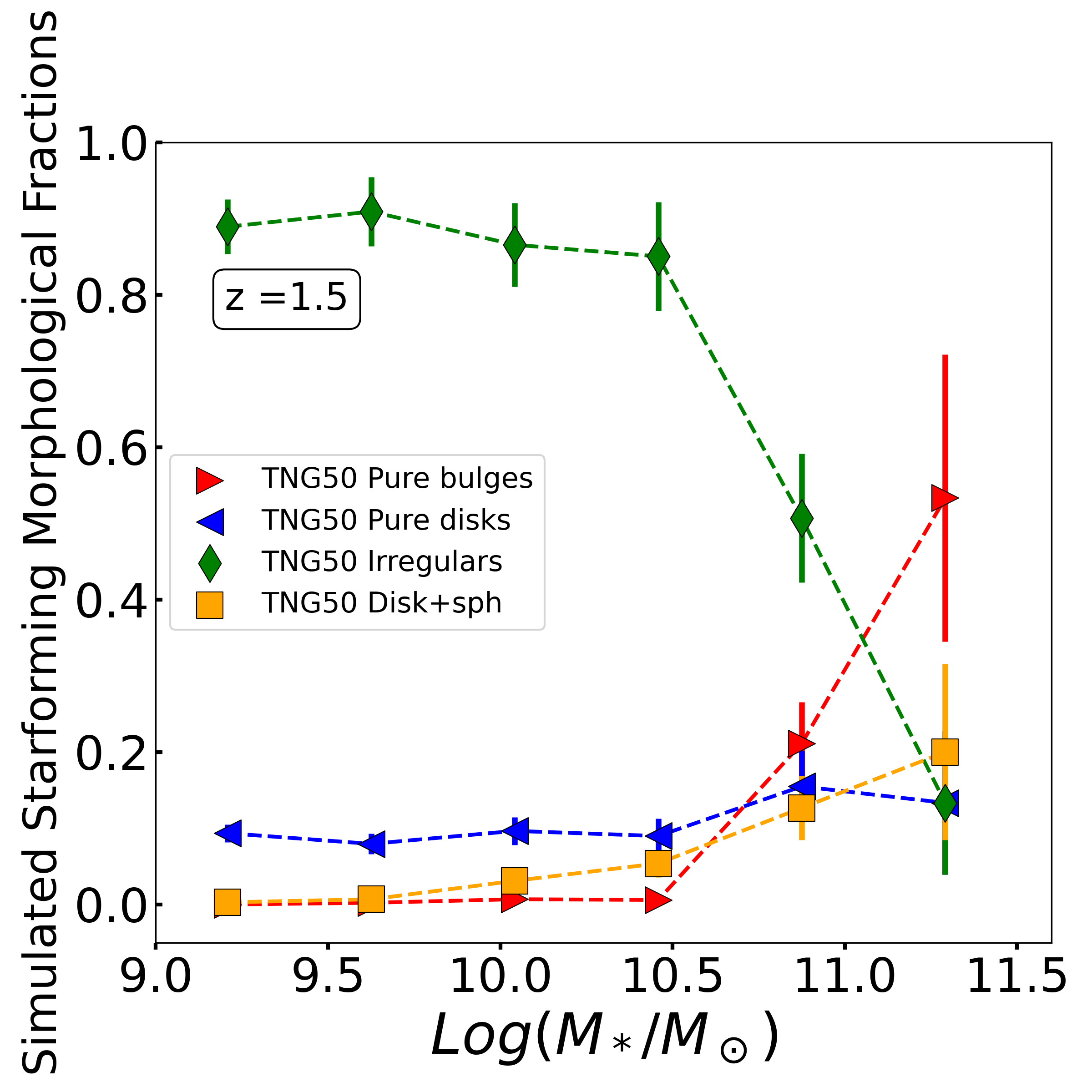}}
    \qquad
    \subcaptionbox{}{\includegraphics[width=0.3\textwidth]{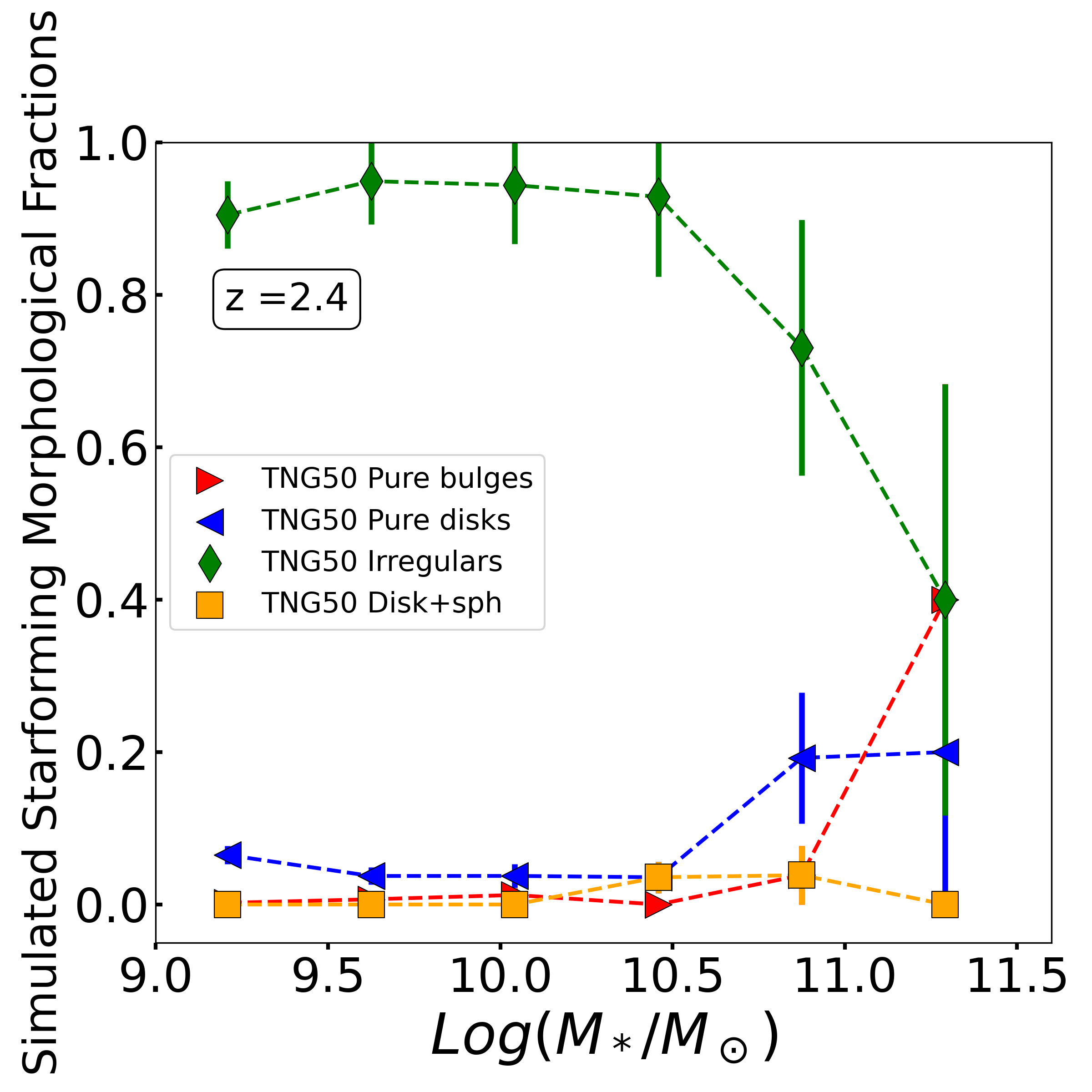}}
    
        \vspace{-5 pt}
    \caption{Morphological fractions of {\it star-forming} galaxies as a function of stellar mass in three snapshots ($z=0.5$, $1.5$, and $2.4$, from left to right). The top row shows observations from CANDELS, the bottom one results from the TNG50 simulation. The different lines indicate different morphological types, as labelled. The observed general trends of morphological abundances of star-forming galaxies as a function of mass and redshift are reasonably well reproduced by TNG50.}%
    \label{fig:mabundance_SF}%
\end{figure*}

\begin{figure*}
    \centering
    \vspace{-12 pt}
    
    \subcaptionbox{}{\includegraphics[width=0.3\textwidth]{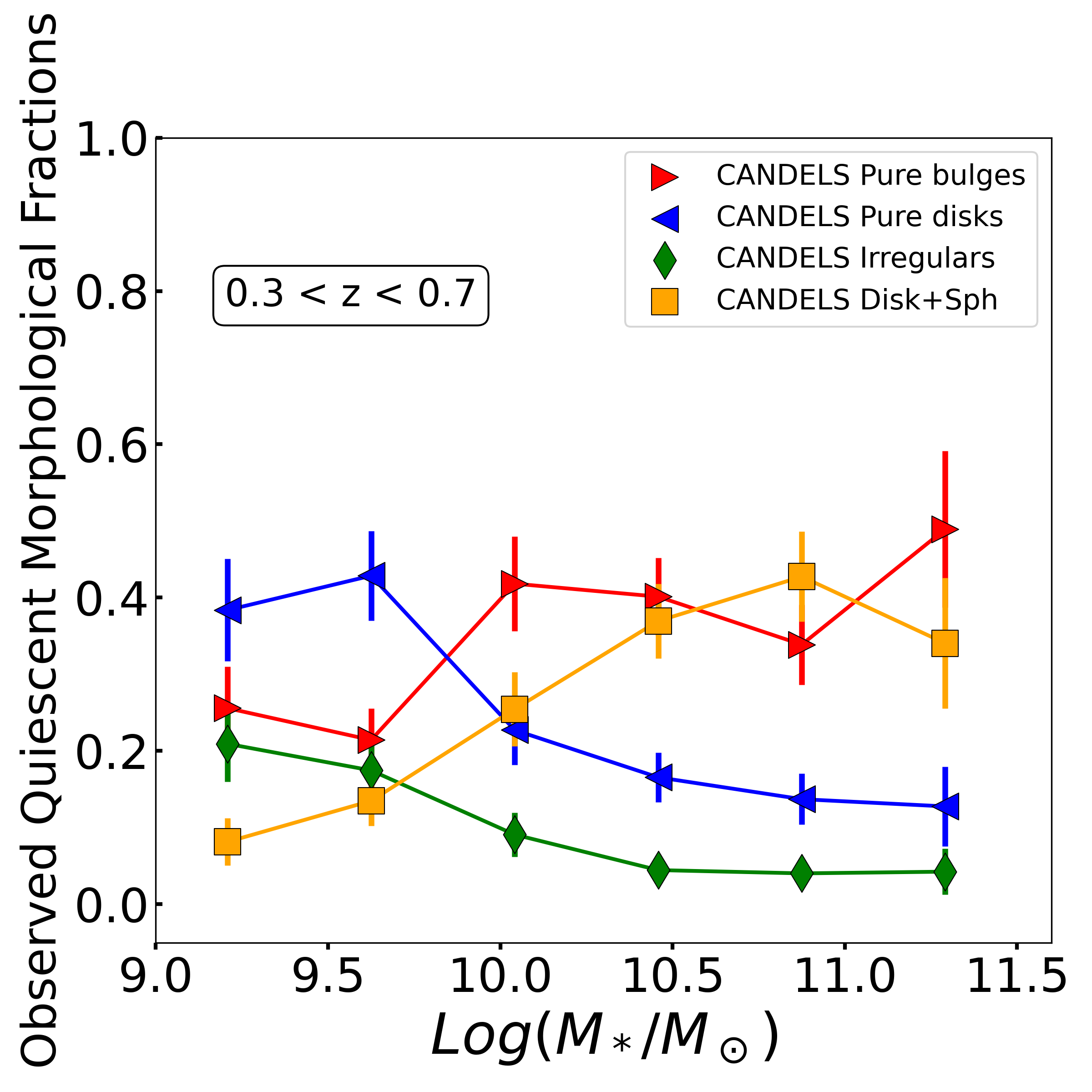}}%
     \qquad
      \subcaptionbox{}{\includegraphics[width=0.3\textwidth]{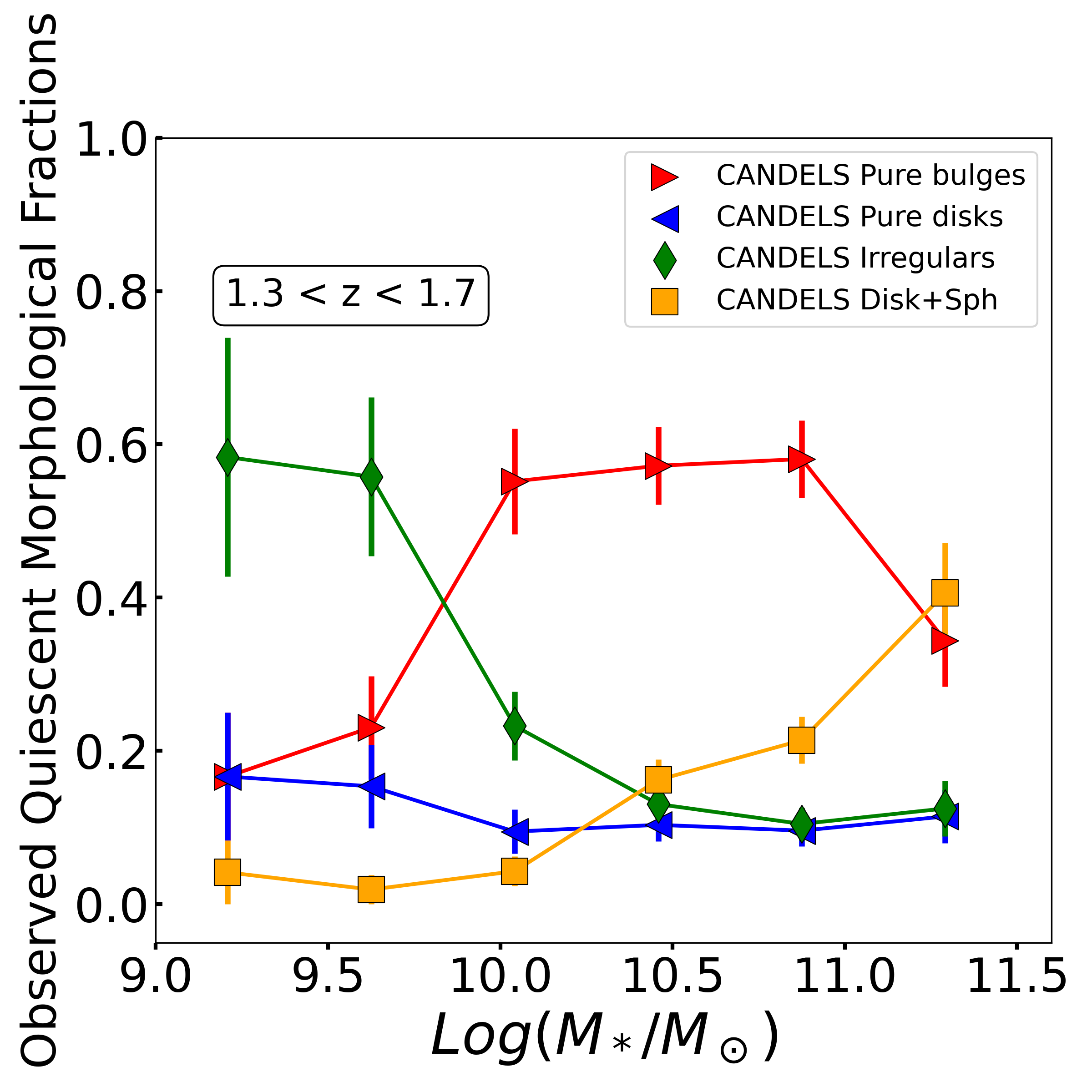}}%
     \qquad
    \subcaptionbox{}{\includegraphics[width=0.3\textwidth]{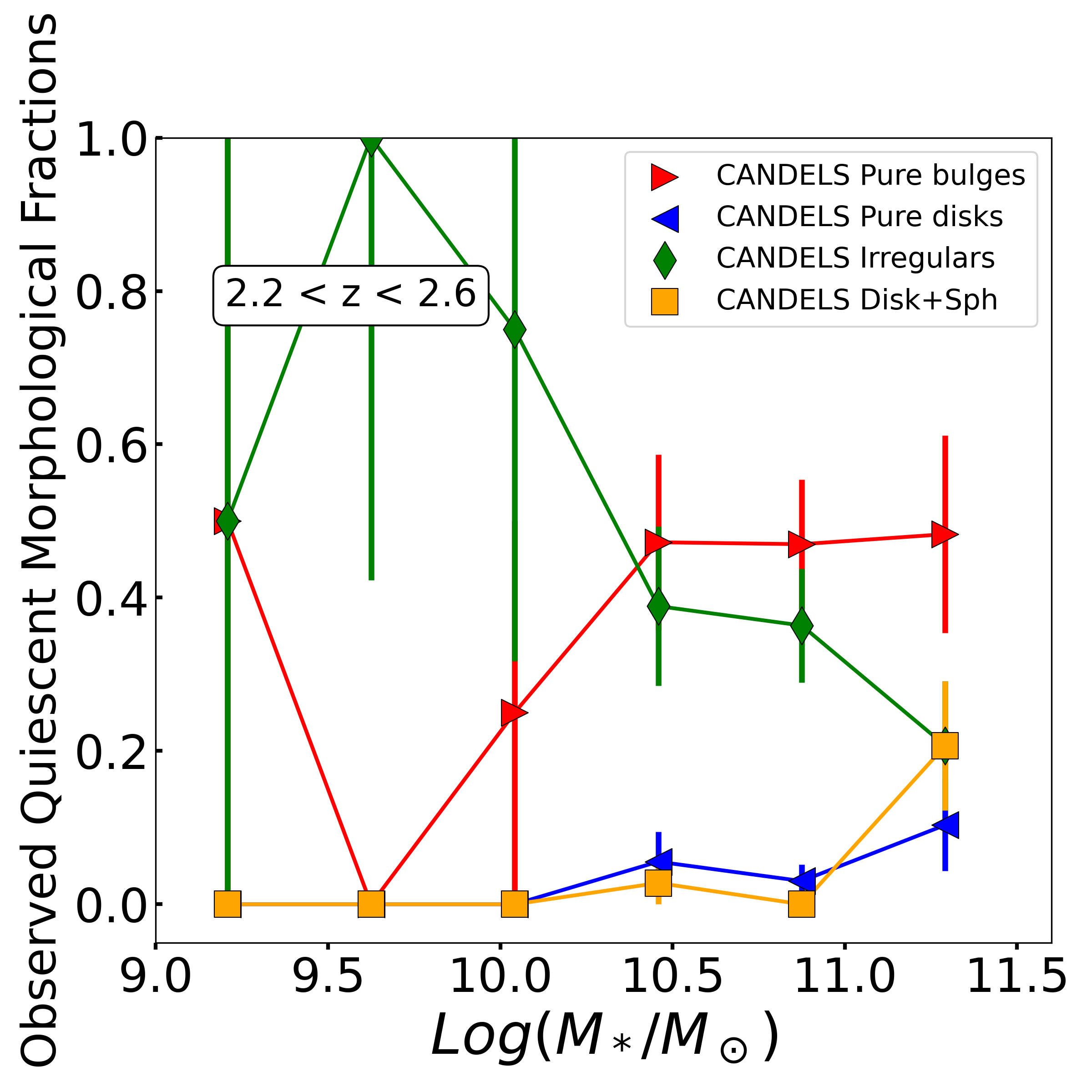}}%
    \qquad
    \subcaptionbox{}{\includegraphics[width=0.3\textwidth]{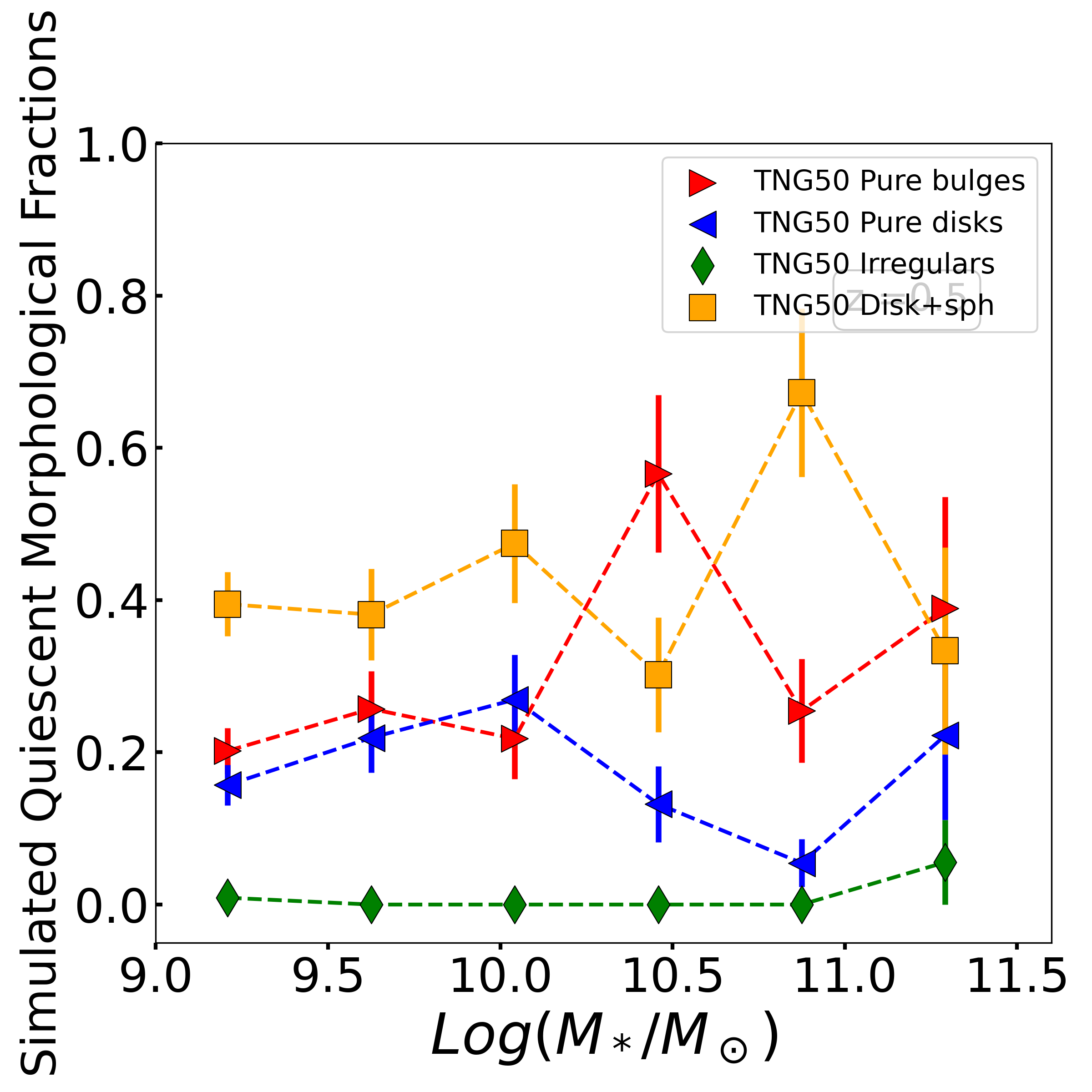}}%
    \qquad
    \subcaptionbox{}{\includegraphics[width=0.3\textwidth]{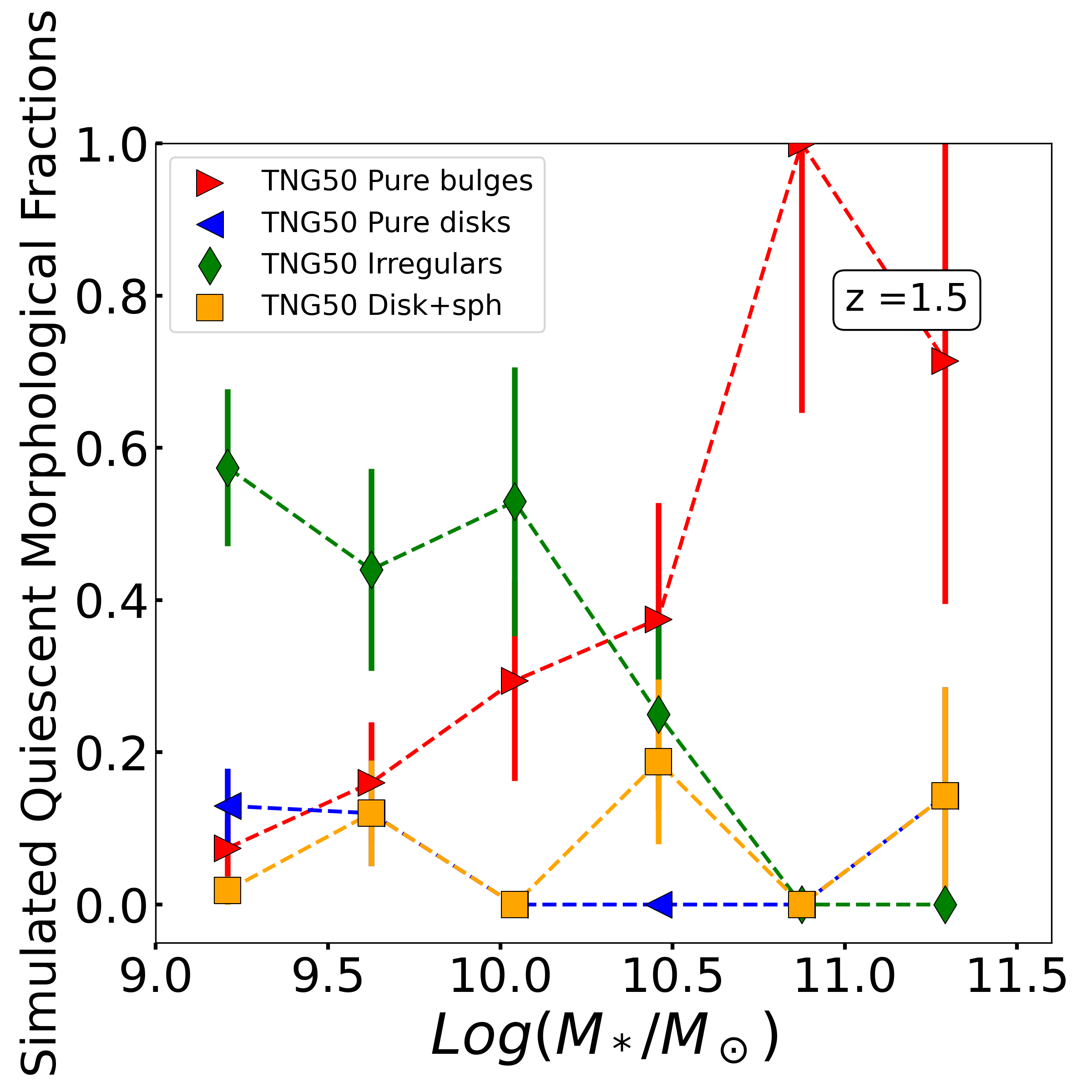}}
    \qquad
    \subcaptionbox{}{\includegraphics[width=0.3\textwidth]{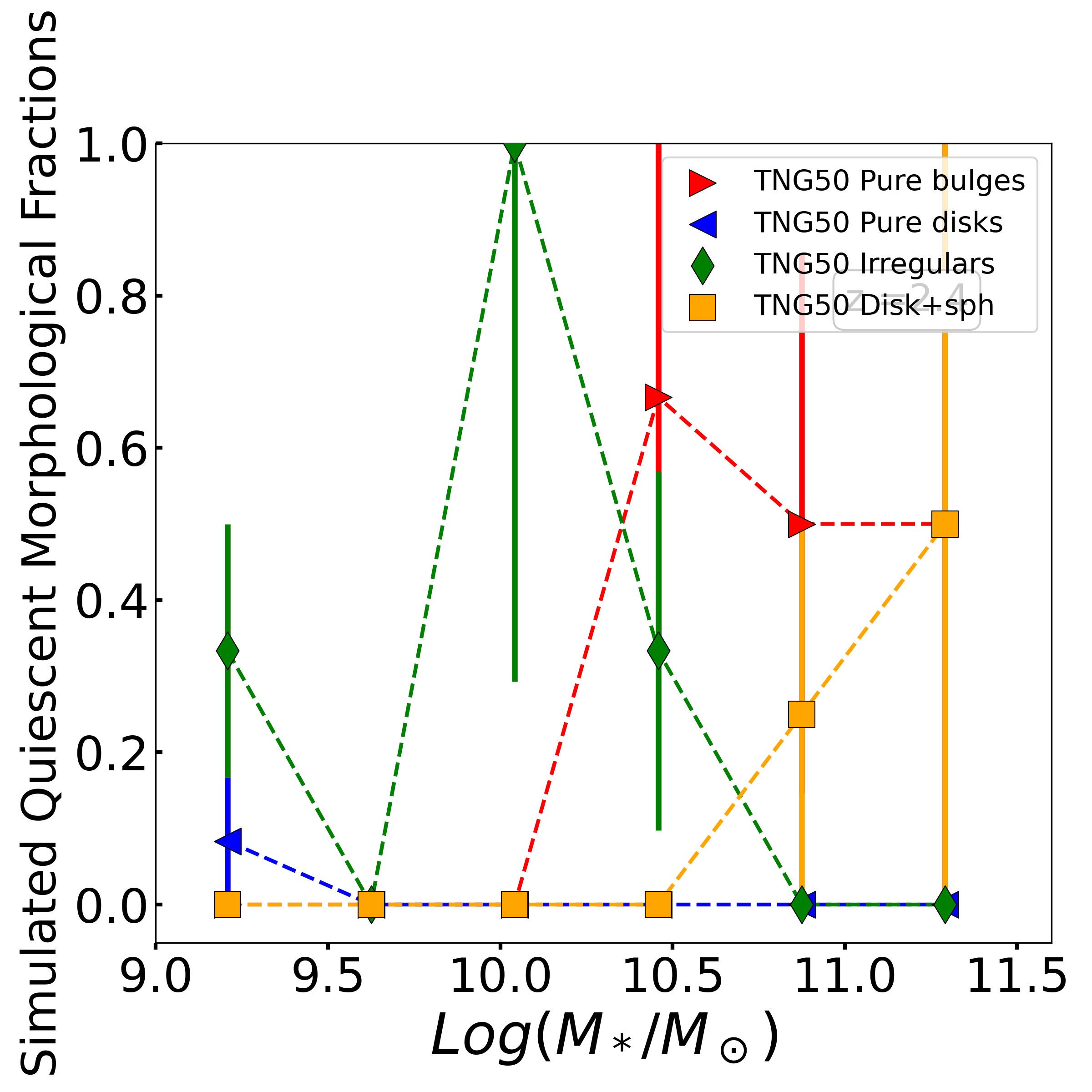}}

      \vspace{-5 pt}
    \caption{Morphological fractions of {\it quiescent and green valley} galaxies  as a function of stellar mass in three snapshots; annotations as in Fig.~\ref{fig:mabundance_SF}. The CANDELS general trends of morphological abundances of quiescent galaxies as a function of mass and redshift are well reproduced by TNG50, but for a lower fraction in the latter of passive spheroids at $z\gtrsim1$ than observed.}%
    \label{fig:mabundance_Q}%
\end{figure*}

Figures~\ref{fig:mabundance_SF} and~\ref{fig:mabundance_Q} show the evolution of the morphological fractions for star-forming and quiescent galaxies respectively. 

As for the whole sample, observed and simulated star-forming galaxies present a similar evolution of the morphological abundances. Above $z\sim1$, the population is dominated by irregular/clumpy galaxies at all stellar masses in both datasets. More symmetric disks start to be the most abundant class at $z<1$. Bulge dominated star-forming galaxies remain very marginal at all redshifts except at the very high mass end which confirms that the main sequence is dominated by galaxies with $n\sim1$ profiles. 

\begin{figure}
    \centering
 
        \includegraphics[width=\columnwidth]{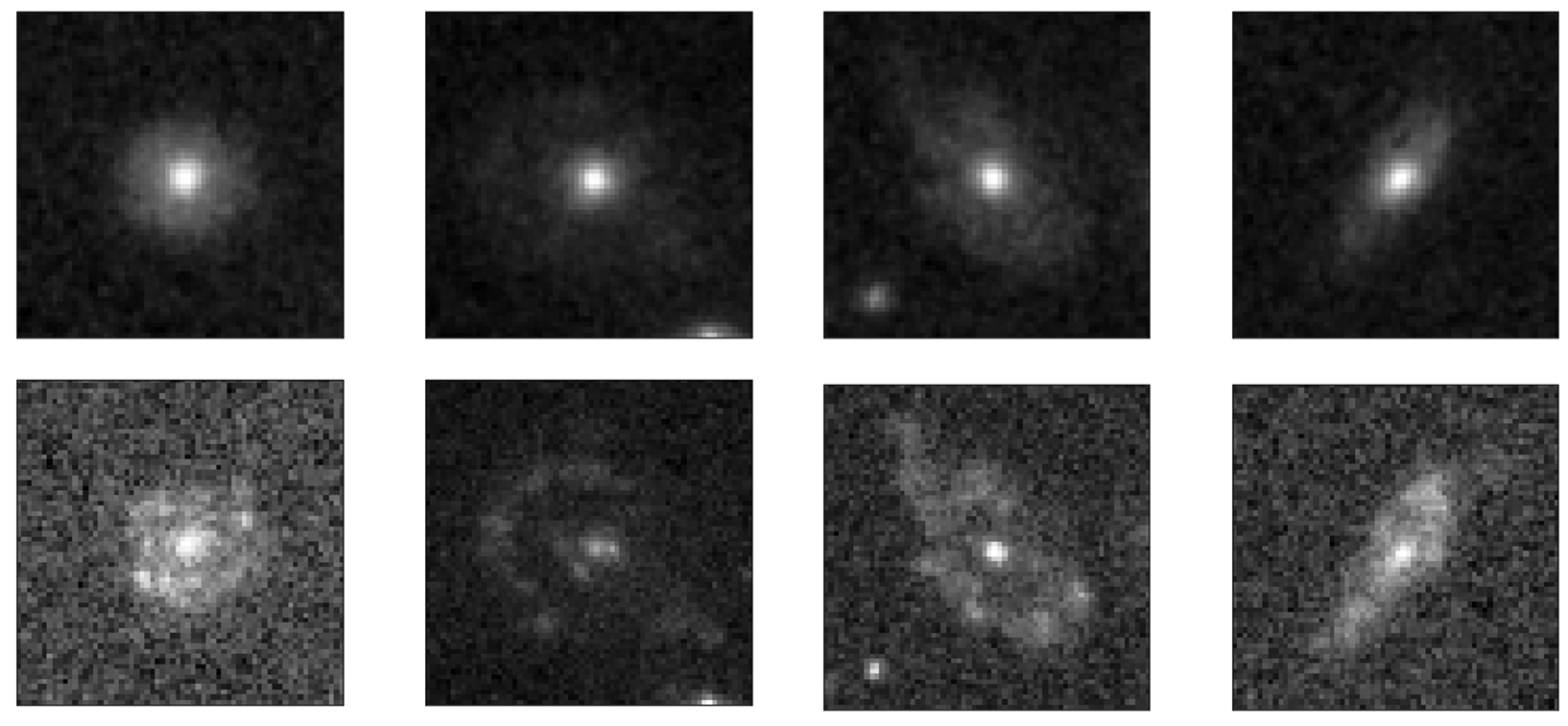}
   
    \caption{Examples of quiescent galaxies in TNG50 classified as irregulars. The top row shows galaxies imaged in the WFC3 F160W filter, and the bottom row corresponds to ACS F775W. The white bands show a star-forming ring around a bulge component, which is likely driving the CNN classification.}
    \label{fig:qirr}
\end{figure}

The majority of bulge dominated systems are in the quiescent population as expected, and this is again well captured by the simulation. At $z<1$, $\sim70\%$ of massive quiescent galaxies are bulge dominated. The differences between simulations and observations are more visible for quiescent galaxies. In particular, we see the fraction of passive spheroids above $z>1$ is  smaller in TNG50 than in CANDELS. This is especially visible in the low mass bin, where the majority of simulated quiescent galaxies have a disk or irregular morphology while in the observations, $\sim20\%-40\%$ of low mass passive galaxies are spheroids. These discrepancies might be partially explained by the fact that we have included green valley galaxies. As shown in figure~\ref{fig:SSFR-MSTAR}, the relative fraction of green valley galaxies in the simulation is larger than in the observations, which could contribute to the larger fraction of irregular systems. Nevertheless, we have visually inspected these low mass irregular quiescent galaxies in the simulation (Figure~\ref{fig:qirr}) . In most cases the optical bands show a compact core with a star-forming clumpy ring around, which likely drives the CNN classification into an irregular system. Using zoom-in numerical simulations, a recent work by~\cite{2020MNRAS.496.5372D} has pointed out that long-lived rings can form around massive quiescent cores when the timescale for inward mass transport for a ring is slower than the replenishment by accretion and the interior depletion by star formation. The authors argue that this condition is usually fulfilled if the cold to total mass ratio interior to the ring is smaller than 0.3, which is plausible for massive galaxies at $z<2$ as the ones studied here. Similar mechanisms might be in place here and might require further investigation.

\section{Evolution of the mass-size and mass-central density scaling laws} 
\label{sec:scaling}
We now compare the stellar mass-size and the stellar mass - central density relations of observed and simulated galaxies. We use the semi-major axis of the best fit Sersic model as a size estimator. For the simulations it is derived using statmorph as described in section \ref{sec:statmorph}. In the CANDELS sample, we use the values reported by \cite{vanderwel2014ApJ...788...28V}. 

\subsection{Mass-size relations}

\begin{figure*}
    \centering
    \vspace{-10 pt}
    \subcaptionbox{}{\includegraphics[width=0.3\textwidth]{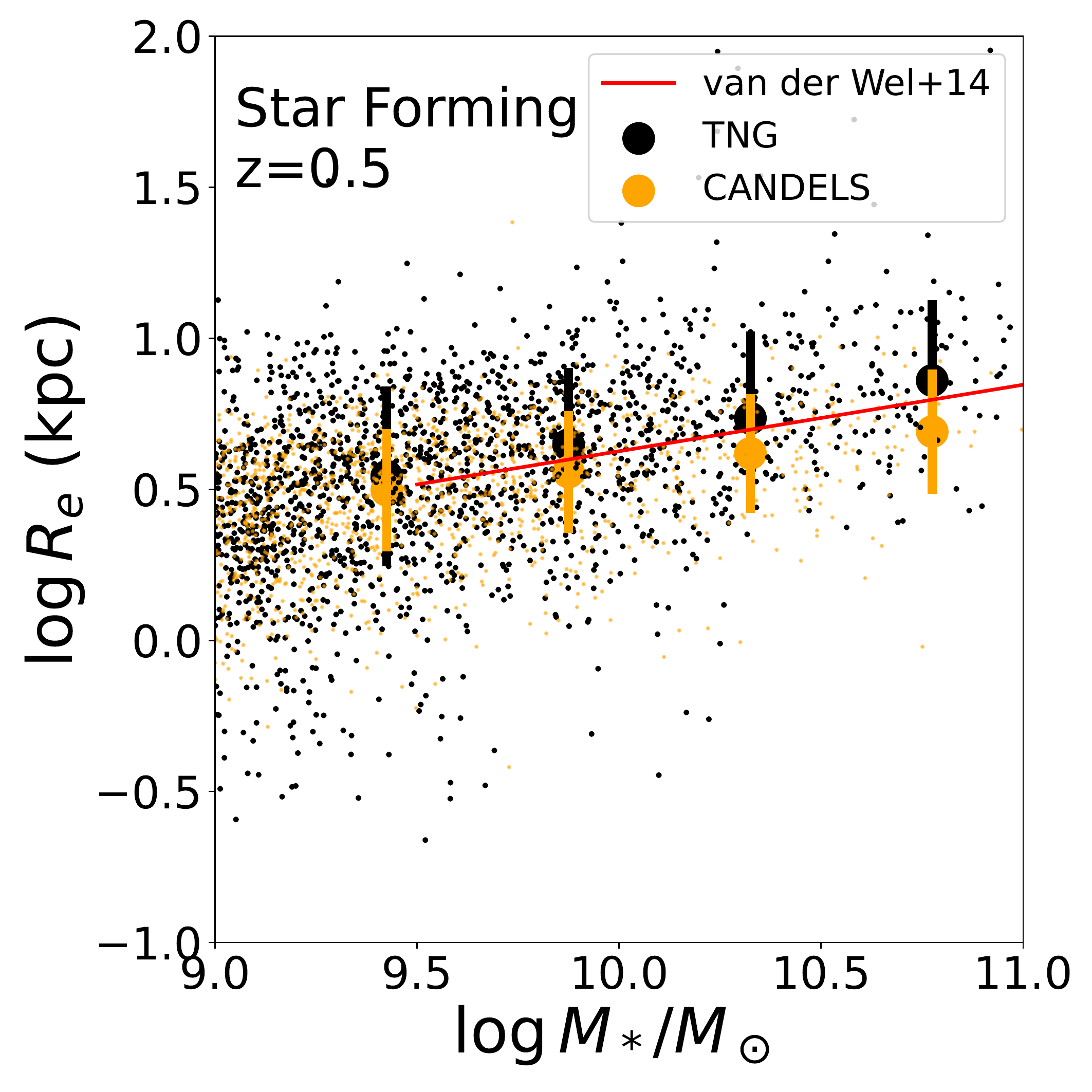}}%
     \qquad
      \subcaptionbox{}{\includegraphics[width=0.3\textwidth]{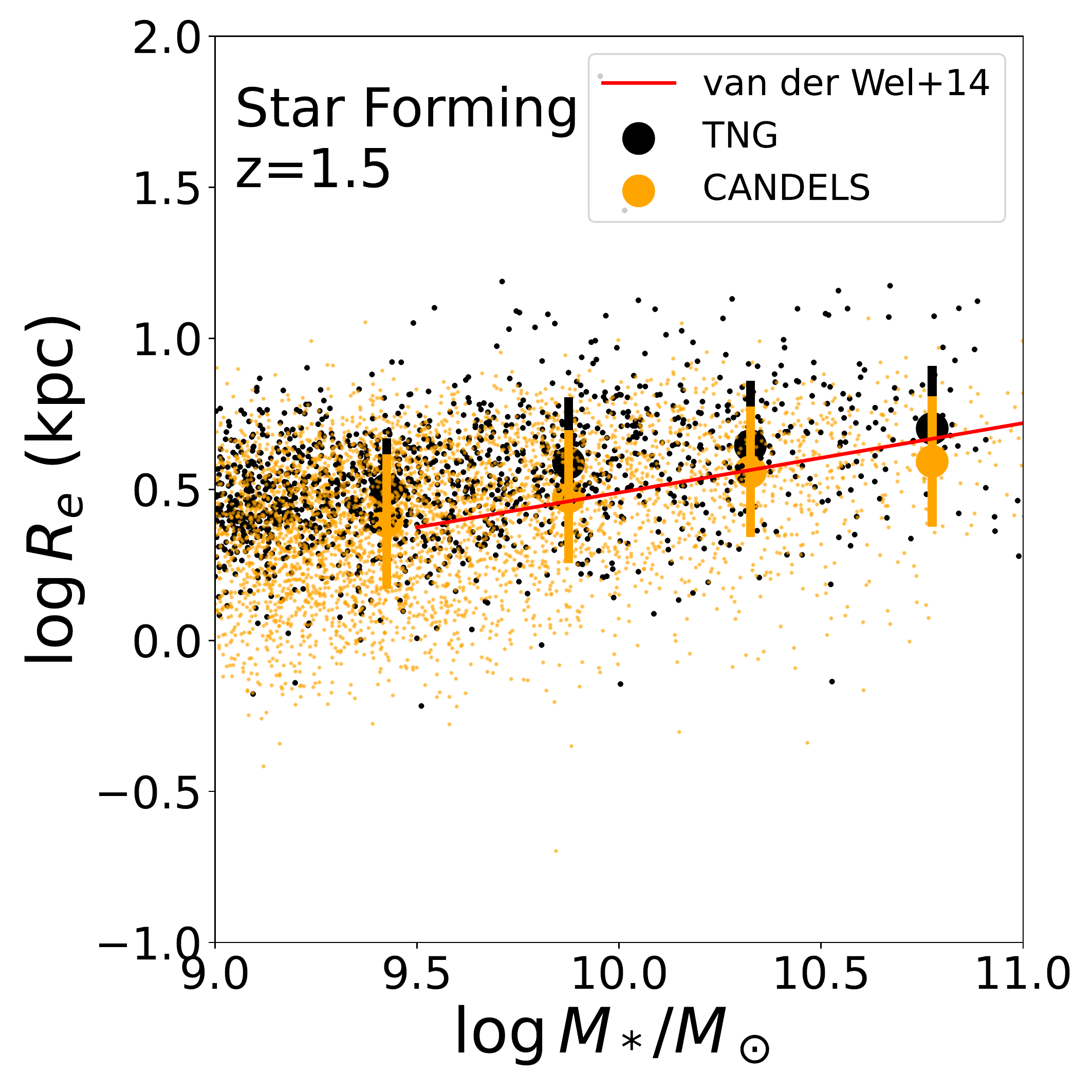}}%
     \qquad
    \subcaptionbox{}{\includegraphics[width=0.3\textwidth]{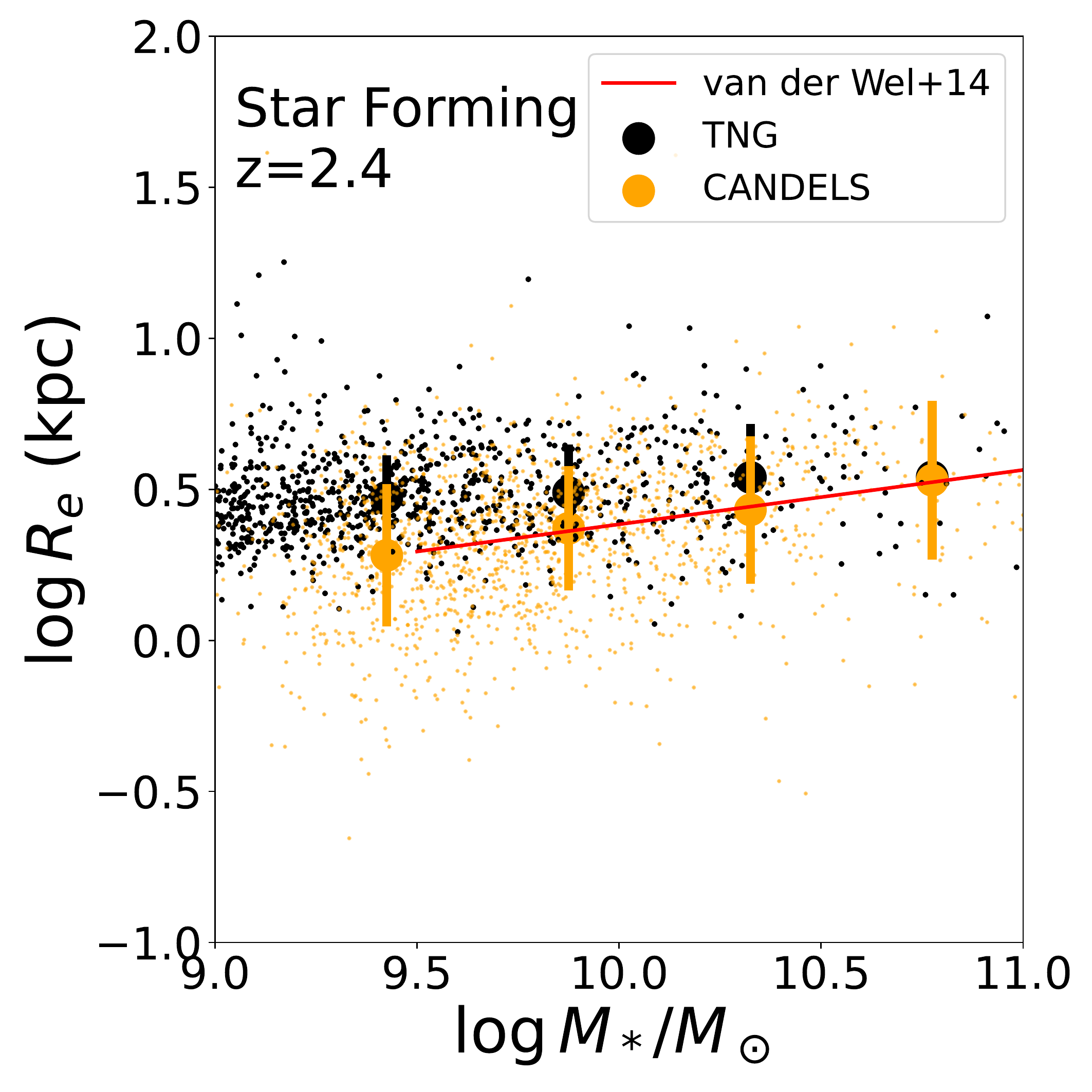}}%
    \qquad
    \subcaptionbox{}{\includegraphics[width=0.3\textwidth]{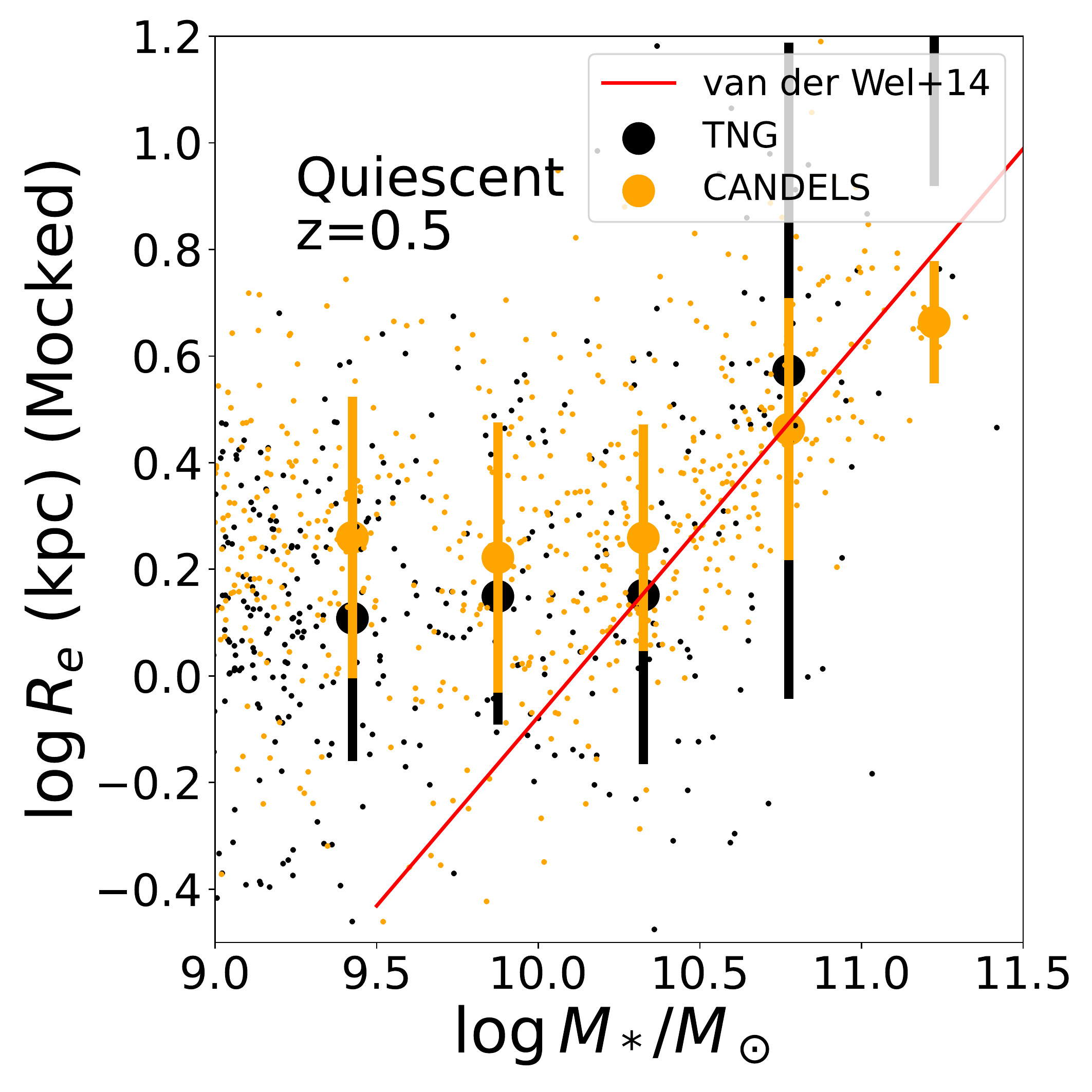}}%
    \qquad
    \subcaptionbox{}{\includegraphics[width=0.3\textwidth]{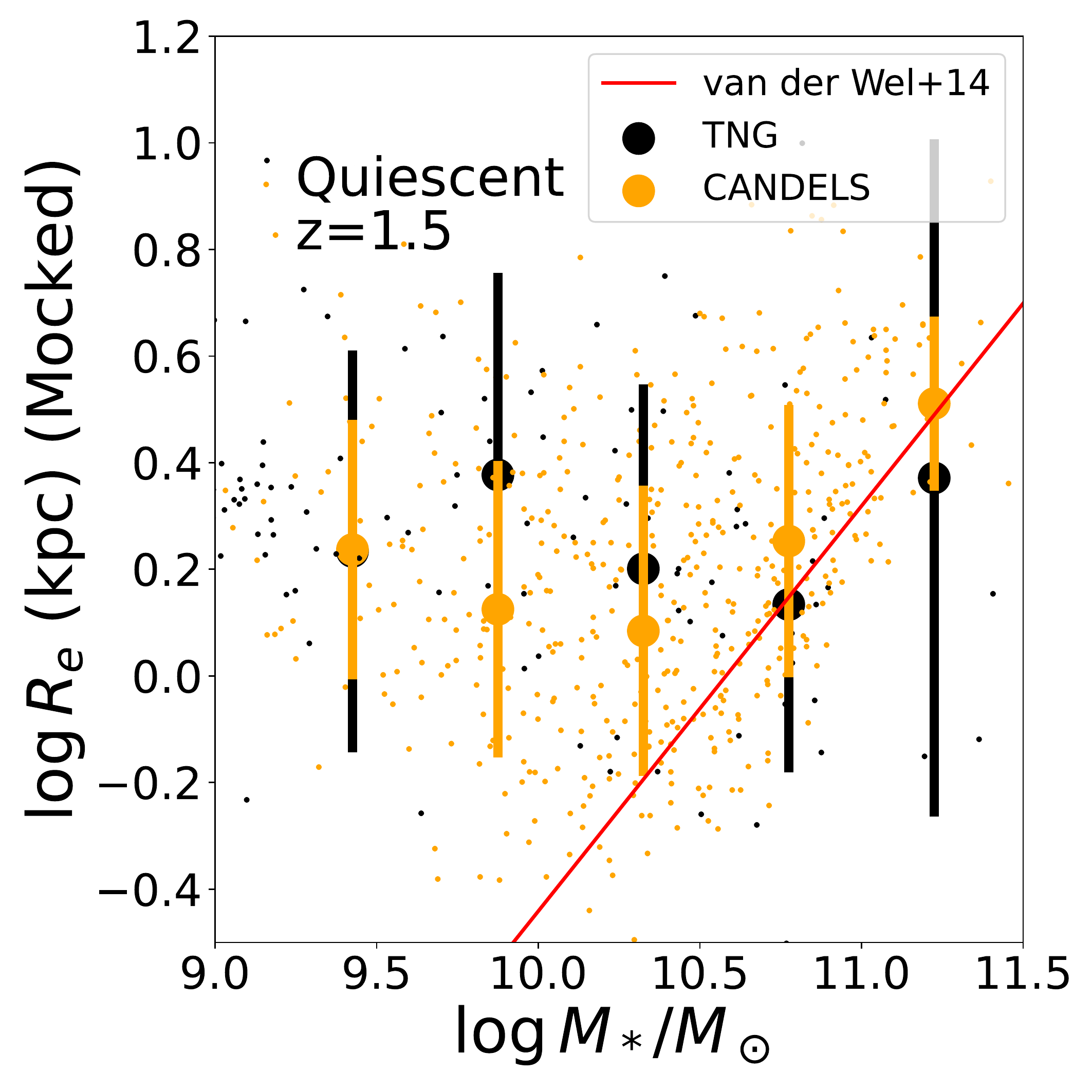}}
    \qquad
    \subcaptionbox{}{\includegraphics[clip,width=0.3\textwidth]{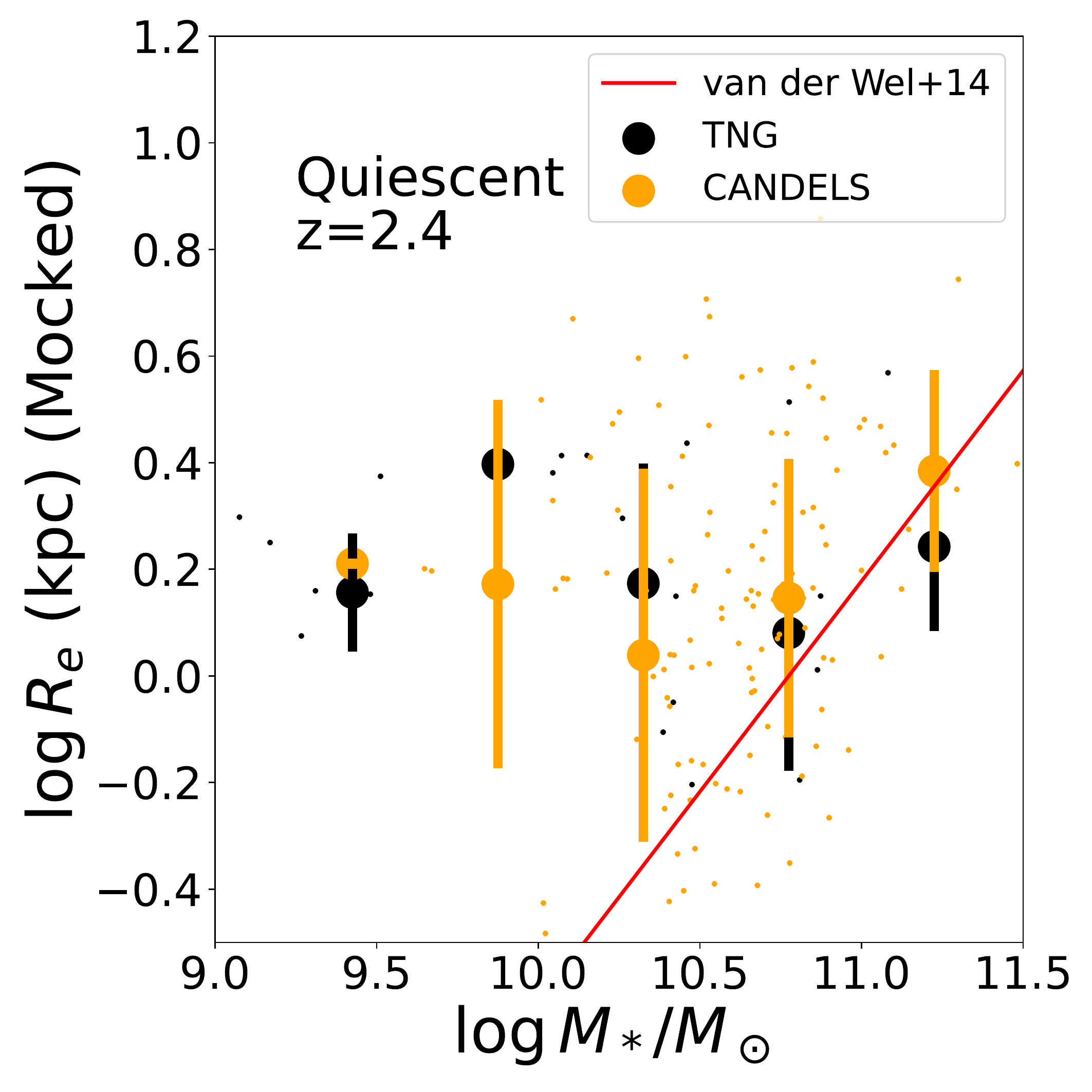}}

      \vspace{-5 pt}
    \caption{$\log M_*-\log R_e$ relation of observed (CANDELS: yellow dots) and simulated (TNG50: black dots) galaxies in three redshift bins ($z=0.5$, $z=1.5$ and $z=2.4$ from left to right), for star-forming (top) and quiescent galaxies (bottom), separately. The large dots with the error bars indicate the median size and scatter of the galaxies, in bins of stellar mass. The red solid lines indicate the power law fits to observations published in \protect\cite{vanderwel2014ApJ...788...28V}. Overall, TNG50 reproduces well the zeropoints, slopes and scatters on the observations.}
    \label{fig:mass-size}%
\end{figure*}

Figure \ref{fig:mass-size} shows the $\log M_*-\log R_e$ relations for star-forming and quiescent observed and simulated galaxies in three redshift bins ($z=0.5$, $z=1.5$ and $z=2.4$). As done in the previous subsections, we do not show all snapshots for the sake of clarity, but the trends are identical.  We observe a generally good agreement in the slope of the observed and simulated mass-size relations across all redshifts for both quiescent and star-forming galaxies. The median galaxy effective radii of observed and simulated galaxies fall well within a 1$-\sigma$ confidence interval of each other in each $\log M_*$ bin. The scatters are also comparable. The results are more noisy for the high redshift quiescent population given the low number statistics of simulated galaxies reported in the previous sections. 

The IllustrisTNG model has been somehow chosen to reproduce the size distribution of galaxies at $z\sim0$ (see~\citealp{2018MNRAS.474.3976G,2018MNRAS.473.4077P} for more details). However, it is not guaranteed that the size evolution matches observations. Our results show that the simulation properly captures the observed trends of both quiescent and star-forming galaxies since $z\sim3$ and quantitatively confirms the qualitative agreement identified by \cite{2019MNRAS.490.3196P} between TNG50 star-forming galaxies and various observational datasets.

\begin{figure*}
    \centering
    \vspace{-10 pt}
    \subcaptionbox{}{\includegraphics[width=0.3\textwidth]{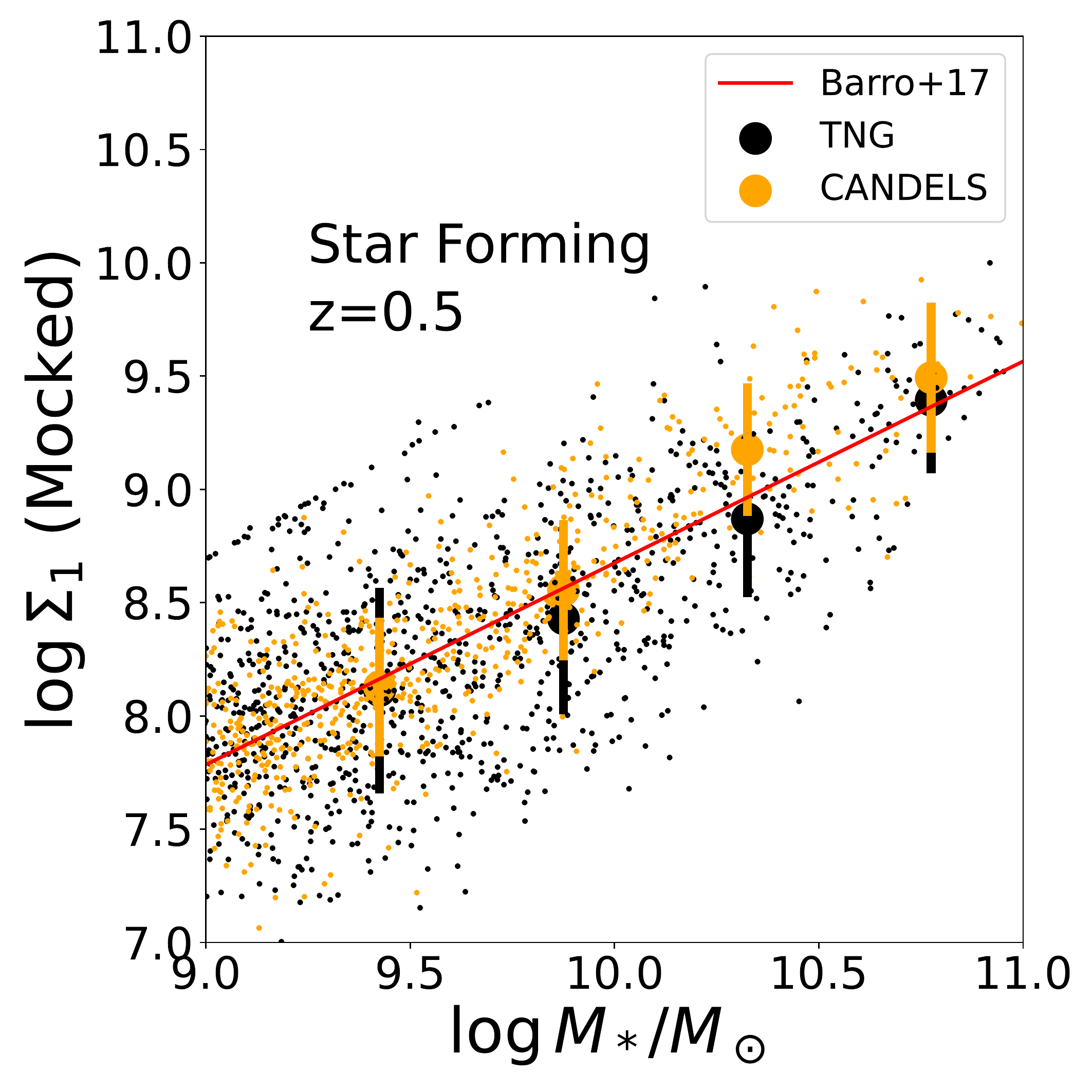}}%
     \qquad
      \subcaptionbox{}{\includegraphics[width=0.3\textwidth]{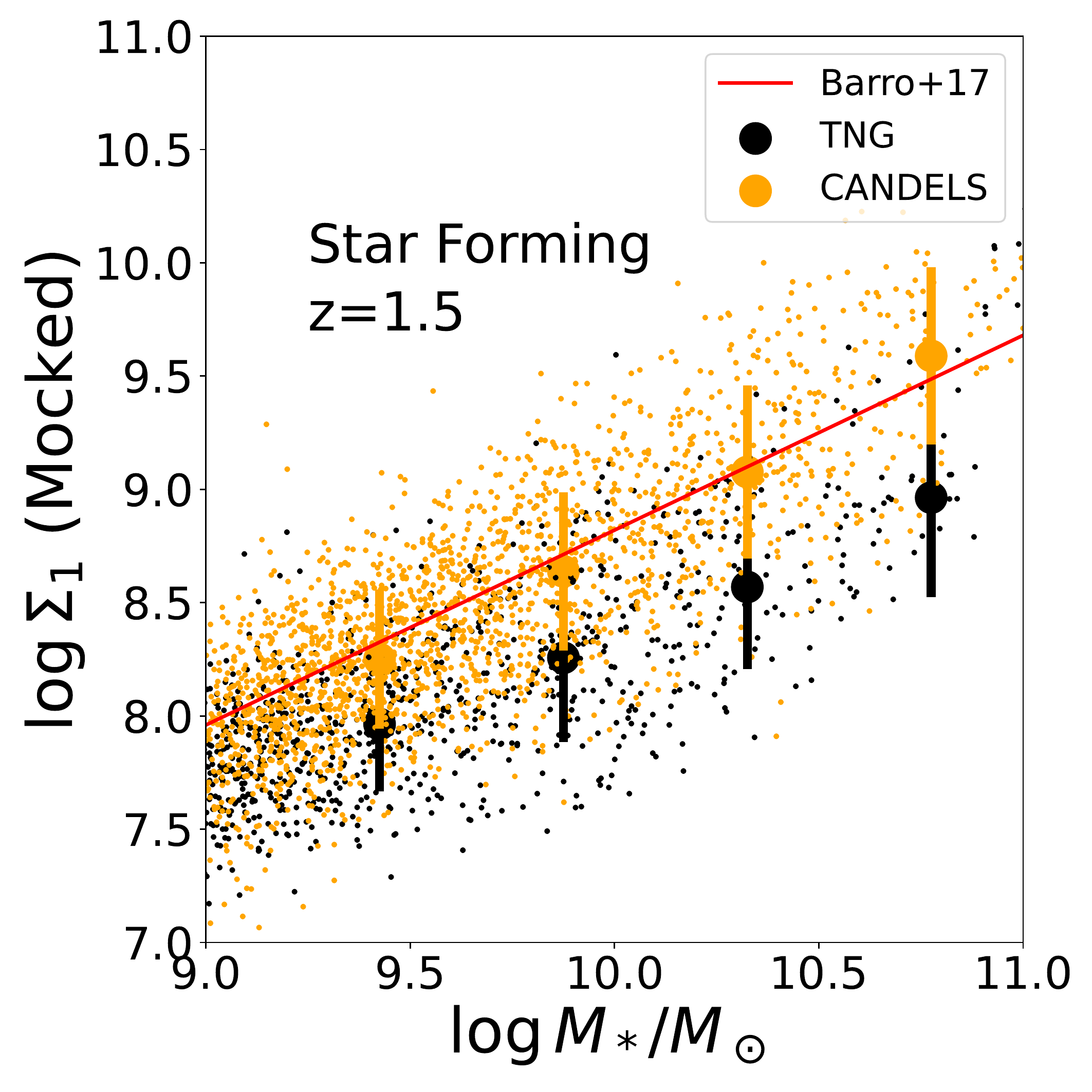}}%
     \qquad
    \subcaptionbox{}{\includegraphics[width=0.3\textwidth]{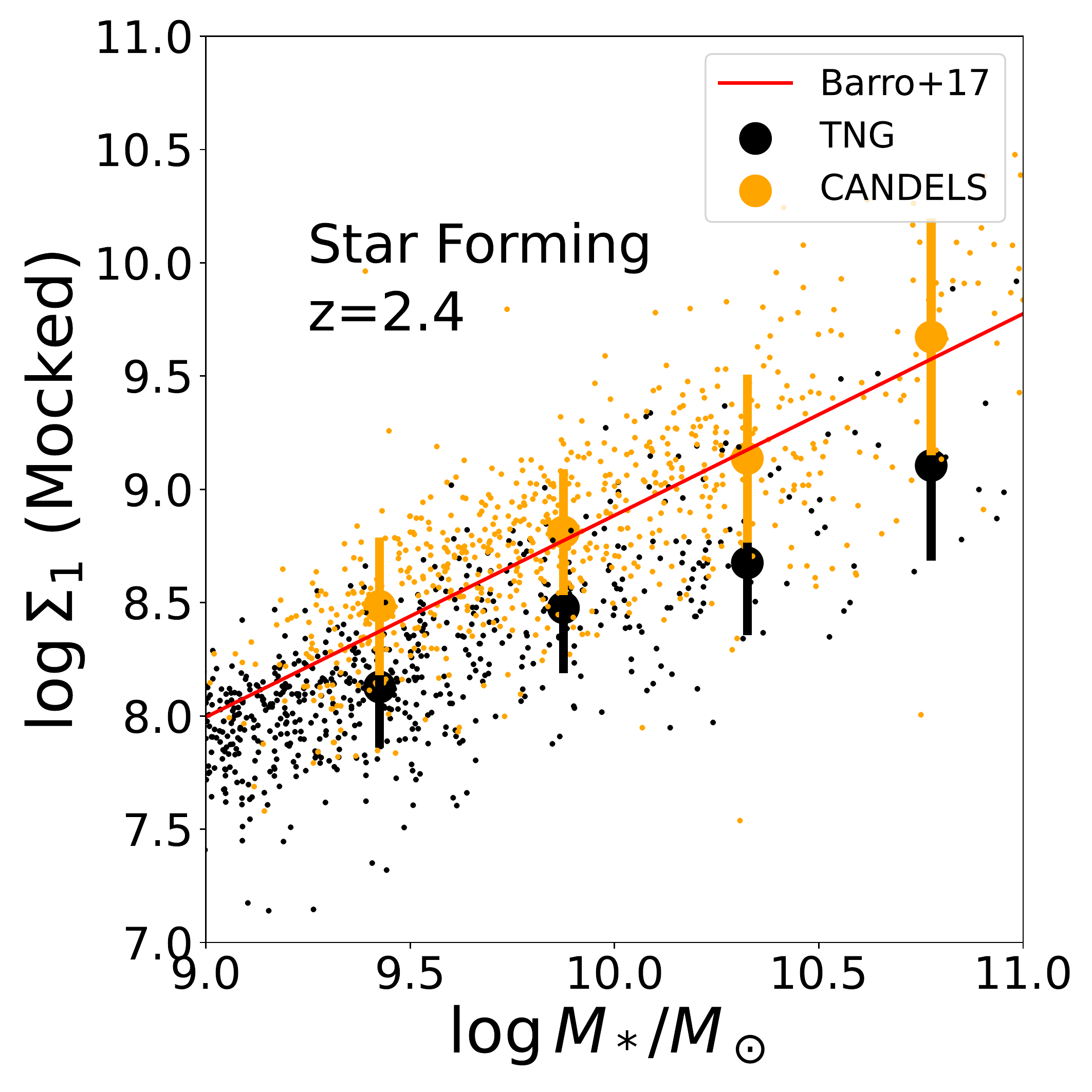}}%
    \qquad
    \subcaptionbox{}{\includegraphics[width=0.3\textwidth]{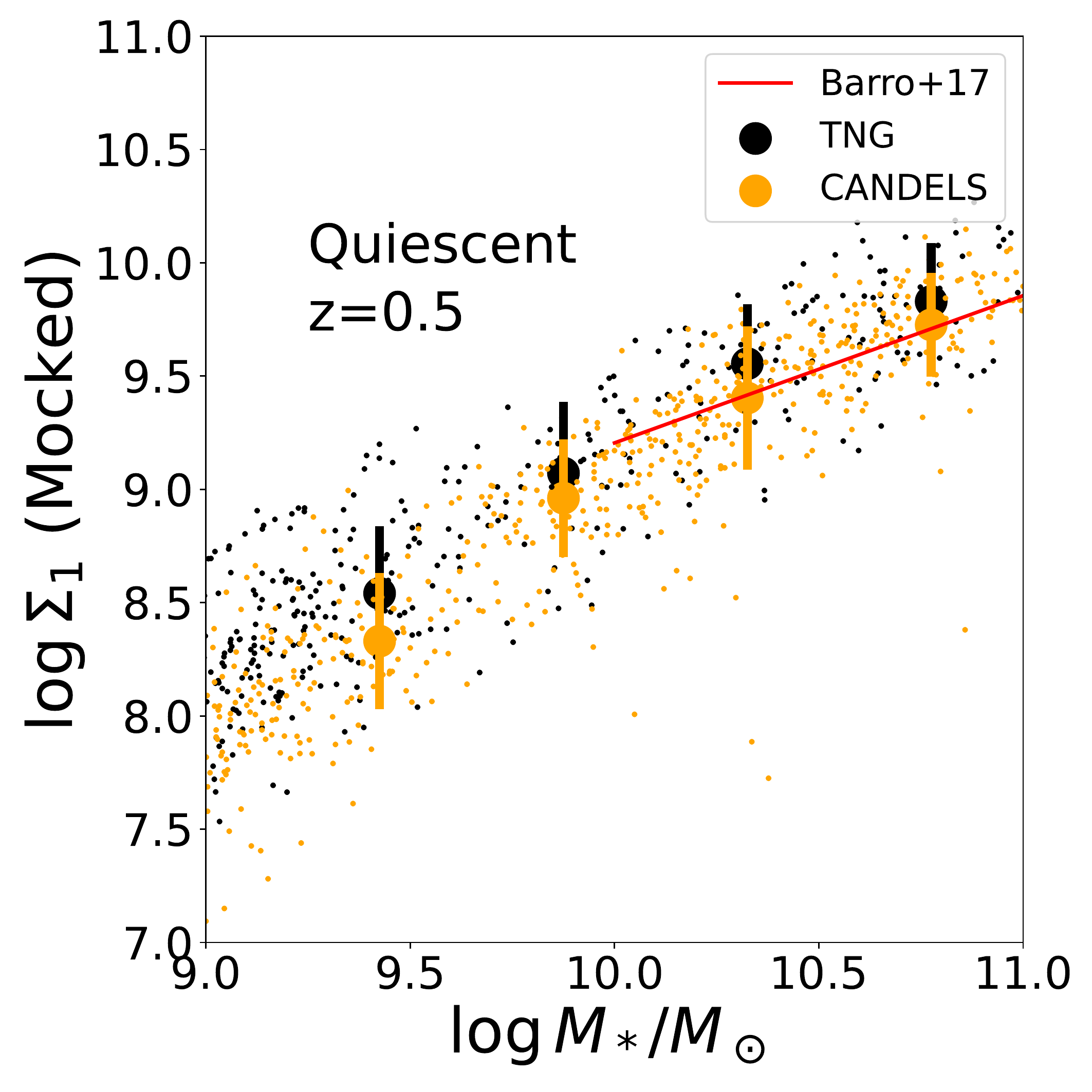}}%
    \qquad
    \subcaptionbox{}{\includegraphics[width=0.3\textwidth]{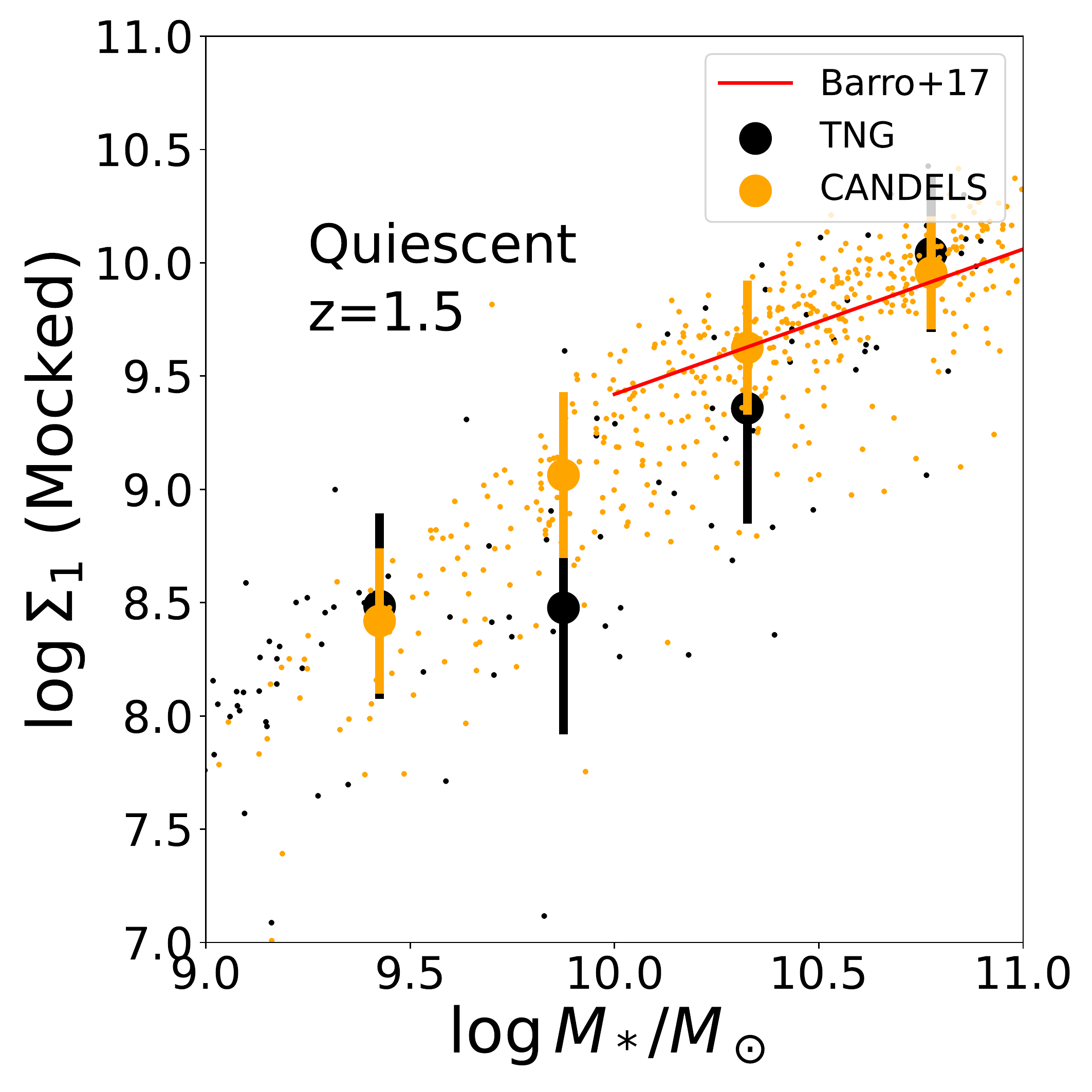}}
    \qquad
    \subcaptionbox{}{\includegraphics[width=0.3\textwidth]{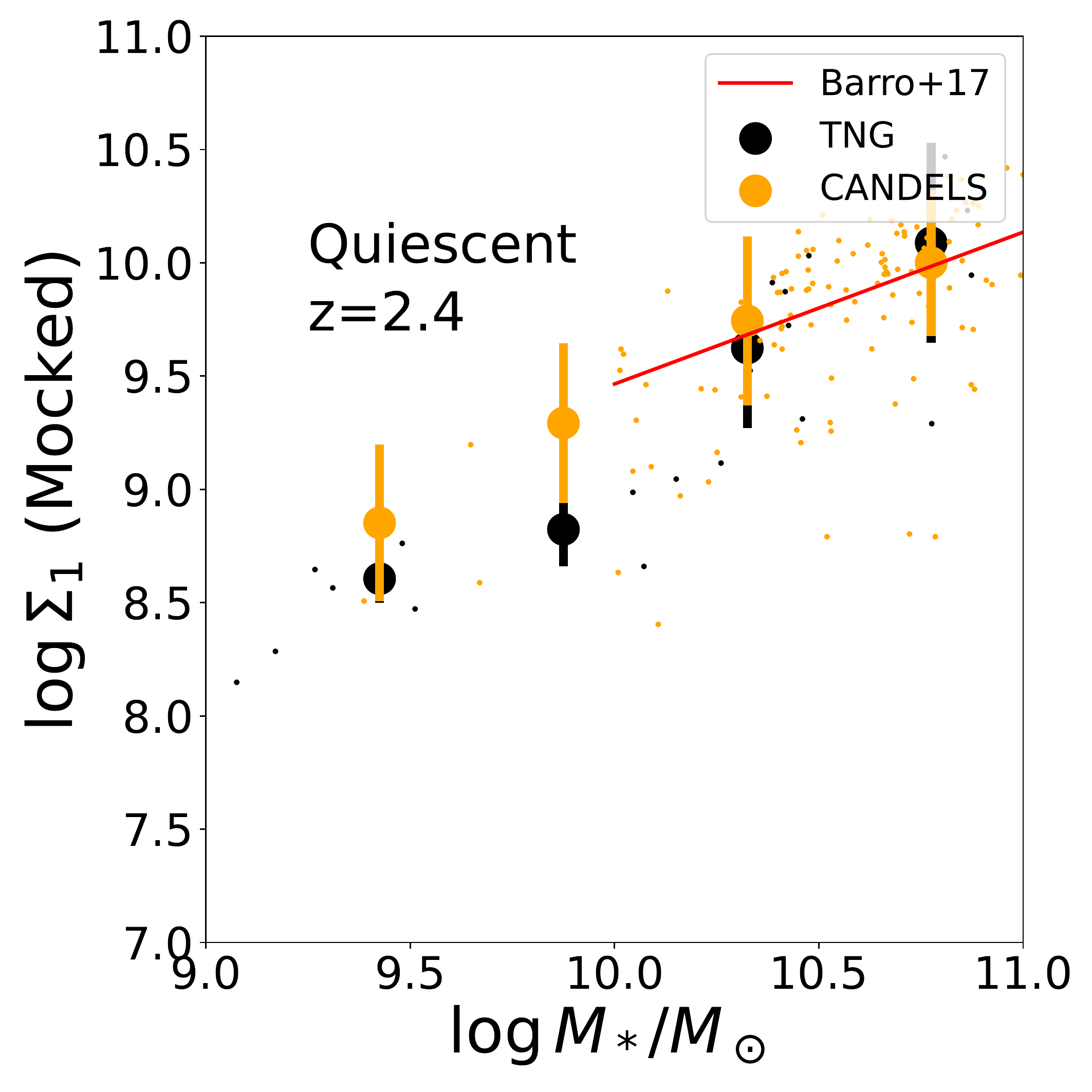}}

      \vspace{-5 pt}
    \caption{$\log M_*-\log\Sigma_1$ relation of  observed (CANDELS: yellow dots) and simulated (TNG50: black dots) galaxies in three redshift bins. Annotations are as in Figure~\ref{fig:mass-size}. The red solid lines indicate the power law fits to observations published in \protect\cite{2017ApJ...840...47B}. Overall, TNG50 reproduces well the zeropoints, slopes and scatters of the observations although TNG50 quiescent (star-forming) galaxies appear to be slightly denser (less dense) than observations.}%
    \label{fig:mass-sigma1}%
\end{figure*}

Figure~\ref{fig:mass-size} also shows the best power law fits from~\cite{vanderwel2014ApJ...788...28V}. It can be noticed that for quiescent galaxies, the published fit seems to be slightly inconsistent with the binned size-mass relation for CANDELS galaxies. There might be several reasons for that. Firstly, the selection of quiescent galaxies is different. ~\cite{vanderwel2014ApJ...788...28V} uses the UVJ diagram while we use here an offset from the Main Sequence. Secondly, the fit in ~\cite{vanderwel2014ApJ...788...28V} is only valid for quiescent galaxies more massive than $10^{10} M_\odot$. In figure~\ref{fig:mass-size}, most of the deviation is seen at lower masses as expected. Finally,~\cite{vanderwel2014ApJ...788...28V} performs a joint fit of both star-forming and quiescent galaxies by maximizing the likelihood, and assumes for that some fraction of contamination between the two populations. We believe that the adopted fitting method itself can also create some discrepancies.

\subsection{Mass-central density relations}

Another important observational result of the last years is that quiescent galaxies present a tight sequence in the $\log\Sigma_1-\log M_*$ plane, $\Sigma_1$ being the stellar mass density in the central kpc~\citep{2017ApJ...840...47B,2021arXiv210105820S}.  

Figure \ref{fig:mass-sigma1} shows the aforementioned  relation for star-forming and quiescent galaxies in different redshift bins as done for the mass-size relation. The central stellar density here, $\Sigma_1$ - mocked, is obtained from the SKIRT images by replicating the same procedure as done in observations (see Section~\ref{sec:statmorph} and Appendix~\ref{app:sigma1}). We only include in this figure objects with $b/a>0.5$ as done in the observations. We observe again a reasonably good match between the scaling relation of observed and simulated galaxies, with the TNG50 simulation reproducing both the evolution and the slope of both quenched and star-forming galaxies well. Namely, the main observational result of past years, which is that quiescent galaxies have larger $\Sigma_1$ values than star-forming galaxies of the same mass, is also reproduced in TNG50 -- this is the case also in TNG100 (for $\Sigma_2$), as recently shown at $z\sim0$ \citep{2021MNRAS.tmp..841W}.


 There are however some noticeable differences. Star-forming TNG50 galaxies above $z=1$ tend to have lower values of $\Sigma_1$ at fixed mass than observed ones while the opposite is observed for the quiescent galaxies. The differences are within the $1\sigma$ scatter but the mean values are systematically offset. This implies that the dynamic range of $\Sigma_1$ is a bit larger in the simulation than in the observations and hence the difference in the scaling laws of quiescent and star-forming galaxies is more pronounced. Resolution effects and differences in the Sersic fits could partially explain the difference. We explore in Appendix~\ref{app:sigma1} the impact of observational effects on the measurement of $\Sigma_1$. We notice however that the main goal of this work is to explore the morphological properties of simulated galaxies in the observational plane, so we will use the mock-observed $\Sigma_1$ values throughout the paper.

\section{The build up of stellar scaling relations, black hole growth and quenching in TNG50}\label{sec:discuss}

The previous sections have shown that the TNG50 simulation reproduces reasonably well both the morphological abundances and the scaling relations of massive galaxies in the redshift range $0.5<z<3$ despite some differences. It is therefore justified to analyze the evolutionary tracks of galaxies in the different planes to understand how these relations are built up in the simulation across cosmic time.

Our main goal is to understand how a galaxy transitions from the star-forming scaling relations to the quiescent ones and how this affects its structure and morphology. In the following, we focus only on massive galaxies ($10.5<\log(M_*/M_\odot)<11$) at $z=0.5$ and follow them back in time in the analyzed snapshots. There are several reasons for this choice. First, this stellar mass range typically corresponds with the scale at which galaxies quench in the TNG model \citep{2017MNRAS.465.3291W,2018MNRAS.475..624N} and quiescent galaxies start to dominate the stellar mass function (e.g. ~\citealp{2020MNRAS.493.1888T}); this is in fact the scale where both in observations and in TNG there is a comparable number of star-forming and quiescent galaxies, at least at low redshifts \citep{2019MNRAS.485.4817D}. Second, since massive galaxies at $z=0.5$ have typically a stellar mass of $\sim10^9~M_\odot$ or larger at $z\sim3$, it allows us to track them down in the \emph{observed} surveys using our mass selected sample.

\subsection{Quenching in the IllustrisTNG model}
\label{sec:Q_TNG}

We start by summarizing the main findings from previous published works regarding the quenching of star formation in TNG. It has been shown that it is the kinetic channel of the SMBH feedback that is responsible for the halting of the star formation in massive galaxies ($\gtrsim$ a few $10^{10} M_\odot$ galaxies), namely the SMBH-driven winds that are invoked when the accretion rates of the SMBH are low \citep{2017MNRAS.465.3291W, 2018MNRAS.475..624N, 2020MNRAS.493.1888T, Donnari2021}. The SMBH-driven winds in the TNG simulations are both ejective and preventative~\citep{2020MNRAS.499..768Z}. They are ejective in that they trigger quenching by removing gas from the star-forming regions of galaxies~\citep{2020MNRAS.493.1888T} or even by reducing the overall gas mass within haloes~\citep{2020MNRAS.491.4462D,2020MNRAS.499..768Z}; they are preventative as they heat up the gas within and around galaxies~\citep{2020MNRAS.499..768Z}, increasing its entropy and its cooling times also in the outer reaches of the halo: this prevents it from fuelling subsequent star formation. The inside-out quenching ensuing from this picture and triggered by the ejective outflows from SMBHs in TNG50 is supported by the shape of the sSFR profiles in quiescent galaxies observed with 3D-HST at $z\sim1$ \citep{NelsonE2021}. Importantly, the low-accretion SMBH feedback also limits the availability of gas for SMBH gas accretion and hence growth so that SMBHs grow either very rapidly while exercising thermal mode feedback at high-accretion rates, at high redshifts and low masses (for SMBH, stellar, and halo masses of $\lesssim 10^8~M_\odot$, i.e. $\lesssim 10^{10.5}~M_\odot$, $\lesssim 10^{12}~M_\odot$, respectively) or more slowly via SMBH-SMBH mergers at lower redshifts and larger masses \citep{2018MNRAS.479.4056W, 2021MNRAS.501.2210T}.

\subsection{The building up of scaling laws in TNG50}

With this in mind, we begin by exploring how the progenitors of star-forming and quiescent galaxies evolve in the stellar structural planes and how their apparent morphologies transform over time. 

\begin{figure*}
    \centering
    \vspace{-10 pt}
    \subcaptionbox{}{\includegraphics[width=0.45\textwidth]{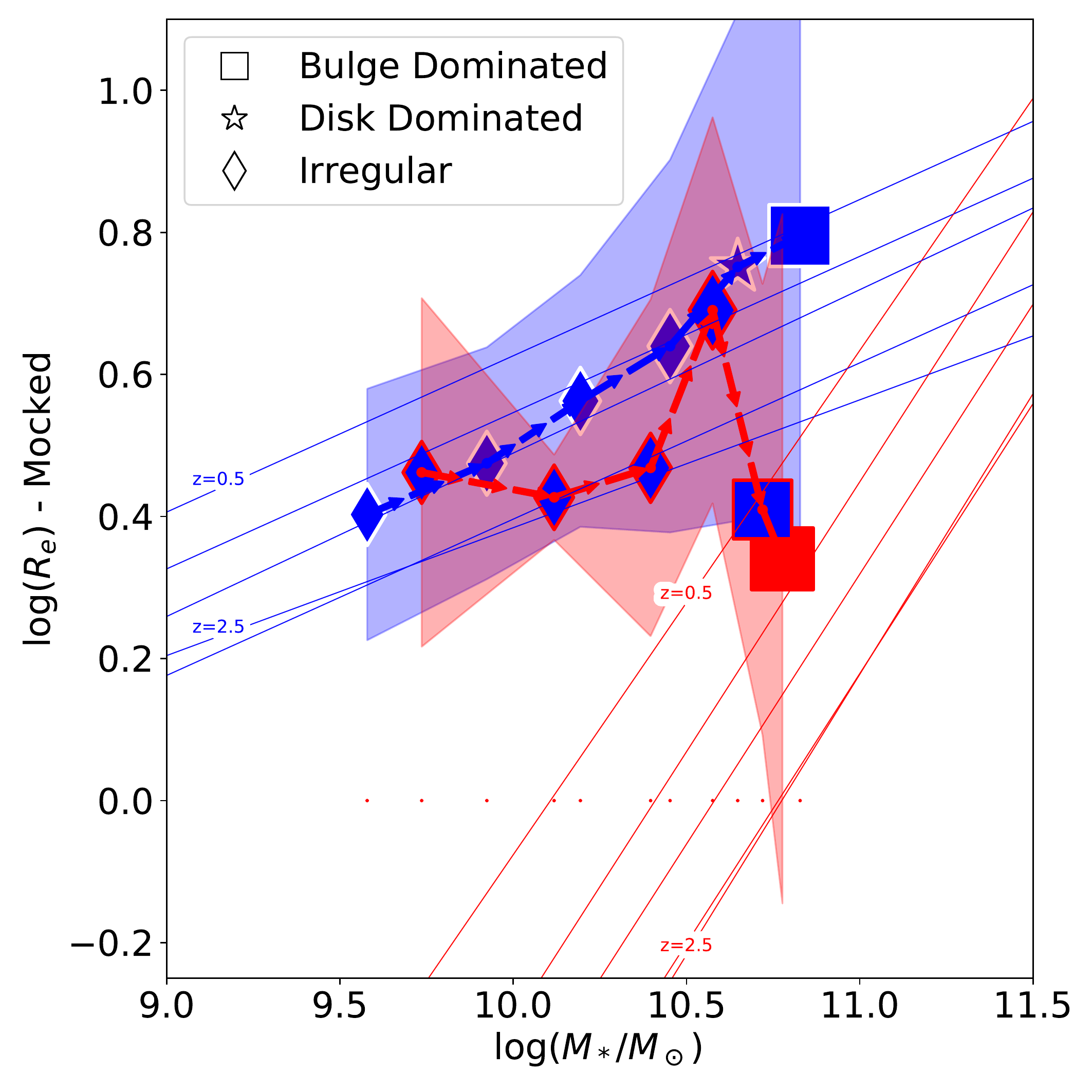}}%
     \qquad
      \subcaptionbox{}{\includegraphics[width=0.45\textwidth]{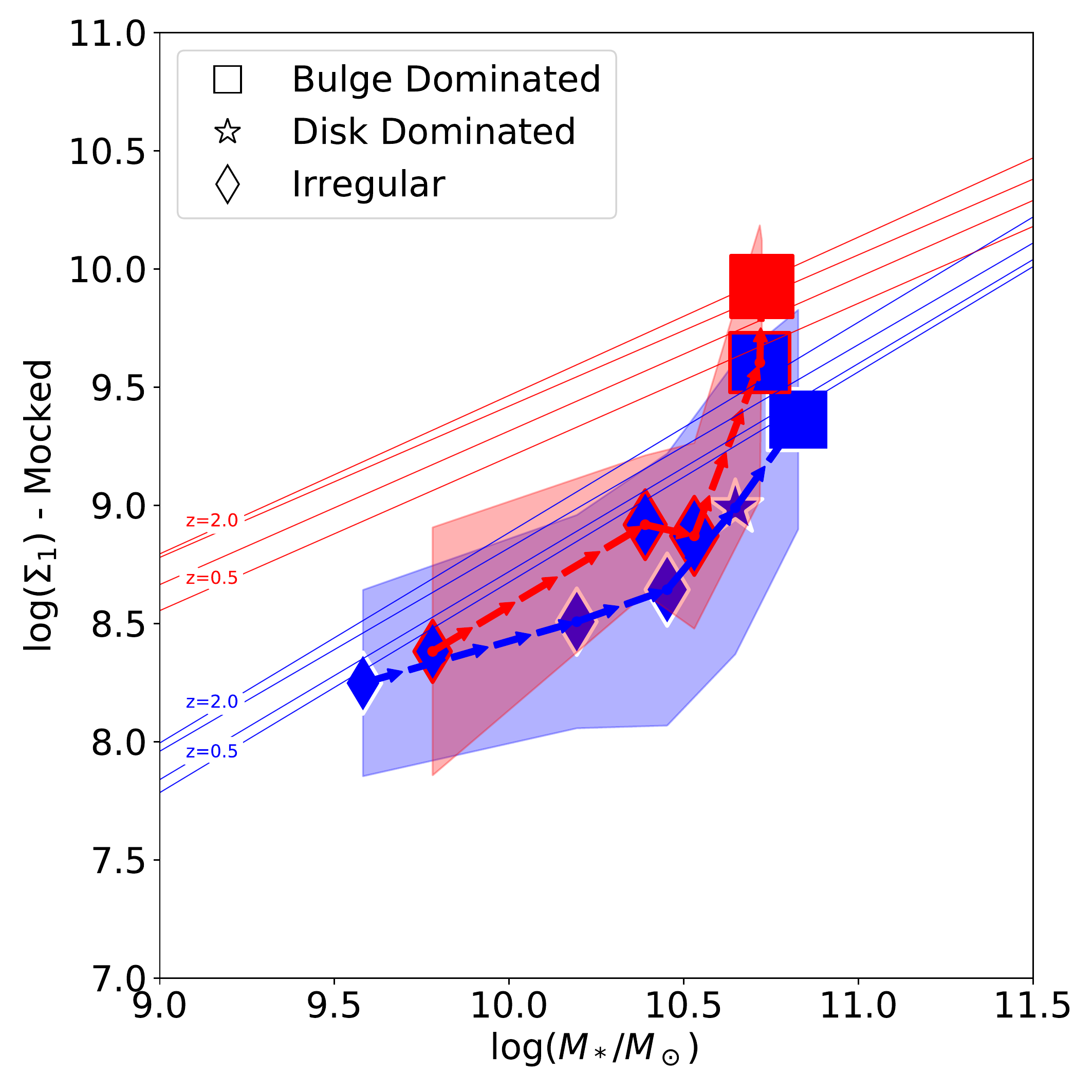}}%
      \qquad
      \vspace{-5 pt}
    \caption{Median evolutionary tracks in the $\log M_*-\log R_e$ (left) and $\log M_*-\log \Sigma_1$ (right) planes of the progenitors of galaxies with stellar masses between $10^{10.5}$ and $10^{11}$ at $z=0.5$, in the TNG50 simulation. The red (blue) arrows indicate the median evolutionary tracks from $z=3$ of quiescent (star forming) progenitors and the symbols the median values at the 6 snapshots considered ($z=3, 24, 2.0, 1.5, 1$, and $z=0.5$). The blue and red shaded regions show the scatter for star-forming and quiescent progenitors respectively. The symbols with a red edge indicate the progenitors of quiescent galaxies. The different symbols indicate the median CNN estimated morphology. The colors show if the median progenitor is star-forming (blue) or quenched (red) at a particular redshift. Symbol size is proportional to the SMBH mass. The blue and red solid lines show the best fit relations for star-forming and quenched galaxies in observed galaxies from the CANDELS survey at different redshifts. }%
    \label{fig:tracks_mstar_re}%
\end{figure*}

Figure~\ref{fig:tracks_mstar_re} shows the median evolution of the progenitors of massive galaxies (i.e. $10.5<\log M_*/M_\odot<11$) at $z=0.5$ in the $log M_*-\log R_e$ and $\log M_*-\log\Sigma_1$ planes, along with the observational ridgelines of star-forming and quenched galaxies~\citep{vanderwel2014ApJ...788...28V,2017ApJ...840...47B}. We divide the sample between the progenitors of galaxies that are quenched at $z=0.5$ ($\Delta\log \rm{sSFR}<-1$) and those which are still in the main sequence ($\Delta\log \rm{sSFR}>-0.45$). We emphasize that the median stellar masses at $z=0.5$ of both samples are very similar.  We also indicate in Figure~\ref{fig:tracks_mstar_re} the median visual morphological type at every snapshot derived from the SKIRT images as explained in Section~\ref{sec:morphs}. For simplicity, we group galaxies into bulge dominated (spheroids, bulge+disk), disk dominated (disks) and irregulars.

First of all, it can be seen that in TNG50, on average, massive galaxies ($10.5<\log(M_*/M_\odot)<11$) at $z=0.5$ tend to quench at late times ($z<1$).  Although the quenching process may start earlier, the median value of $\Delta \log \rm{sSFR}$ is lower than $-1$ only for the $z=0.5$ snapshot. As shown in section~\ref{sec:morphs}, the fraction of massive quiescent galaxies above $z\sim1$ is rather low in the simulation which necessarily implies that the majority of massive galaxies quench at later epochs.

The figure also shows that, on average, in the case of the size-mass relation, galaxies evolve along the ridgeline while they are star forming, increasing both stellar mass, $R_e$ and $\Sigma_1$. However, in the case of the central density-mass relation, star-forming galaxies evolve below it. On the other hand, the evolutionary tracks of quiescent galaxies differ substantially from the average relations at fixed cosmic epoch.
Interestingly, a global apparent morphological transformation takes place for both star-forming and quiescent galaxies above a characteristic stellar mass of $\sim 10^{10.5}$ solar masses. Below that mass, the median apparent optical rest-frame stellar morphology is predominantly irregular. Above that mass, it becomes a symmetric disc or bulge dominated system more comparable to the types populating the local Hubble Sequence. This morphological change is reflected in a change in the slope of the $\log M_*-\log\Sigma_1$ track. The central density increases faster than the stellar mass, which also points towards the building up of a bulge component. It is important to notice that this transformation is observed in both populations: galaxies that quench and galaxies that remain star-forming at $z=0.5$.See also~\cite{2019MNRAS.487.5416T} for a detailed discussion of the relation between morphological transformations and quenching in the TNG simulation. 

However, even if the physical mechanisms driving morphological transformations seem comparable for galaxies that quench and those which have not yet, they appear to be more dramatic for the former. As a matter of fact, Figure~\ref{fig:tracks_mstar_re} shows that quiescent galaxies experience an important decrease of their observed light-weighted projected effective radius before quenching which moves the galaxy down in the $\log M_* - \log R_e$ plane and up in the $\log M_* - \log \Sigma_1$ plane, where the observational quiescent ridgelines lie. This decrease of the mock-observed effective radius is not seen in the star-formation population that instead continues to evolve along the SF ridgeline. This is also interesting because it indicates that quiescent galaxies in TNG50 reach the quenched ridge line in the mass size plane following a \emph{vertical} track with a fast increase of the \emph{observed} central density and a decrease of the effective radius.

\subsection{SMBHs and the structural evolution of galaxies}

At this point, it is interesting to investigate what are the causes of these morphological transformations and how they relate to quenching. 

One possibility is that they are triggered by external stochastic processes such as mergers. The merger history of the progenitors of star-forming and quiescent galaxies appears to be very similar though, both in terms of major ($>1:4$) and minor ($<1:4$) mergers with practically no mergers (minor or major) happening in the last Gyr for both populations. Only one galaxy in our sample has had a major merger since redshift two. It is thus unlikely that mergers are responsible of the increase in central density we observe in Figure~\ref{fig:tracks_mstar_re}. This is also consistent with the findings of~\cite{Weinberger2018,2021MNRAS.tmp..841W} for TNG.

Another possibility is that the morphological and structural transformations are connected to the SMBH growth. As described in Subsection~\ref{sec:Q_TNG}, kinetic feedback from the SMBH is responsible for quenching star formation in central massive galaxies in the TNG model and it typically starts being dominant above a stellar mass of about $10^{10.5}$ which also corresponds with the characteristic stellar mass for morphological transformations we extract in Figure~\ref{fig:tracks_mstar_re}.

  \begin{figure*}
    \centering
    \vspace{-10 pt}
    \subcaptionbox{}{\includegraphics[width=\columnwidth]{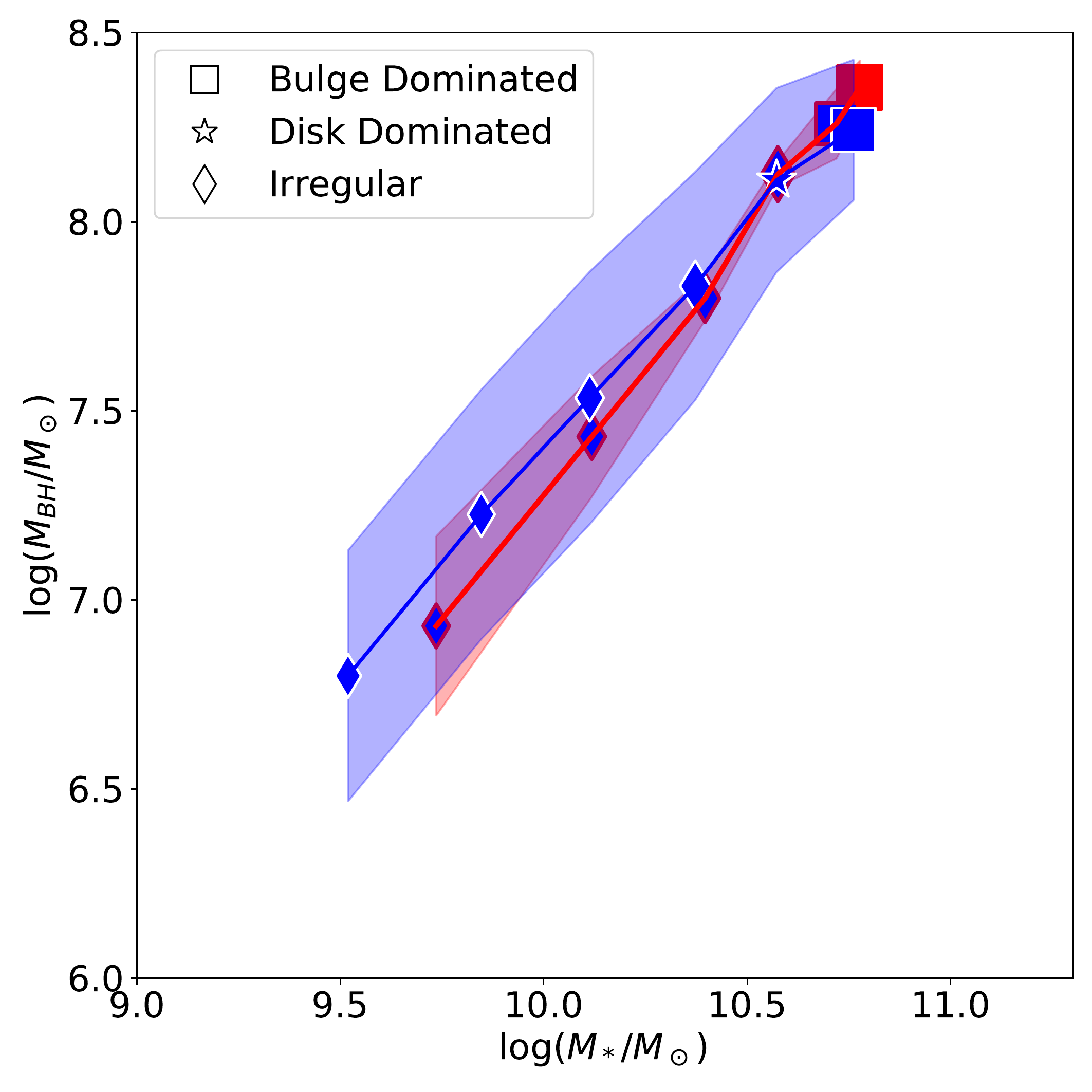}}%
     \qquad
      \subcaptionbox{}{\includegraphics[width=\columnwidth]{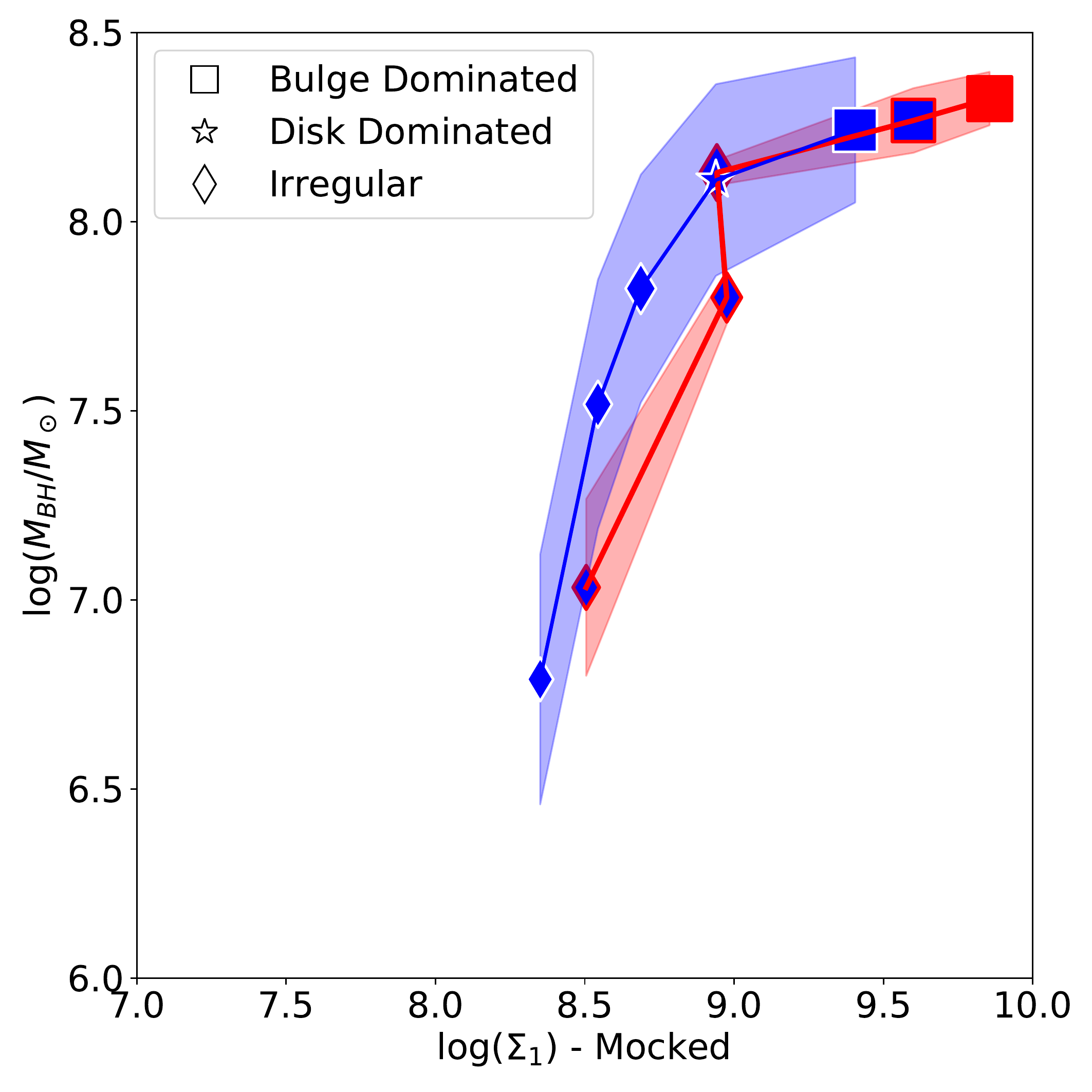}}%
     
      \qquad

        \vspace{-10 pt}
    \caption{Median SMBH growth as a function of stellar mass (left) and $\Sigma_1$ (right) for the progenitors of galaxies with stellar masses between $10^{10.5}$ and $10^{11}$ at $z=0.5$, in the TNG50 simulation. Annotations are as in Fig.~\ref{fig:tracks_mstar_re}. The filling colors of the symbols indicate if the median progenitor is star-forming (blue) or quenched (red) at a given redshift.}%
    \label{fig:tracks_sigma1_bh}%
\end{figure*}

We investigate in Figure~\ref{fig:tracks_sigma1_bh} the SMBH growth history in the progenitors of star-forming and quiescent galaxies in TNG. We show the median evolutionary tracks in the $\log\Sigma_1-\log M_{\rm BH}$  and $\log M_*-\log M_{\rm BH}$  planes. We use here mock-observed $\Sigma_1$ values derived from the SKIRT images. Both star-forming and quiescent galaxies follow similar tracks.  The $\log\Sigma_1-\log M_{\rm BH}$ relation shows that both SMBHs and $\Sigma_1$ in TNG grow at low mass: in this regime, the SMBHs are in fast accretion mode which implies the availability of gas in the central regions. However, the growth of $\Sigma_1$ is less steep than the stellar mass. We speculate that this could be partly driven by 
 SN feedback efficiently suppressing the building up of a bulge component i.e. of a high-density stellar core (e.g.~\citealp{2015MNRAS.452.1502D,2020arXiv201209186L}). Even though also in TNG SN feedback has been shown to reduce the growth of SMBHs \citep{2021MNRAS.503.1940H,2021MNRAS.501.2210T}, the latter is not completely prevented and during this phase galaxies exhibit a disky/irregular morphology. Because we show observationally mocked values, biases in the observational methods that over estimate the values of $\Sigma_1$, especially at high redshift, can also contribute to decrease the apparent evolution of $\Sigma_1$ (See Appendix~\ref{app:sigma1}). 

We then see in the right panel of Figure~\ref{fig:tracks_sigma1_bh} that the morphological transformation seen in Figure~\ref{fig:tracks_mstar_re} corresponds to a decrease in the growth rate of the SMBH. This is a signature that the SMBH has gone into a low accretion mode. 
This transition may arise when SN feedback becomes inefficient at preventing gas cooling, as the galaxy becomes sufficiently massive \citep{2017MNRAS.465...32B, 2021MNRAS.501.2210T}. Within the TNG model, it has been shown that this transition corresponds to the time when the energy ever injected via low-accretion kinetic feedback mode equals and then surpasses the binding energy of the gas within the galaxy \citep{2020MNRAS.493.1888T}. Surprisingly, this transition also corresponds with an increase of the central stellar density even if less gas to form stars is in principle available. 

\begin{figure}
	\includegraphics[width=\columnwidth]{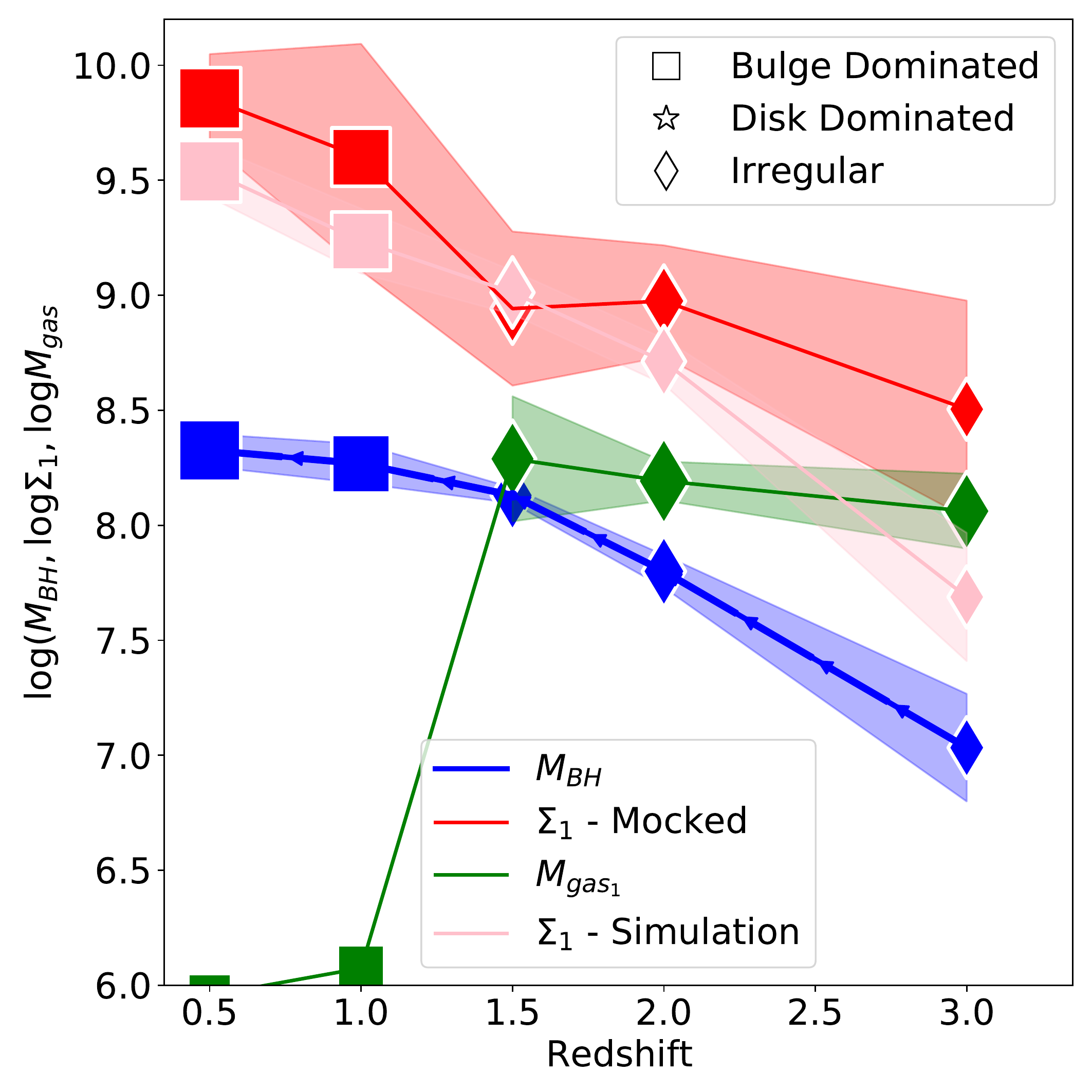}
		\includegraphics[width=\columnwidth]{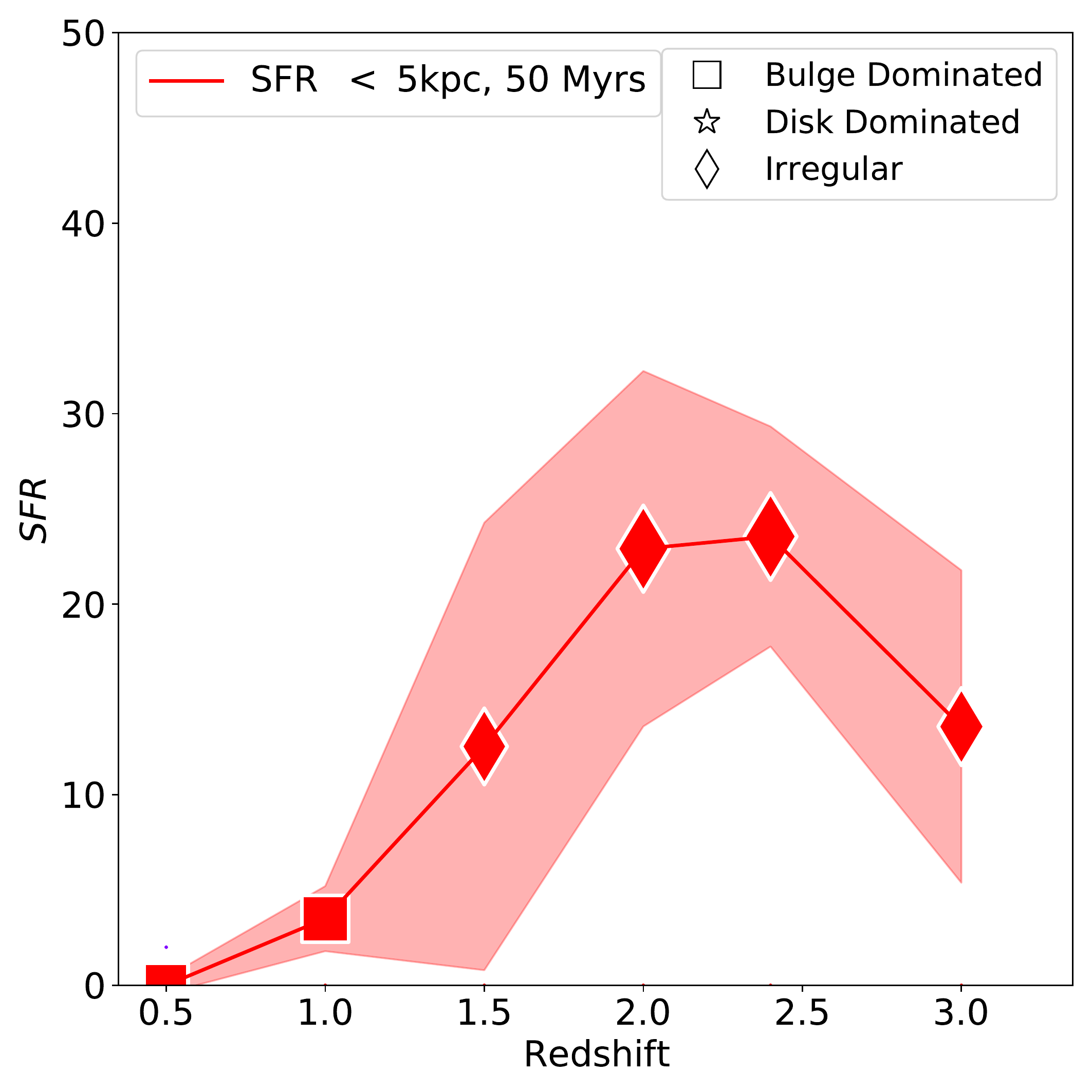}
    \caption{Median evolutionary tracks as a function of redshift of various galactic components of the progenitors of TNG50 quiescent galaxies with stellar masses between $10^{10.5}$ and $10^{11}$ at $z=0.5$, as those in Figs.~\ref{fig:tracks_mstar_re} and \ref{fig:tracks_sigma1_bh}. In the top panel, we show SMBH mass (blue), mock-observed central stellar mass density (red), and gas mass in the central kpc (green); in the bottom, SFRs in the central regions of galaxies. The symbols indicate the median apparent morphological type. The pink lines and symbols indicate the values of $\Sigma_1$ measured directly from the simulation output.} 
    \label{fig:tracks_vs_z}
\end{figure}
We further investigate this behavior by exploring the time evolution of the SMBH mass, $\Sigma_1$, gas mass in the central kpc, and central star formation rate in the progenitors of massive quiescent galaxies in Figure~\ref{fig:tracks_vs_z}. The appendix~\ref{app:ind_tracks} includes individual tracks of galaxies. We confirm that, for TNG50 massive galaxies selected at $z\sim0.5$, the SMBH grows mostly at $z>1$, while the galaxy has an irregular stellar morphology. During this phase, $\Sigma_1$ increases up to $\sim10^9$ $M_\odot$kpc$^{-2}$. 
At $z\lesssim1.5$, the gas in the center starts to be rapidly depleted which is translated into a decrease of the SFR. At the same time a morphological transformation is triggered, i.e. increase of observed central density and change of the rest-frame optical morphology. The decrease of gas pushes the SMBH into a low accretion mode, which turns on kinetic feedback and prevents more gas to be accreted. We notice again that $\Sigma_1$ continues growing at $z<1$, even when the gas fraction is low. The increase, therefore, does not seem to be driven by star formation.

\begin{figure}
	\includegraphics[width=\columnwidth]{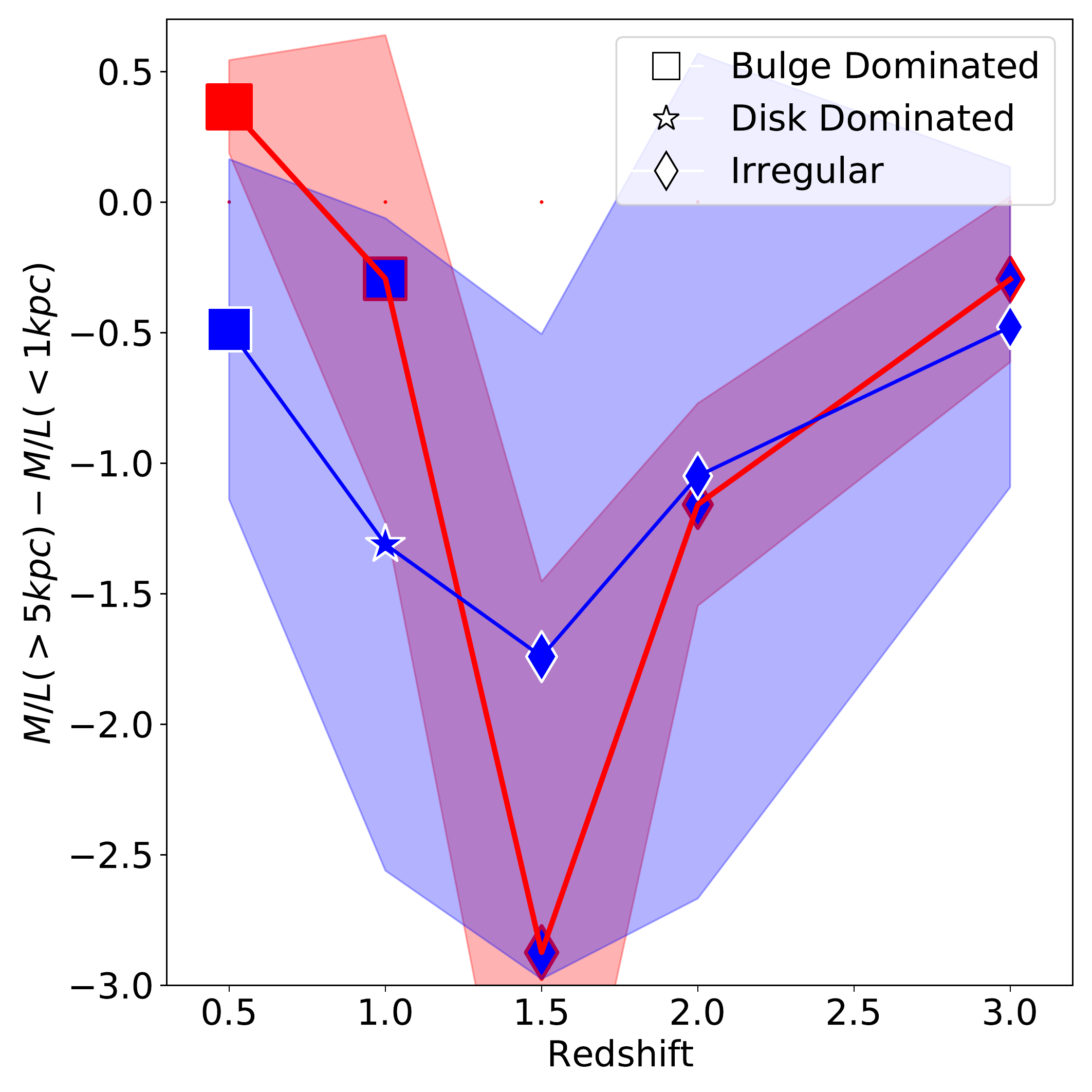}

    \caption{Evolution of the difference between M/L in the external parts ($>5$ kpc) and the inner parts ($<1$ kpc) of the progenitors  of star-forming and quiescent massive galaxies at $z=0.5$, according to TNG50. Annotations are as in Fig.~\ref{fig:tracks_mstar_re}. The evolution of the M/L difference is steeper for the progenitors of quiescent galaxies at $z<1$ at the offset of quenching. It suggests that fading contributes to enhance M/L gradients and therefore to the observed morphological changes. } 
    \label{fig:MoverL_vs_z}
\end{figure}

There are several factors that could explain the increase of $\Sigma_1$ even when the SFR is decreasing. Some redistribution of stars given the change in the gravitational potential can contribute to this effect. It can also be enhanced by observational effects. Figure~\ref{fig:tracks_vs_z} also shows the evolution of $\Sigma_1$ measured directly from the simulation output by computing the stellar density in the central kpc. The evolution is steeper at high redshift and shallower at low $z$ than for the observationally-derived quantity. Because of a variety of observational effects at high redshift, the \emph{mock-observed} $\Sigma_1$ values tend to be over estimated (see Appendix~\ref{app:sigma1}). At low redshift, because of the decrease in SFR, the fading of the stellar populations can also contribute to an apparent increase of the stellar density and a change in the observed morphology.  Figure~\ref{fig:MoverL_vs_z} shows the evolution of the difference between the outer and inner M/L ratios. At $z<1$, the 
evolution of M/L gradients for the progenitors of quiescent galaxies is steeper than for progenitors of star-forming galaxies. It suggests that fading also contributes to the observed morphological transformations at a characteristic stellar mass. A similar conclusion was reached by~\cite{2013ApJ...776...63F} in SDSS.  Appendix~\ref{app:sigma1} explores more in detail the differences between mocked and simulation-based measurements of $\Sigma_1$.

\subsection{Why do some galaxies quench earlier than others?}
An interesting question is why some galaxies quench earlier than others of the same stellar mass. We can see in Figure~\ref{fig:tracks_sigma1_bh} that in TNG SMBHs grow slightly faster in the progenitors of quiescent galaxies in that the slope of the stellar mass-SMBH  relation is steeper. Quiescent galaxies have therefore slightly more massive SMBHs at their centers which, within the TNG model, is most certainly the reason why they quench earlier.

\begin{figure*}
    \centering
    \vspace{-5 pt}
    \subcaptionbox{}{\includegraphics[width=0.46\columnwidth]{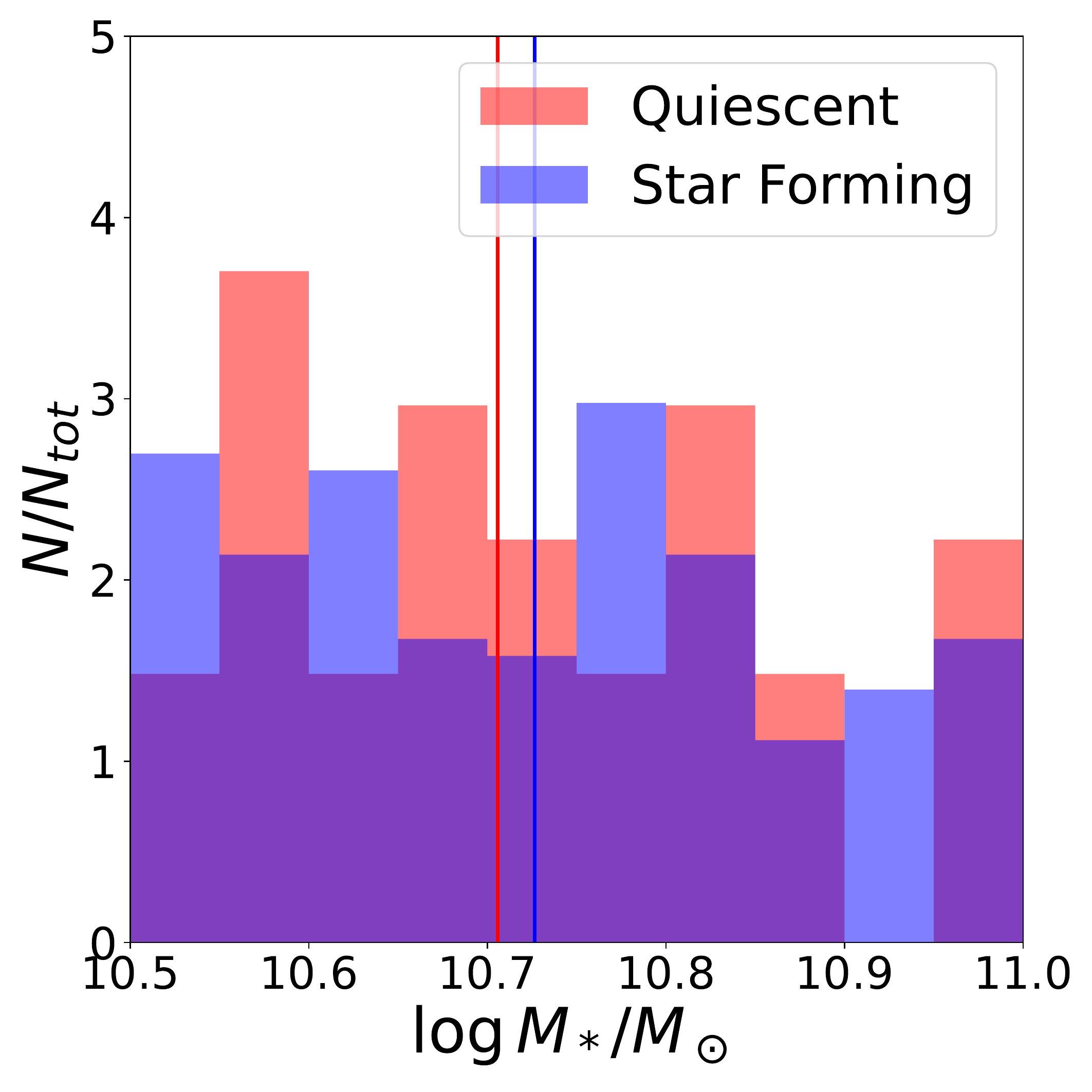}}%
     \qquad
      \subcaptionbox{}{\includegraphics[width=0.46\columnwidth]{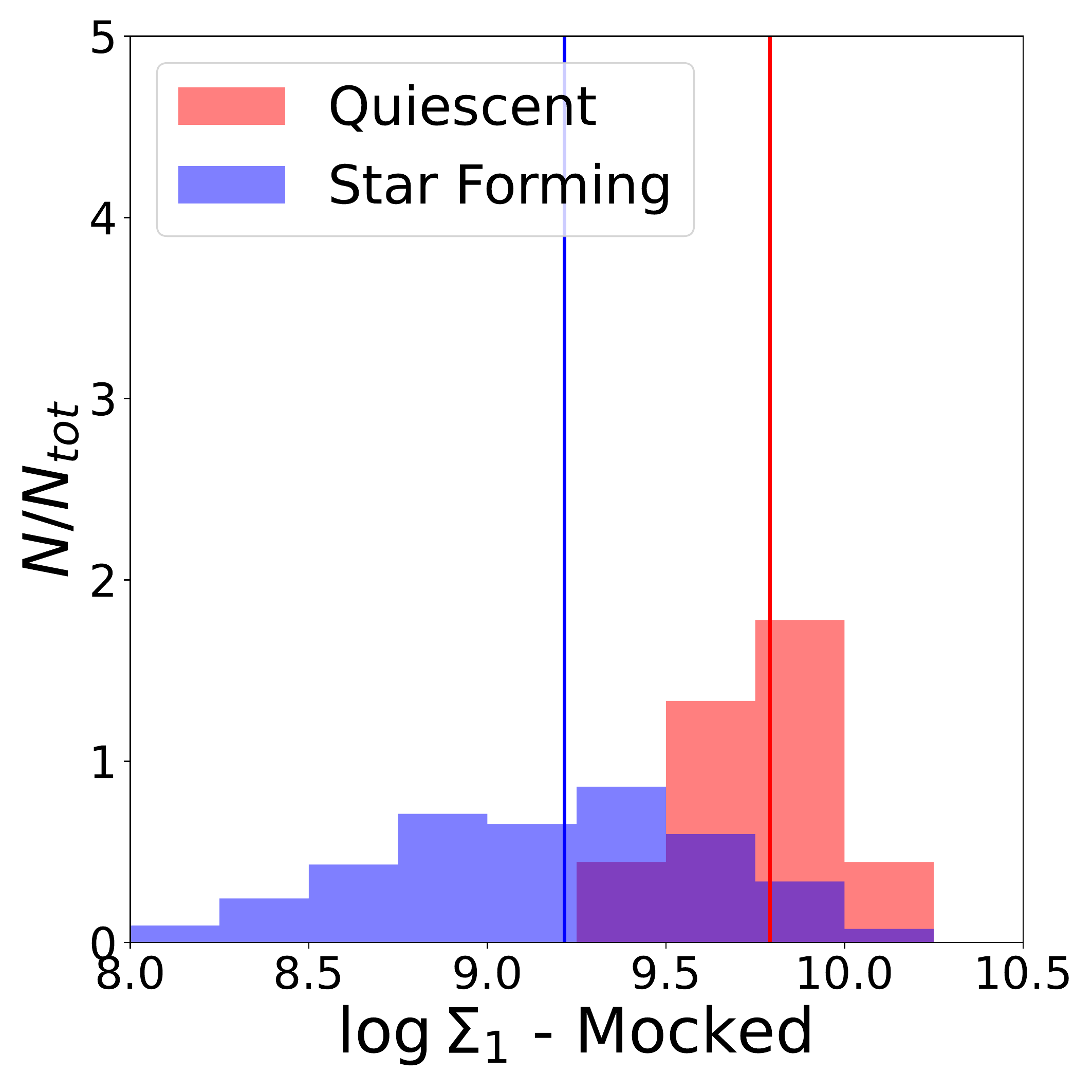}}%
     \qquad
      \subcaptionbox{}{\includegraphics[width=0.46\columnwidth]{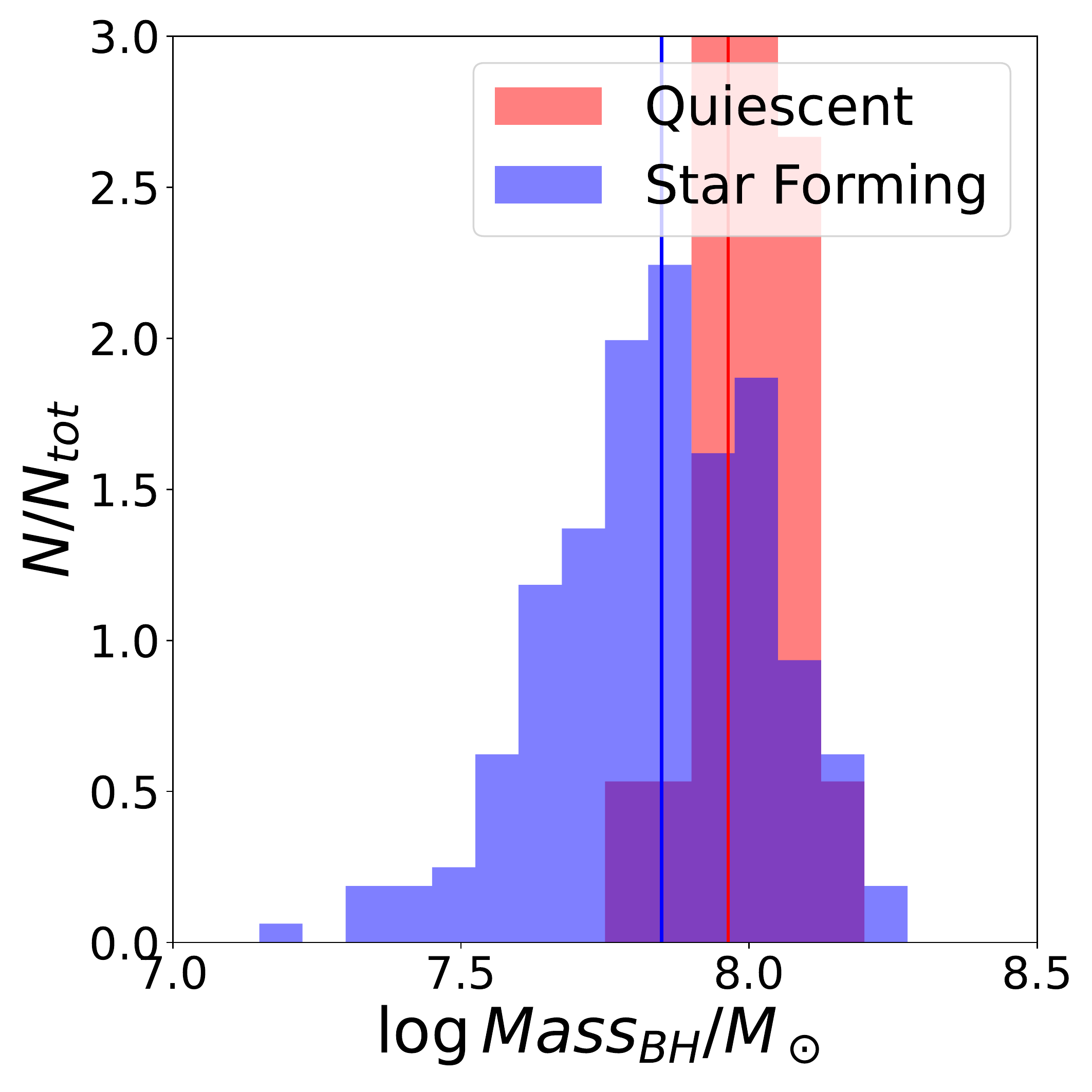}}%
     \qquad
      \subcaptionbox{}{\includegraphics[width=0.46\columnwidth]{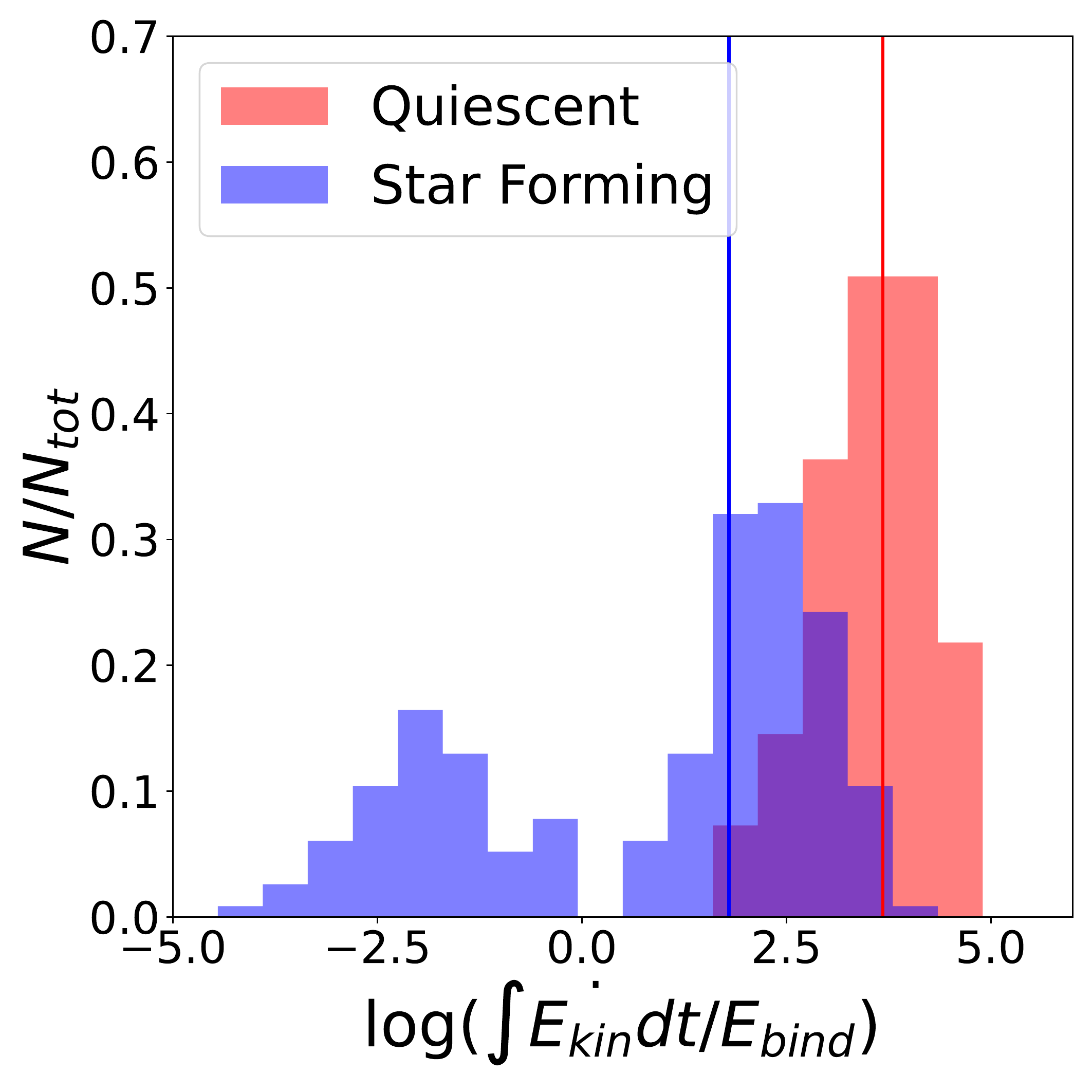}}%

      \vspace{-5 pt}
    \caption{ Distributions of (a) stellar mass, (b) mocked central stellar density, (c) black hole masses and (d) ratio of total kinetic energy and gas binding energy for star-forming (blue histograms) and quiescent (red histograms) galaxies at $z=0.5$ with stellar masses in the range $(10^{10.5}-10^{11})$ solar masses. The vertical lines show the median values.  }%
    \label{fig:energy}%
\end{figure*}

We explore this further in Figure~\ref{fig:energy}, which shows the distributions at $z=0.5$ of stellar masses, black hole masses, central stellar densities and the ratio of total kinetic energy injected by the SMBH over the gas binding energy for massive quiescent and star-forming galaxies. The latter is computed using the same approach as in~\cite{2020MNRAS.493.1888T}. We compute the total time-integrated amount of black hole-driven wind energy that has been released into the gas particles at each time-step ($\int\dot{E_{kin}} dt$) and divide by the gravitational binding energy of the gas within 5kpc defined as:
 $$ E_{\rm bind} = \frac{1}{2}\sum_{g(r<5kpc)} \phi_g m_g$$
 where $m_g$ is the mass of a gas cell, $\phi_g$ is the gravitational potential felt at this cell position. We use a fixed $5$ kpc aperture for the binding energy as it roughly corresponds to $2$ effective radii for galaxies in our sample.

 Figure~\ref{fig:energy} confirms that the stellar mass distribution of the star-forming and quiescent populations in TNG50 is very similar. This is expected by construction as we selected them to be in a narrow stellar mass bin. Galaxies that have already quenched have however larger central mass densities by almost a factor of $\sim10$ as already reported in Figure~\ref{fig:tracks_mstar_re} as well as SMBHs $\sim 30\%$ more massive. The most striking difference though resides in the ratio between injected kinetic energy and binding energy. The ratio is a factor of a $\sim100$ larger in quiescent galaxies than in star-forming galaxies of the same stellar mass. This confirms that kinetic feedback is the main cause for quenching. A similar trend was already reported at $z=0$ by~\cite{2020MNRAS.493.1888T}.

A logical subsequent question is why some galaxies experience a faster growth of their central SMBHs than others of similar stellar mass, ending up in an earlier quenching. A possibility is that the evolution depends on the galaxy properties. In fact, Figure~\ref{fig:tracks_mstar_re} shows that galaxies that quench tend to be more compact (smaller effective radius and larger $\Sigma_1$ values) than galaxies that don't, even during the star-forming phase. This is interesting because it suggests that small star-forming galaxies tend to quench earlier than extended ones because their SMBHs grow more efficiently. More compact sizes implies indeed that the density of gas is higher and therefore allows for a faster growth of the SMBH. 
A similar conclusion has been recently reached by~\cite{2020arXiv201108198G} with the TNG100 simulation, but using a different approach and focusing on galaxies selected at $z=2$. Those authors find that extended galaxies have lower SMBH masses and therefore a weaker or delayed kinetic-mode feedback, which in turn delays quenching of star formation in these systems. The work by~\cite{2021MNRAS.tmp..841W} with TNG100 at lower redshifts also suggests that dense galaxies have a steeper increase of the central density than galaxies with diffuse cores. They claim that this is mostly due to differences in the angular momentum of accreting gas. Finally, the connection between stellar size, compactness, and quenching had also been originally discussed in the context of TNG100 by \citealt{genel2018MNRAS.474.3976G}, who found that the main-sequence high-redshift progenitors of quenched $z=0$ galaxies are drawn from the lower-end of the size distribution of the overall population of main-sequence high-redshift galaxies. 

If the initial size is an important parameter to predict the future evolution of a galaxy, an interesting question that arises is why there are small and large galaxies of the same stellar mass, i.e. the origin of the observed scatter in the mass-size relation (e.g.~\citealp{vanderwel2014ApJ...788...28V}). Some works have highlighted that the galaxy size is tightly correlated with the dark-matter halo virial radius (e.g~\citealp{2013ApJ...764L..31K,2018MNRAS.473.2714S}). This would imply that the scatter in the mass size relation is partly a consequence of the scatter in the stellar mass - halo mass relation. Variations in halo spin parameter and halo concentrations can also contribute to the observed scatter as pointed out by previous works.  It is an interesting question to be addressed in dedicated, future work.

 \subsection{Comparison with other models}

 We have presented how the mocked stellar scaling relations of massive galaxies at intermediate redshifts are built in the TNG50 simulation. We have shown that the evolution of the stellar structure of galaxies in TNG50 is tightly correlated with the growth and feedback of the SMBH. One key prediction of the model is that SMBHs grow predominantly when galaxies are relatively low mass, with $\log M_*/M_\odot < 10.5$, which for massive low-redshift galaxies implies their SMBHs grew mostly at higher redshift. Other works have attempted to explain the scaling laws of star-forming and quiescent galaxies, most recently e.g. \citealt{2021arXiv210105820S} with ZFOURGE and CANDELS data.
 
 The recent empirical BHvH (Black Hole versus Halo) model by \cite{2020ApJ...897..102C} also attempts to describe the evolution of the stellar scaling relations of galaxies from $z\sim3$. In the BHvH model, feedback from the SMBH is assumed in the BHvH to be the main cause for quenching star formation in galaxies, as it is the case for TNG50. The key assumption of the model is that galaxies quench when the total emitted SMBH radiation is a multiple of the halo gas binding energy. This threshold is achieved at lower stellar masses for small galaxies than for large ones, which explains the tilt of the quenched ridgeline in the mass-size plane. One key prediction of the model is that the SMBH rapidly grows when the galaxy starts quenching and enters the Green Valley and when the central stellar density is high. 

\begin{figure}
	\includegraphics[width=\columnwidth]{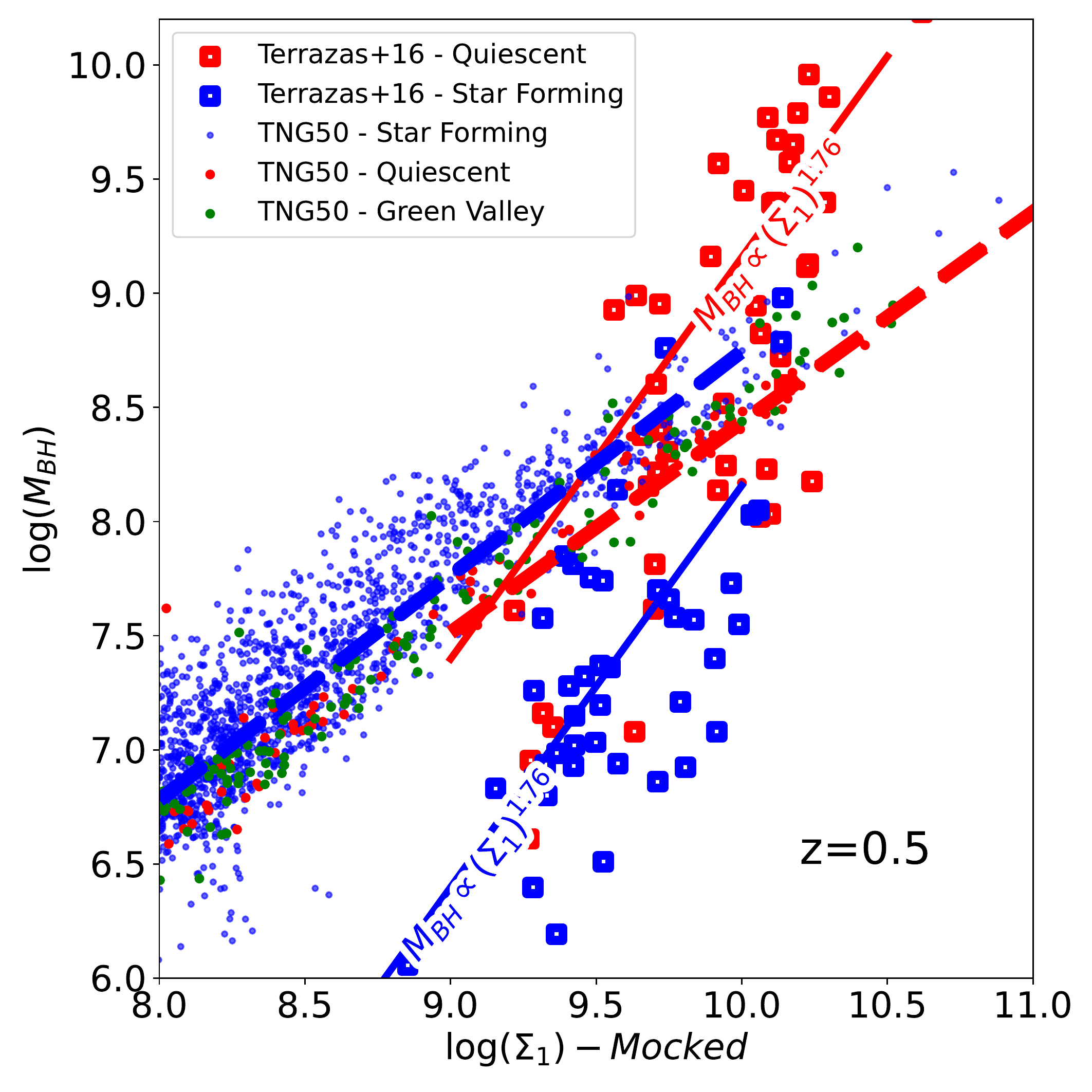}
    \caption{$\log\Sigma_1-\log M_{\rm BH}$ relation in TNG50 at $z=0.5$ for star-forming (blue dots), green-valley (green dots) and quenched galaxies (red dots). The red and blue dashed lines indicate the best fit power law relation for quiescent and star-forming galaxies respectively. The red and blue squares indicate observations from~\protect\cite{2017ApJ...844..170T} of a sample of $z=0$ galaxies with measured SMBH masses. The values of $\Sigma_1$ for each galaxy have been computed using the stellar mass and the best fit relation from~\protect\cite{2017ApJ...840...47B} at $z=0.5$. The red and blue solid lines indicate the relation used in the model by~\protect\cite{2020ApJ...897..102C}, which is calibrated on observations by~\protect\cite{2017ApJ...844..170T}.} 
    \label{fig:sigma1_BH}
\end{figure}

 Both models, TNG50 and BHvH, therefore predict that quenching happens when the total amount of injected energy by the BH is larger than the gas binding energy by some factor. We note however that the definition of gas binding energies are fundamentally different in the two models. Here we considered the energy within 5kpc while the BHvH model refers to the binding energy of the hot halo gas. Moreover, one additional difference between the two models resides in the way BHs grow and how this correlates to structural evolution. As we have seen in the previous subsection, in TNG50, the SMBH grows fast at lower stellar masses together with the central stellar density. The SMBHs  acquire most of their final mass via gas accretion before the stellar core of the galaxy is totally built. In the empirical model, however, the central stellar density is assumed to grow first. Most of the SMBH growth happens when the galaxy has almost acquired its final central density and started quenching. Figure 4 of~\cite{2020ApJ...897..102C} schematically shows that the SMBHs grow by a factor of $\sim 10$ when the galaxy starts leaving the Star Formation Main Sequence while $\Sigma_1$ remains almost constant. A recent work by~\cite{2020arXiv201209186L} using the newHorizon simulation~\citep{2020arXiv200910578D} also finds for a subsample of galaxies that the increase of $\Sigma_1$ precedes the growth of the super massive black hole. The differences with the TNG model might come from the way SN feedback is implemented. In the BHvH model, this is partly motivated by observational constraints from~\cite{2017ApJ...844..170T} in a sample of $z=0$ galaxies with measured SMBH masses. The~\cite{2017ApJ...844..170T} sample shows indeed that quiescent galaxies have more massive SMBHs (by a factor of 10) than star-forming galaxies of the same mass. \cite{2020ApJ...897..102C} also point out that the Milky Way and Andromeda have similar central mass densities but very different SMBH masses.  The measured relation between stellar mass and central density \citep{2017ApJ...840...47B} implies that, at fixed $\Sigma_1$, quiescent galaxies have more massive BHs. This is interpreted by the BHvH model as an evidence of SMBH growth in the green valley with little growth of the central density. Figure~\ref{fig:sigma1_BH} shows the $\log\Sigma_1-\log M_{\rm BH}$ relation for simulated galaxies at $z=0.5$ together with the observational points by~\cite{2017ApJ...844..170T}. We emphasize that the $\Sigma_1$ values are computed from the stellar masses using the best fit scaling law from~\cite{2017ApJ...840...47B}. If one assumed that the observational sample of T17 was representative and unbiased, then TNG50 does not reproduce these observed data points. By construction, the BHvH model fits well the relation. It is worth emphasizing however that the~\cite{2017ApJ...844..170T} sample is not complete and therefore might not be representative of the global population of galaxies. Moreover, the~\cite{2017ApJ...844..170T} sample is at $z=0$ while we are studying $z=0.5$ galaxies. More measurements of BH masses are needed to rule out or not a BH growth model.

   Some previous works have also suggested that stellar cores can be rapidly built trough inflows of gas towards the galaxy centers (i.e. compaction events,~\citealp{2015MNRAS.450.2327Z}). In these simulations without Black Holes, these events are not directly linked to feedback and are typically  triggered by a variety of mechanisms, including mergers, fly-bys or counter rotating streams of incoming gas. This does not seem to be the case in TNG50. However, even if the trigger is different, gas compaction could also be present in TNG50 and could be a plausible mechanism to explain the growth of the SMBH and central stellar densities. We performed a simple test to have a first order indication. \cite{2018ApJ...858..114H} trained a CNN to identify compaction events on mock HST images from the zoom in VELA simulation suite. We used the trained CNN model out-of-the-box to classify the mock images of TNG50 galaxies in three classes as done in the aforementioned work (i.e. pre-compaction, compaction and post-compaction). We then compute the median probability of being in a compaction phase.  We observe that in the snapshots right before quenching is triggered, the galaxies are classified as being in a compaction phase with high confidence. It also corresponds with the peak in gas mass. This suggests that the physical conditions of compaction could be also observed in TNG50 although more work is needed to explore this in detail. It is worth noticing that the same is observed for star-forming galaxies. Therefore the compaction event seems to happen at a critical mass independently of quenching, corresponding with the decrease of efficiency of SN feedback.


\section{Summary and Conclusions}
We have analysed the \emph{mocked} structural and morphological properties of a mass complete ($\log M_*/M_\odot>9$) sample of galaxies between $z=0.5$ and $z=3$ in the TNG50 simulation of the IllustrisTNG project as compared to observations from the CANDELS survey. In order to ensure a fair comparison between observed and simulated samples, we have generated mock HST images of TNG50 galaxies using the radiative transfer code SKIRT. We then have used the images to estimate visual like optical rest-frame morphologies using a CNN trained on observations and structural parameters by fitting a 2D Sersic model to the surface brightness distributions.\\

In the first part of the work, we have compared the observed evolution of morphological fractions and scaling relations in CANDELS and TNG50. Our main findings are as follows:

\begin{itemize}
    \item TNG50 reproduces overall the global observed trends of the morphological fractions with redshift and stellar mass for both quiescent and star-forming galaxies. Star-forming galaxies are dominated by irregular/clumpy systems at $z>1$ and by symmetric disks at lower redshifts. Quiescent galaxies are more bulge dominated at all redshifts, especially at the high-mass end. However, the TNG50 simulation tends to under predict the abundance of low-mass spheroids by $\sim20\%$ at the expense of an over prediction of quiescent irregular systems. We show that these are typically galaxies with dense stellar cores surrounded by star-forming rings. The TNG50 simulation also under predicts the fraction of massive quiescent galaxies, especially at high redshift. 
    
    \item The $\log M_*-\log R_e$ and $\log M_*-\log \Sigma_1$ relations are also well captured by the simulation between $z=0.5$ and $z=3$. Both the slope and normalization of star-forming and quenched simulated galaxies fall within $1\sigma$ of the observed best-fit values. Quiescent galaxies in TNG50 have smaller effective radii than star-forming galaxies of a similar stellar mass and also have higher central mass densities, as reported in several observational works. The difference in $\log\Sigma_1$ between the star-forming and quiescent populations seems to be more pronounced when fitting Sersic models to the mock images to estimate the central density than when the central density is measured from the simulation output directly.
  
\end{itemize}

Based on the reported good agreement, we have then investigated how massive ($\log M_*/M_\odot>10.5$) galaxies at $z=0.5$ change their morphology, quench and evolve in the \emph{observed} scaling laws since $z\sim3$:

\begin{itemize}
    \item On average, star-forming galaxies evolve along the scaling laws, particularly along the mass-size relation, increasing mass, size and stellar density from $z=3$. On the other hand, when TNG50 massive galaxies selected at $z=0.5$  quench, typically at $z<1$, they follow almost downward (upward) vertical tracks in the $\log M_*-\log R_e$ ($\log M_*-\log \Sigma_1$) planes to reach the quiescent ridgelines. 
    
    \item  TNG50 massive galaxies experience on average a morphological transformation from disky irregular/clumpy systems to symmetric Hubble-type systems in the main sequence at a characteristic stellar mass of $\sim 10^{10.5}~M_\odot$. This is translated into a change of slope in the track in $\log M_*-\log \Sigma_1$ plane. 
    
     \item We show that the morphological transformations of massive galaxies in TNG50 are tightly correlated to the activity of their SMBHs. The SMBH grows fast when the galaxy is low mass, and together with the central stellar density. When the SMBH goes into the low-accretion, kinetic feedback mode, the gas is rapidly depleted in the central regions, the galaxy increases its observed central mass density and eventually quenches. We show that fading of the stellar populations is a key contributor to this apparent morphological transformation. The SMBHs grow slightly more efficiently in galaxies that are initially more compact, which therefore end up quenching earlier.
    
    \item The TNG50 model predicts therefore that black holes grow first or in parallel with the central stellar densities of galaxies. At fixed SMBH mass, quiescent galaxies have larger $\Sigma_1$ values. We show that this is potentially in tension with some observational measurements of black hole masses in the local universe. The picture from TNG50 is in contrast with the model proposed by~\cite{2020ApJ...897..102C}, wherein SMBHs grow more during quenching (Green-Valley). Future data can help constrain between these differences in the details of how galaxy quenching and morphological transformation are related.

\end{itemize}

\section*{Acknowledgements}

SV thanks ERASMUS+ for providing funding his position at the Instituto de Astrofisica de Canarias. The TNG50 simulation was realised with compute time granted by the Gauss Centre for Super-computing (GCS) under the GCS Large-Scale Project GCS-DWAR (2016; PIs Nelson/Pillepich). 

\section*{Data Availability}
The IllustrisTNG simulations, including the most recent TNG50, are publicly available and accessible at \url{www.tng-project.org/data} \citep{2019ComAC...6....2N}. Observational data from the CANDELS survey is also publicly available at \url{http://arcoiris.ucolick.org/candels/}.



\bibliographystyle{mnras}
\bibliography{BODY.bib} 

\appendix

\section{Observational effects on the measured galaxy properties}
\label{app:sigma1}

\begin{figure*}
    \centering
    \vspace{-10 pt}
    \subcaptionbox{}{\includegraphics[width=0.3\textwidth]{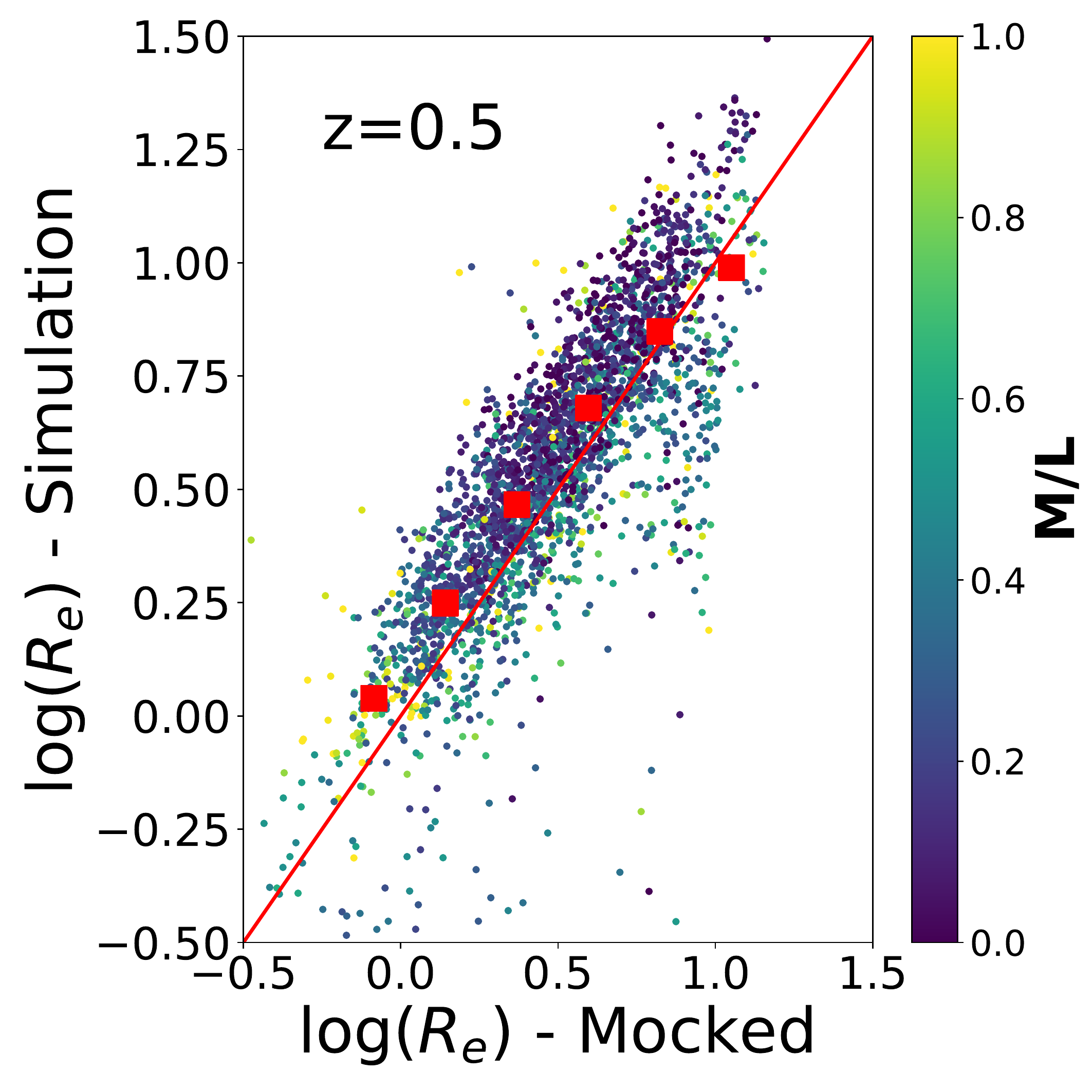}}%
     \qquad
      \subcaptionbox{}{\includegraphics[width=0.3\textwidth]{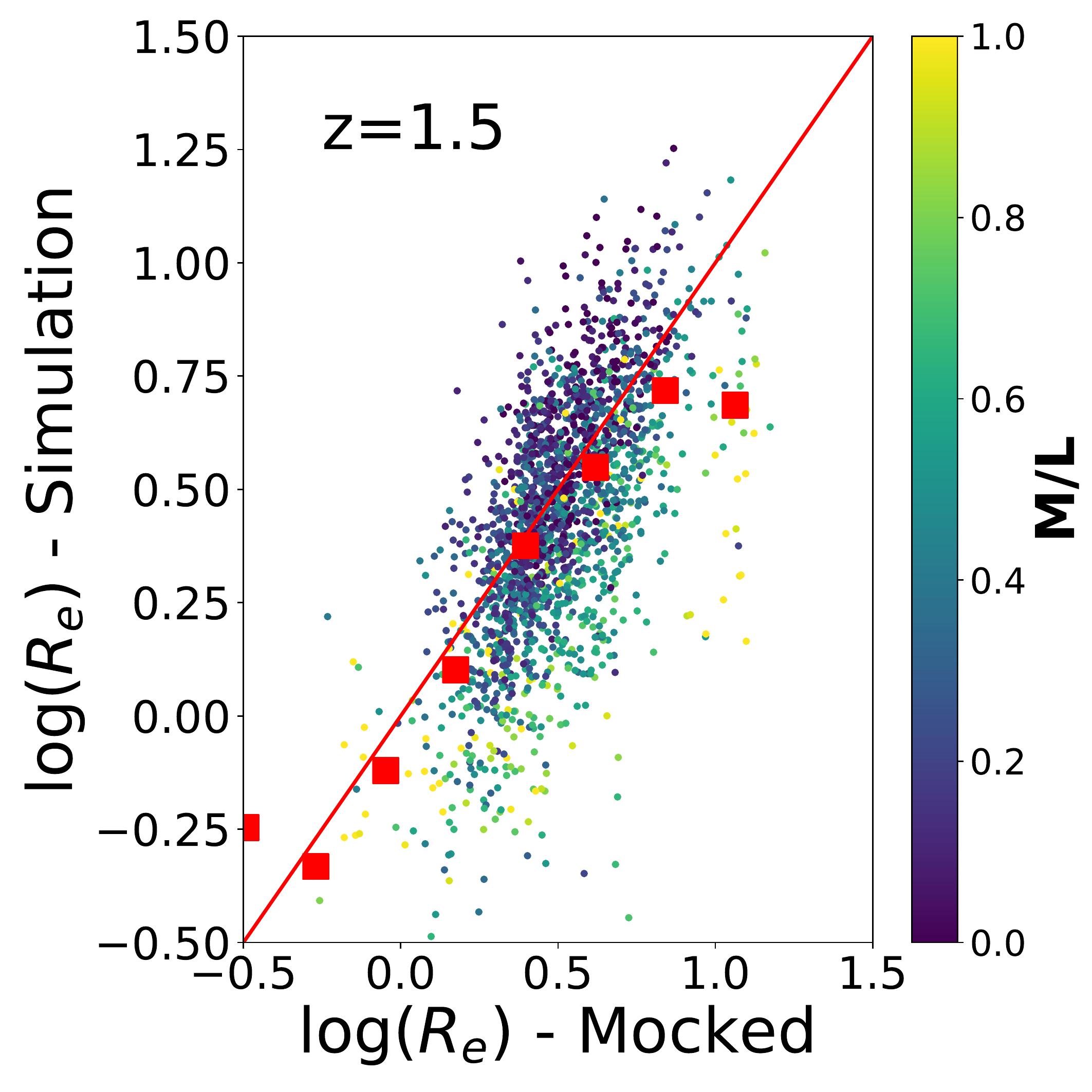}}%
     \qquad
    \subcaptionbox{}{\includegraphics[width=0.3\textwidth]{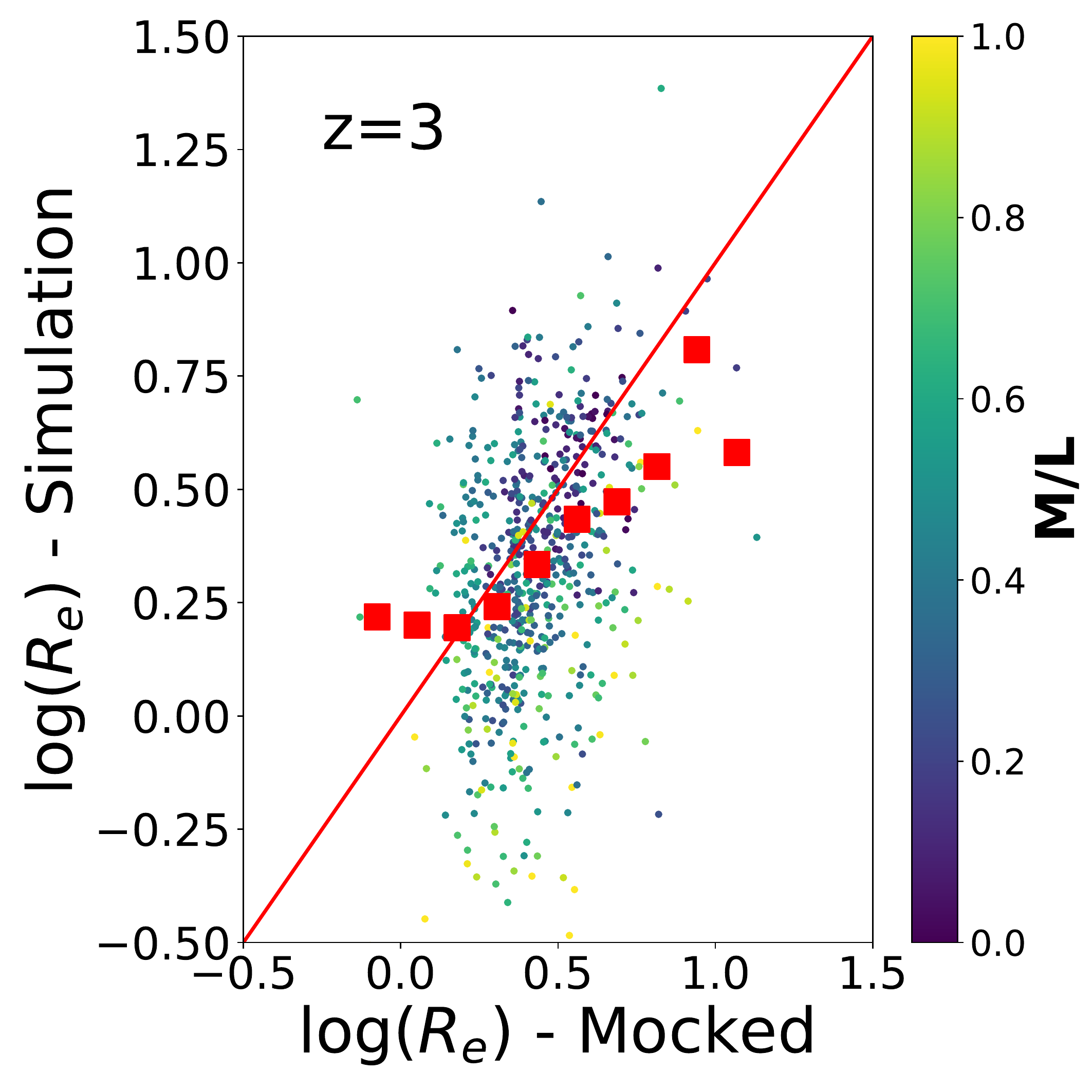}}%
    \qquad
    \subcaptionbox{}{\includegraphics[width=0.3\textwidth]{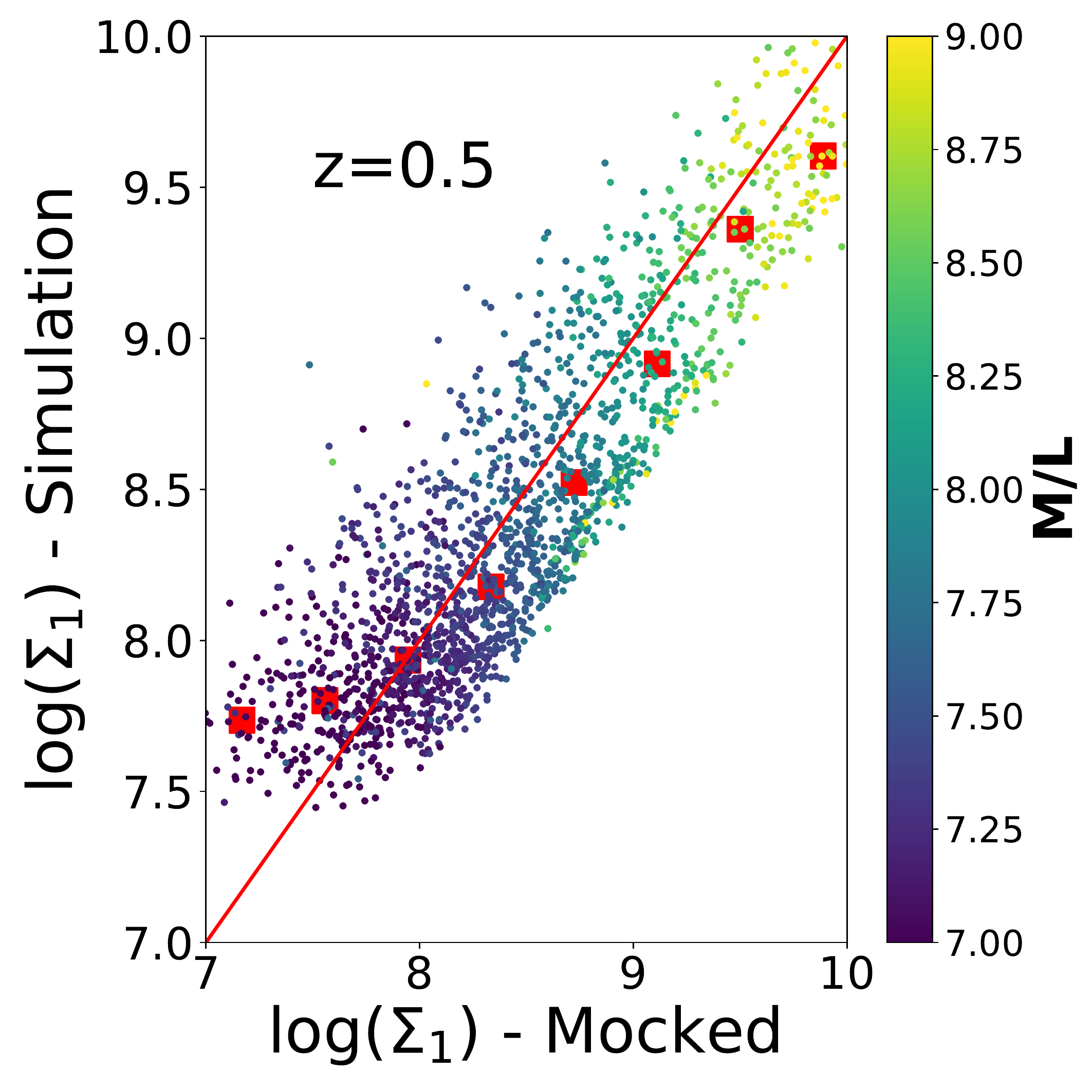}}%
    \qquad
    \subcaptionbox{}{\includegraphics[width=0.3\textwidth]{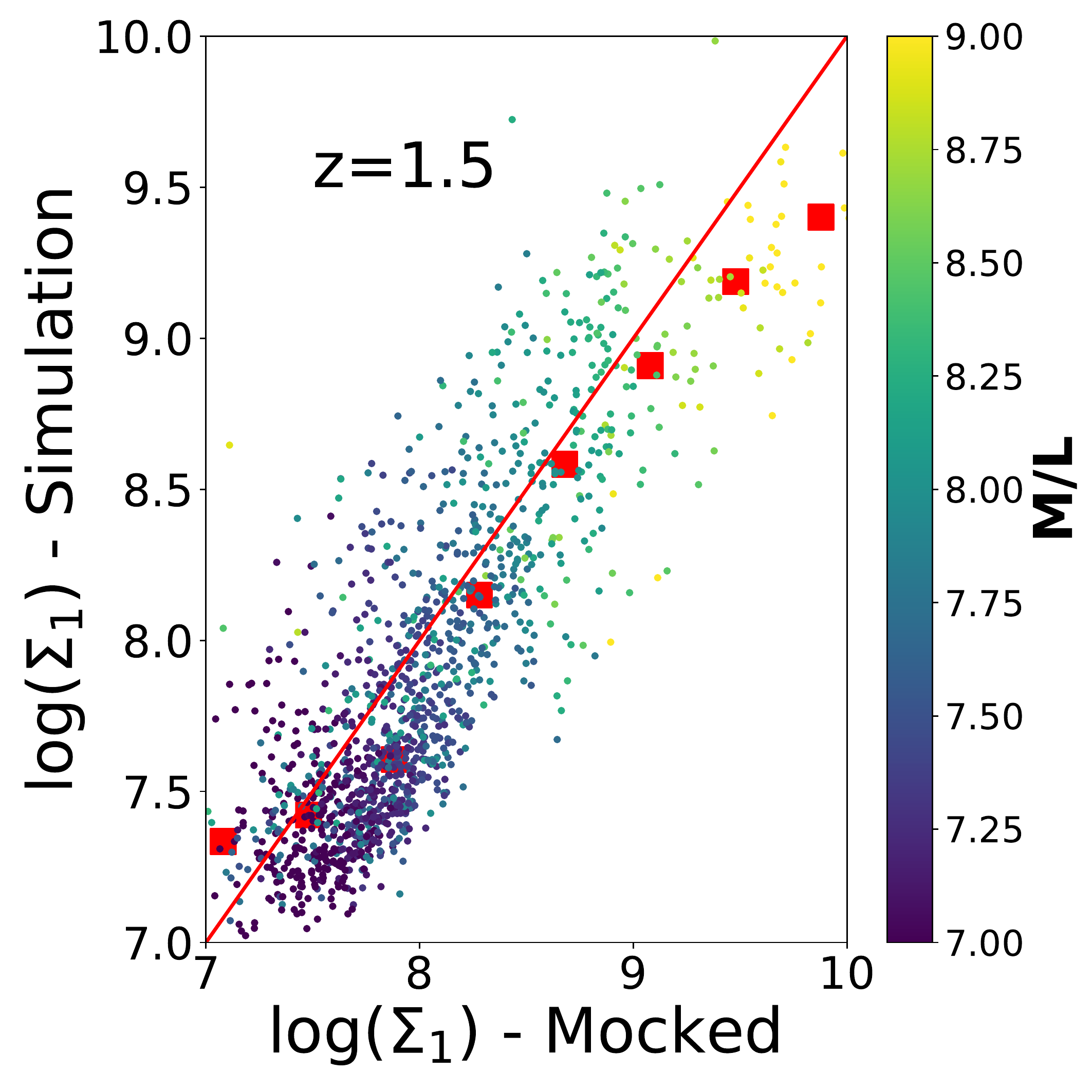}}
    \qquad
    \subcaptionbox{}{\includegraphics[width=0.3\textwidth]{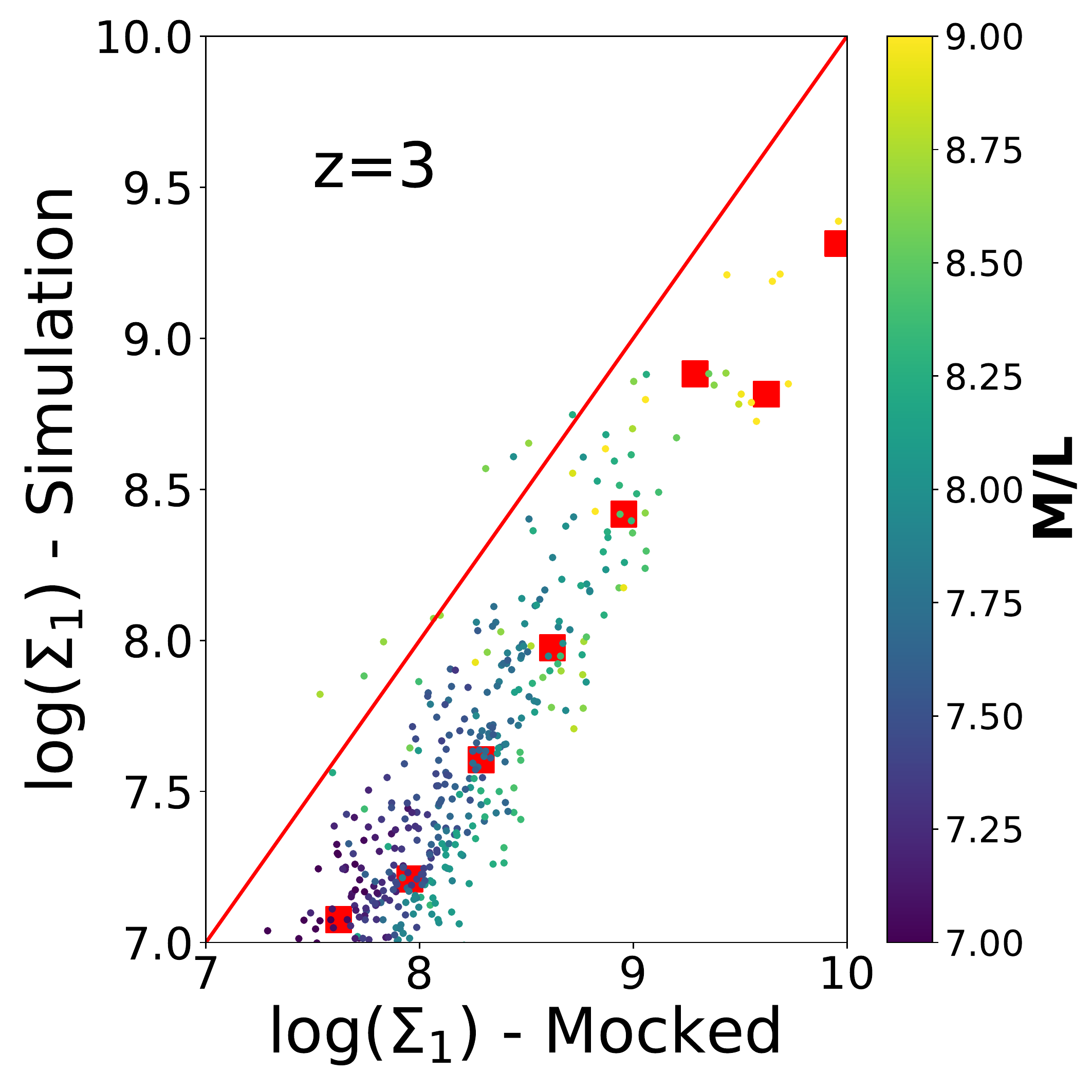}}

      \vspace{-5 pt}
    \caption{Comparison between the mocked and simulated effective radii (top row) and central stellar mass densities (bottom row). The color code shows the mass to light ratio within two effective radii. The large red squares are the median values in different bins. The different panels show different redshift slices as labeled. }%
    \label{fig:s3D_re}%
\end{figure*}

In the previous sections we have applied the same techniques used in observations to mock images including most of the observational effects (e.g. PSF, noise). This enables a consistent comparison with observations. In the simulation, we have also access to the true intrinsic values of the effective radii and central mass densities measured directly on the stellar particles. We can therefore assess the impact of observational effects such as resolution or noise as well as of the conversion from light to mass on the derived properties. In figure~\ref{fig:s3D_re} we show a comparison between the intrinsic 3D effective radius and the central stellar density and the mocked values. We observe a clear correlation between the two quantities. Nevertheless, there are also some noticeable effects. First of all, there is a redshift dependence. At $z>2$, there is a weaker correlation between the 3D half mass size and the measured one. The measured central stellar mass density is also over estimated by a factor of $\sim5$. We speculate that this could be due to a combination of low SNR and spatial resolution. Another contributing factor could be that the Sersic profile may not provide an adequate description of the typical profiles of high-redshift galaxies in the simulation. At lower redshift, although there is a tighter correlation, we see that the mass to light ratio (M/L) has a noticeable impact on the derived effective radii and central stellar densities. For high M/L values, the light based effective radii are overestimated while the opposite effect is seen for central stellar densities.

\begin{figure*}
    \centering
    \vspace{-10 pt}
    \subcaptionbox{}{\includegraphics[width=0.3\textwidth]{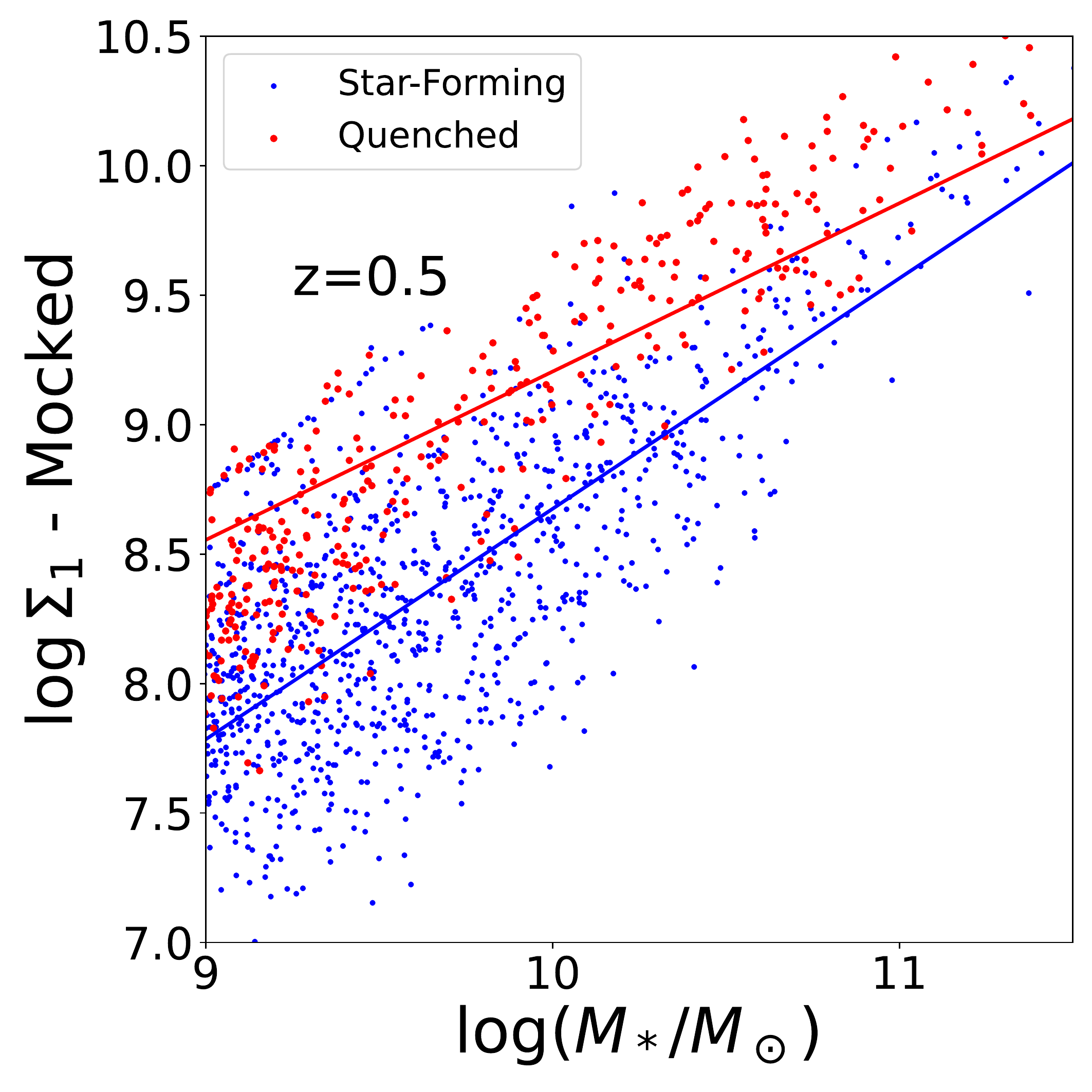}}%
     \qquad
     \subcaptionbox{}{\includegraphics[width=0.3\textwidth]{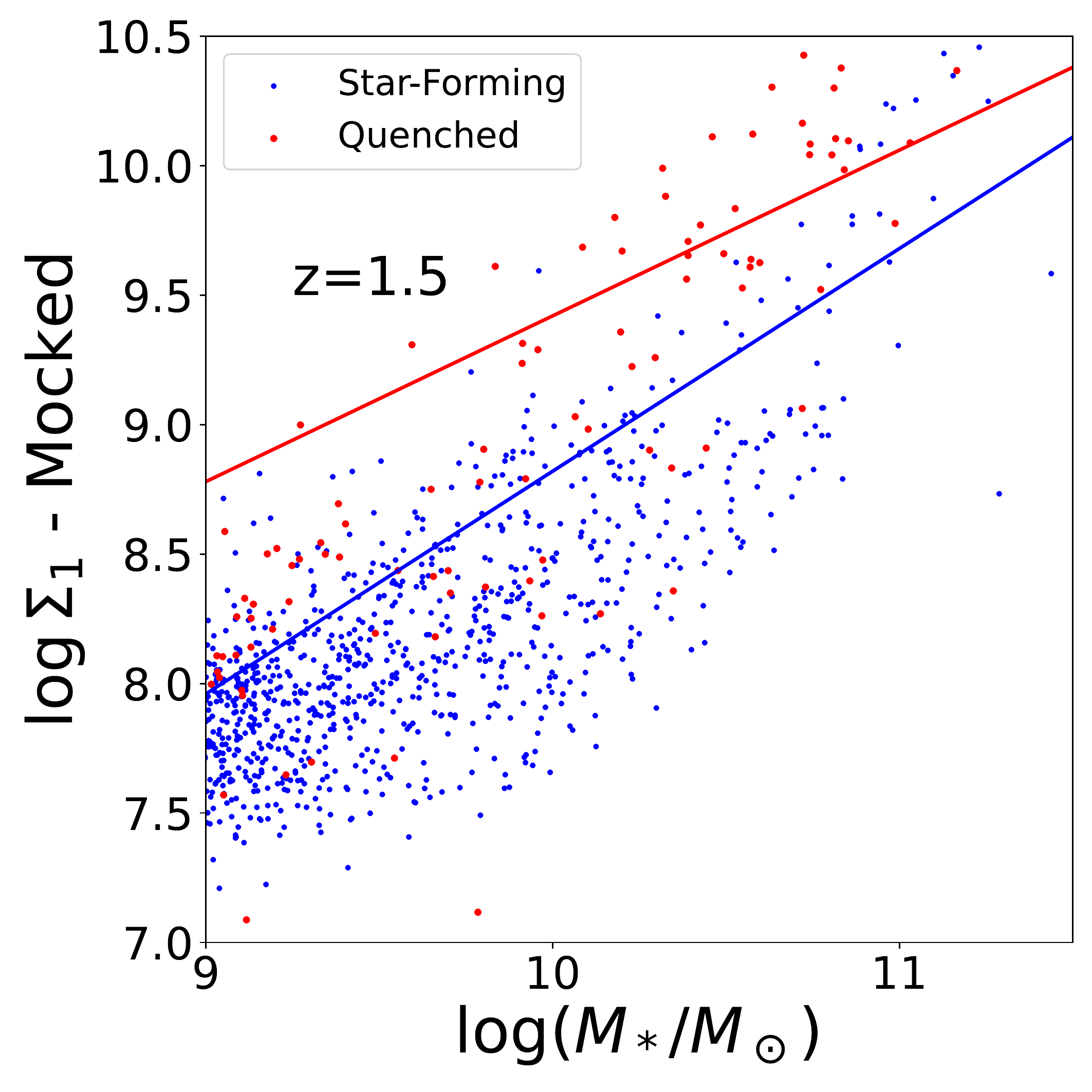}}%
      \qquad
        \subcaptionbox{}{\includegraphics[width=0.3\textwidth]{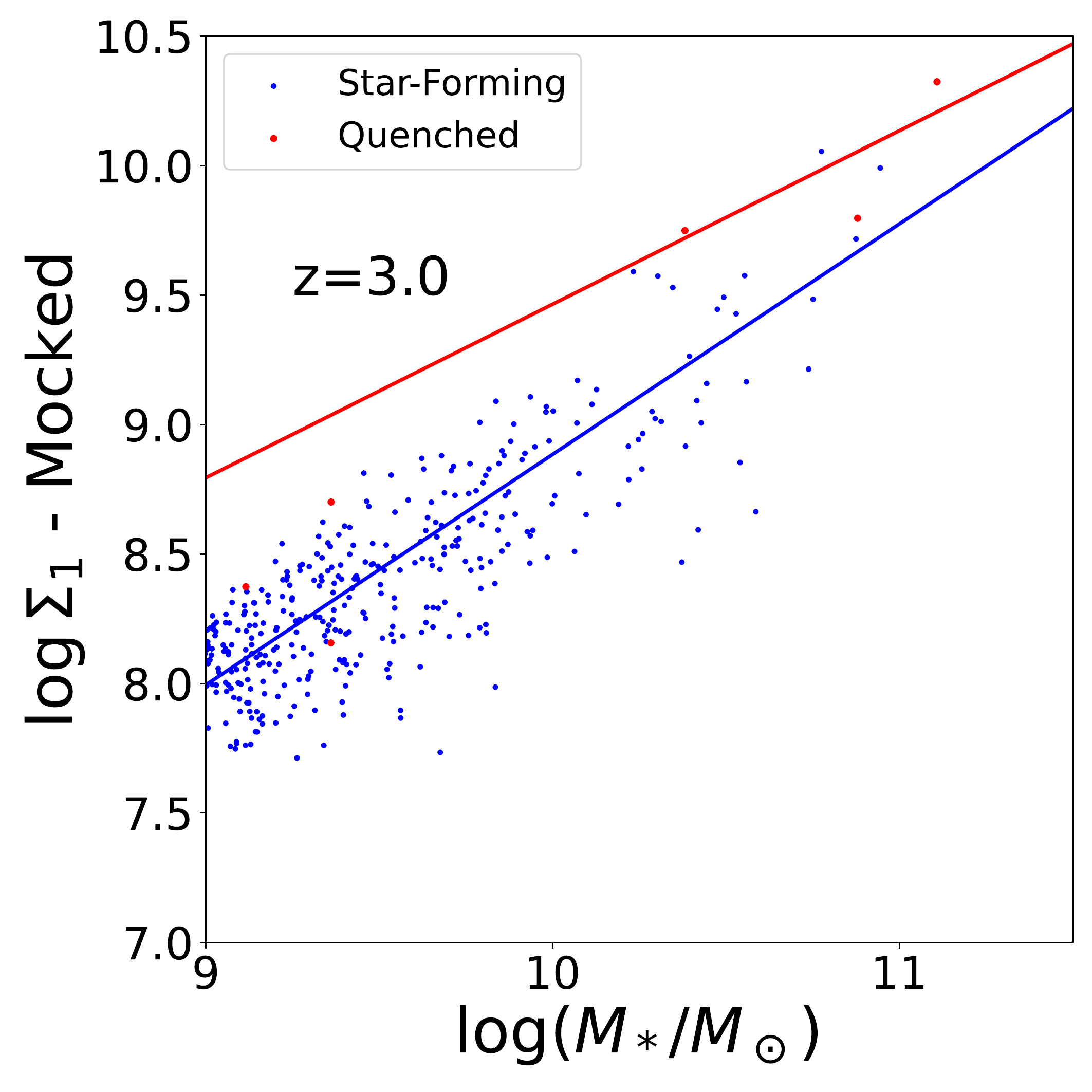}}%
        \qquad
      \subcaptionbox{}{\includegraphics[width=0.3\textwidth]{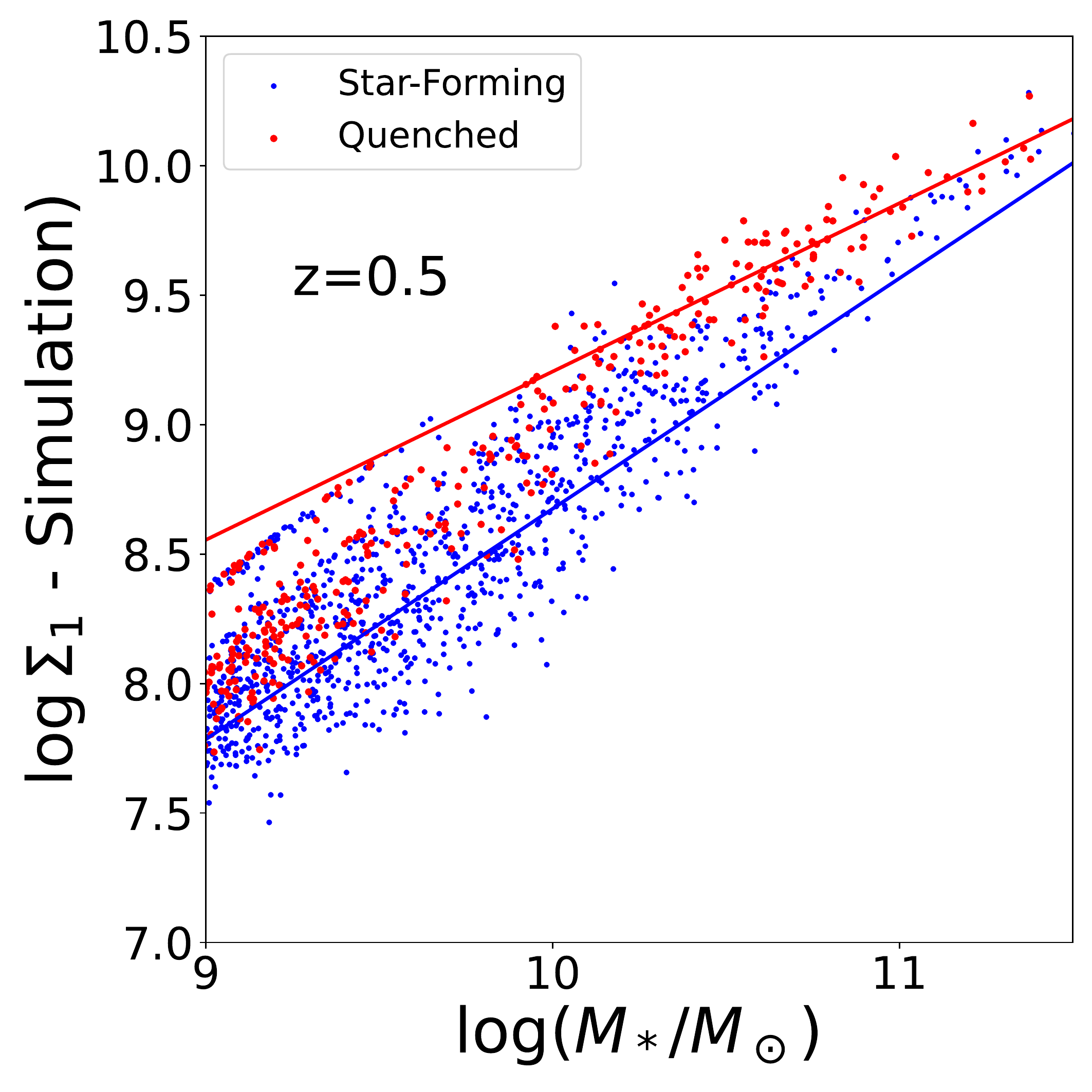}}%
    \qquad
      \subcaptionbox{}{\includegraphics[width=0.3\textwidth]{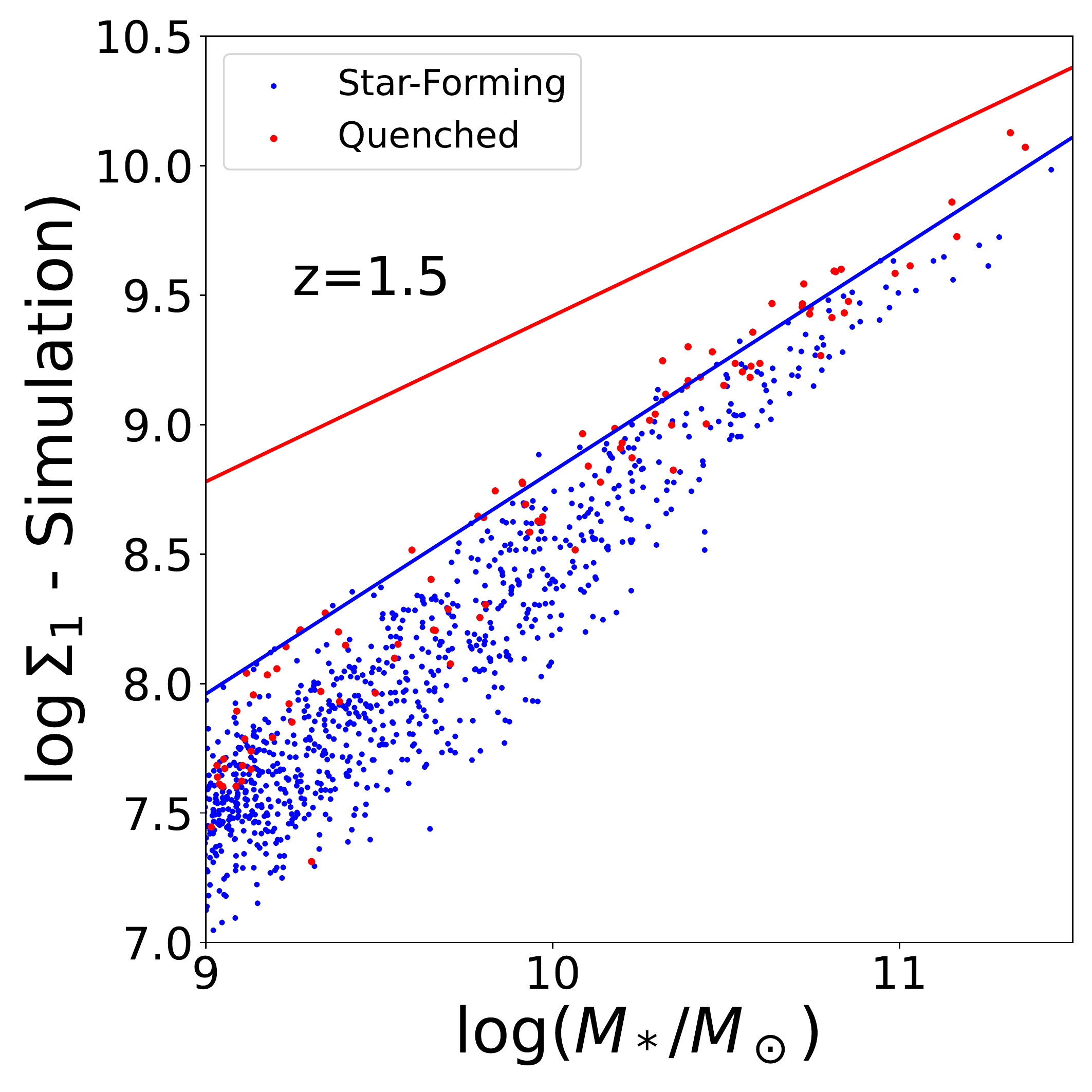}}
        \qquad
      \subcaptionbox{}{\includegraphics[width=0.3\textwidth]{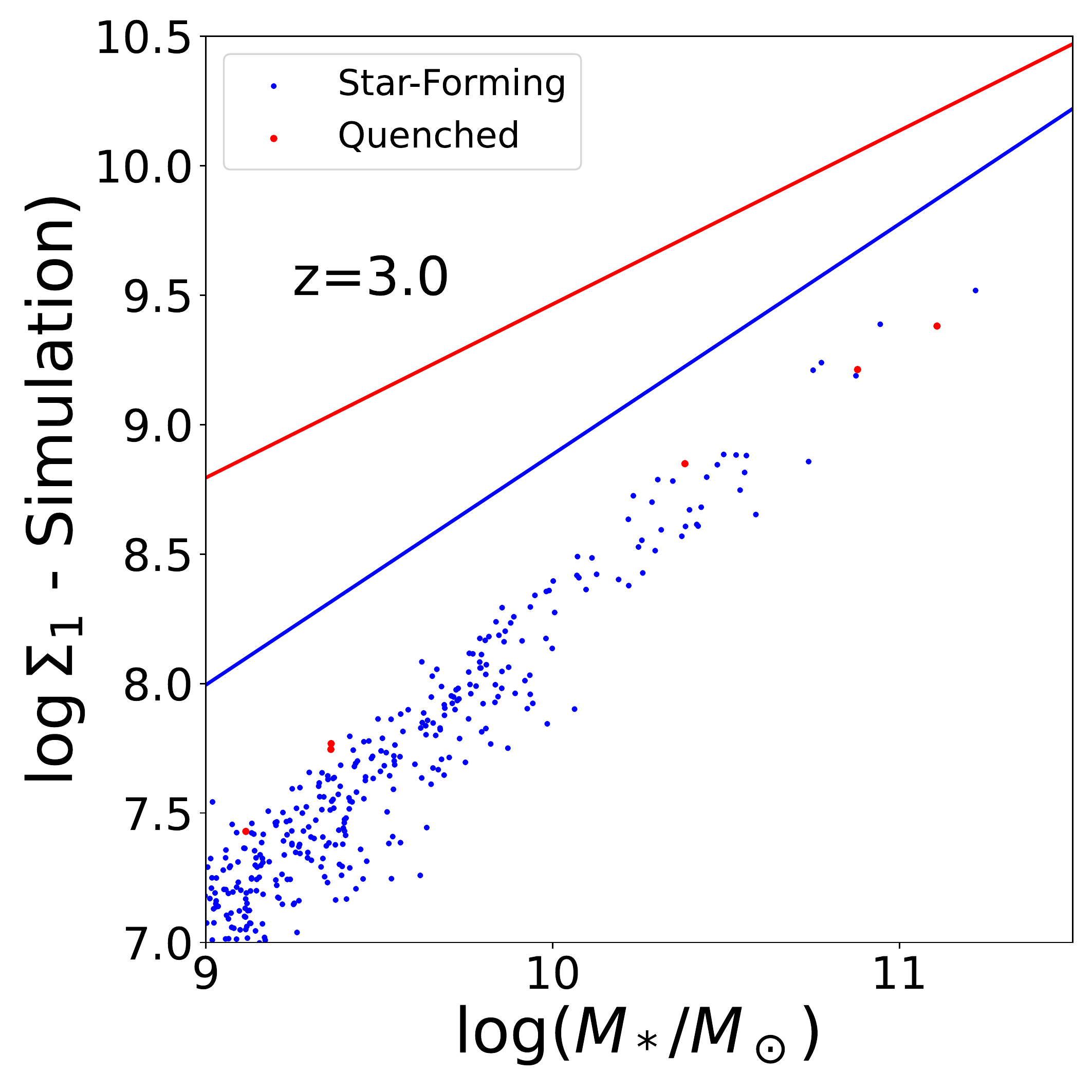}}%

      \vspace{-5 pt}
    \caption{Comparison of the $\log M_*-\log\Sigma_1$ relations when $\Sigma_1$ is computed using the SKIRT images (top row) or the stellar particles from the simulation (bottom row). From left to right, each column shows galaxies at $z=0.5$, $z=1.5$ and $z=3$.} Red and blue dots and lines show quiescent and star-forming galaxies, respectively. The separation between quiescent and star-forming galaxies is enhanced when $\Sigma_1$ is estimated using the images and the best Sersic model. The solid blue and red lines correspond to the observational best fit relations for star forming and quiescent galaxies respectively. %
    \label{fig:sigma1_TNG_Sersic}%
\end{figure*}

These observational biases have some impact on the measured scaling laws. We compare the relation between stellar mass and stellar mass density in the central kpc ($\Sigma_1$) using the two different estimators for $\Sigma_1$ in figure~\ref{fig:sigma1_TNG_Sersic} for two redshift slices.  We show that, when the central mass density is computed from the simulation output, the values of $\Sigma_1$, for quiescent galaxies in particular, are lower and so the difference between star-forming and quiescent galaxies is less pronounced. The effect is enhanced at higher redshifts where galaxies are smaller and with lower S/N. . This is simply a preliminary assessment and requires further investigation. However it illustrates how {\emph observational effects} can impact the observational trends and interpretations.

\section{Individual tracks}
\label{app:ind_tracks}

\begin{figure*}
    \centering
    \vspace{-10 pt}
    \subcaptionbox{}{\includegraphics[width=0.6\columnwidth]{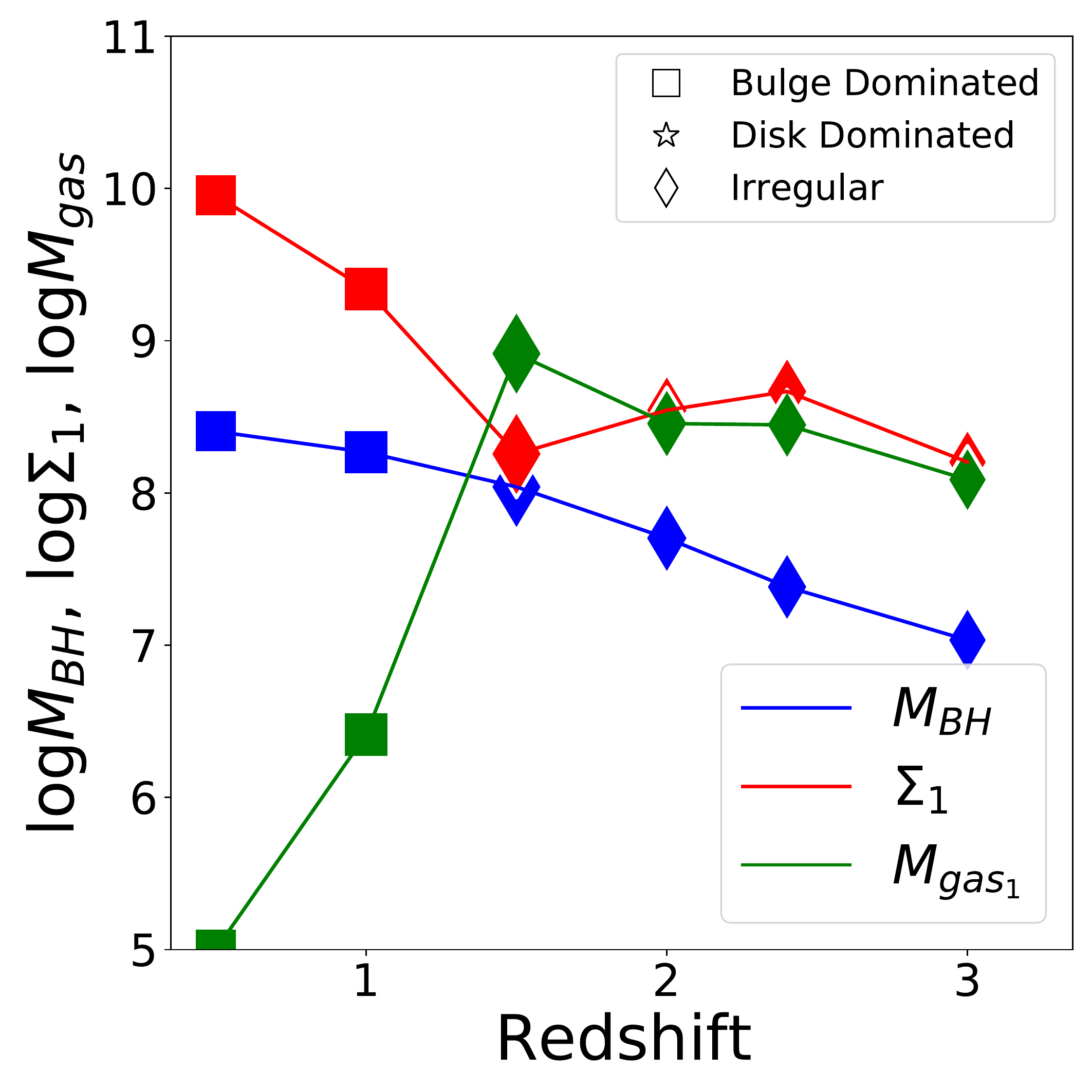}}%
     \qquad
      \subcaptionbox{}{\includegraphics[width=0.6\columnwidth]{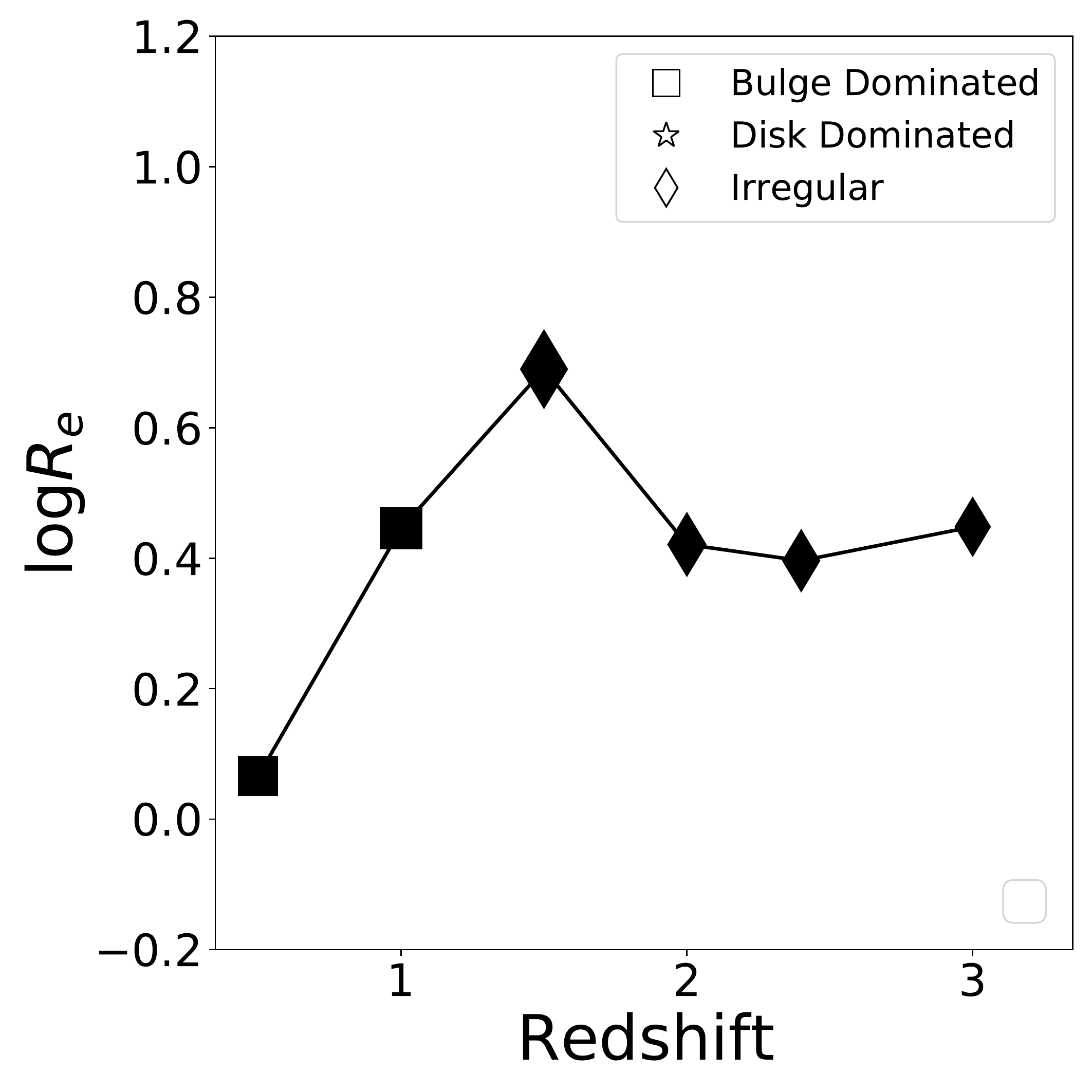}}%
     \qquad
      \subcaptionbox{}{\includegraphics[width=0.6\columnwidth]{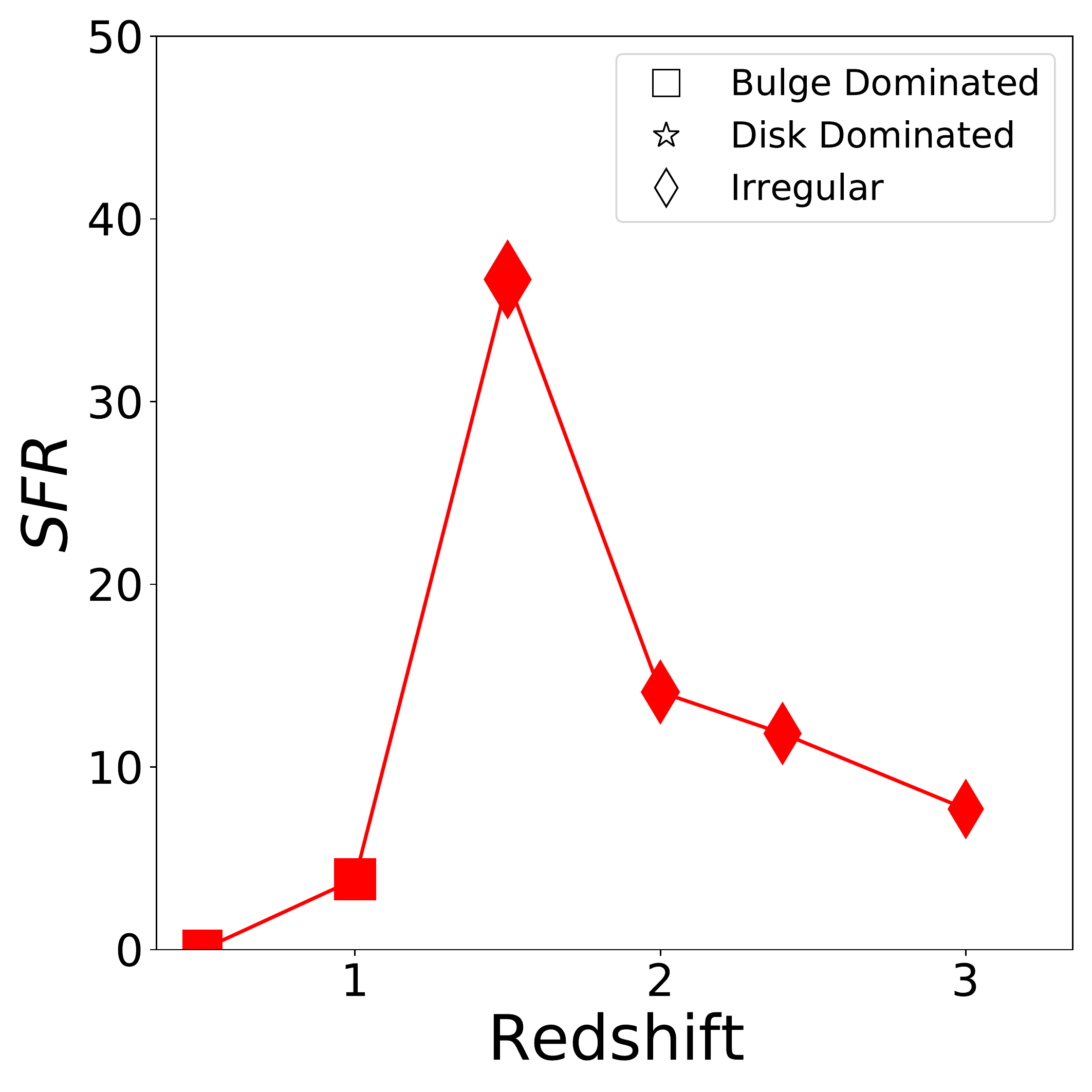}}%
          \qquad
     \subcaptionbox{}{\includegraphics[width=0.6\columnwidth]{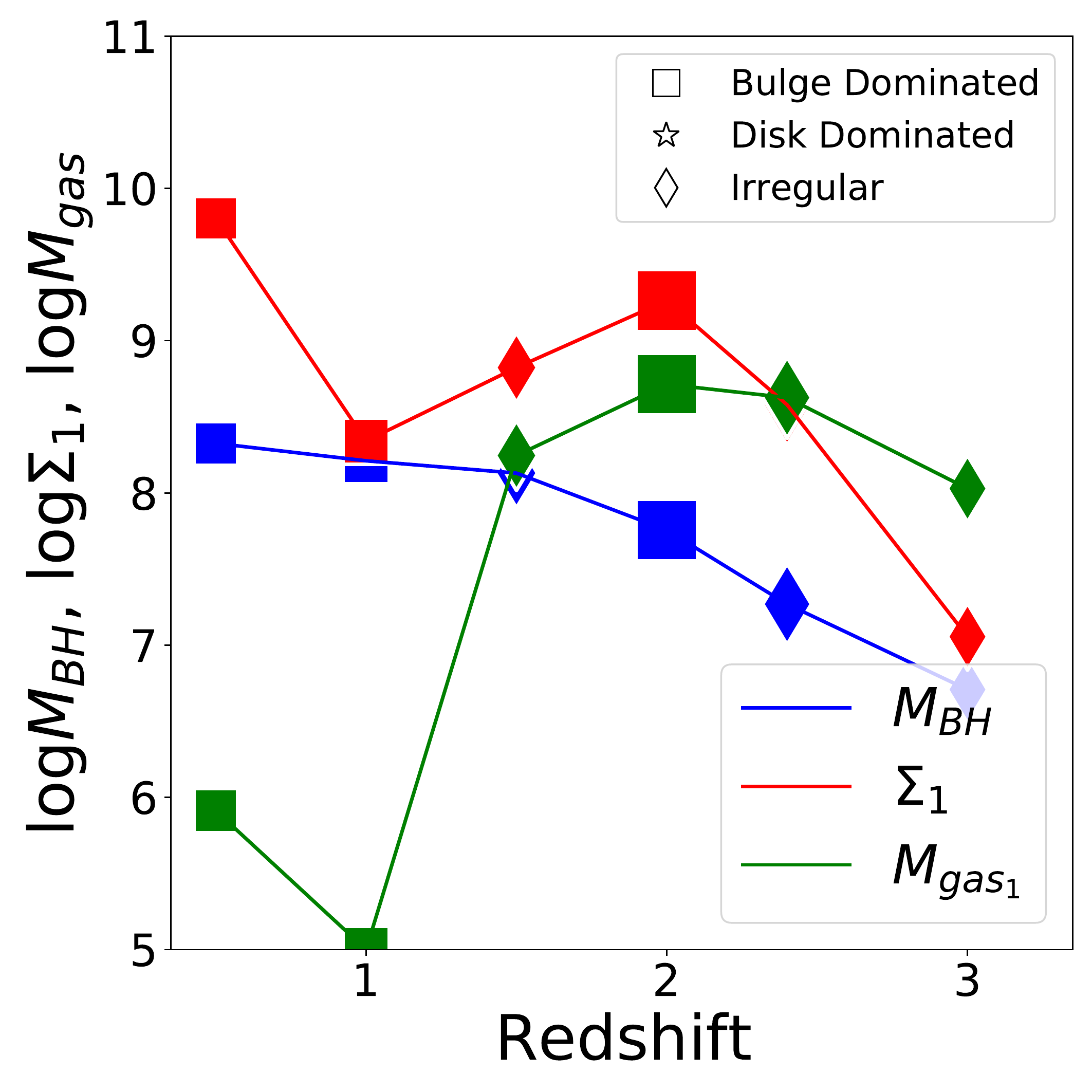}}%
     \qquad
      \subcaptionbox{}{\includegraphics[width=0.6\columnwidth]{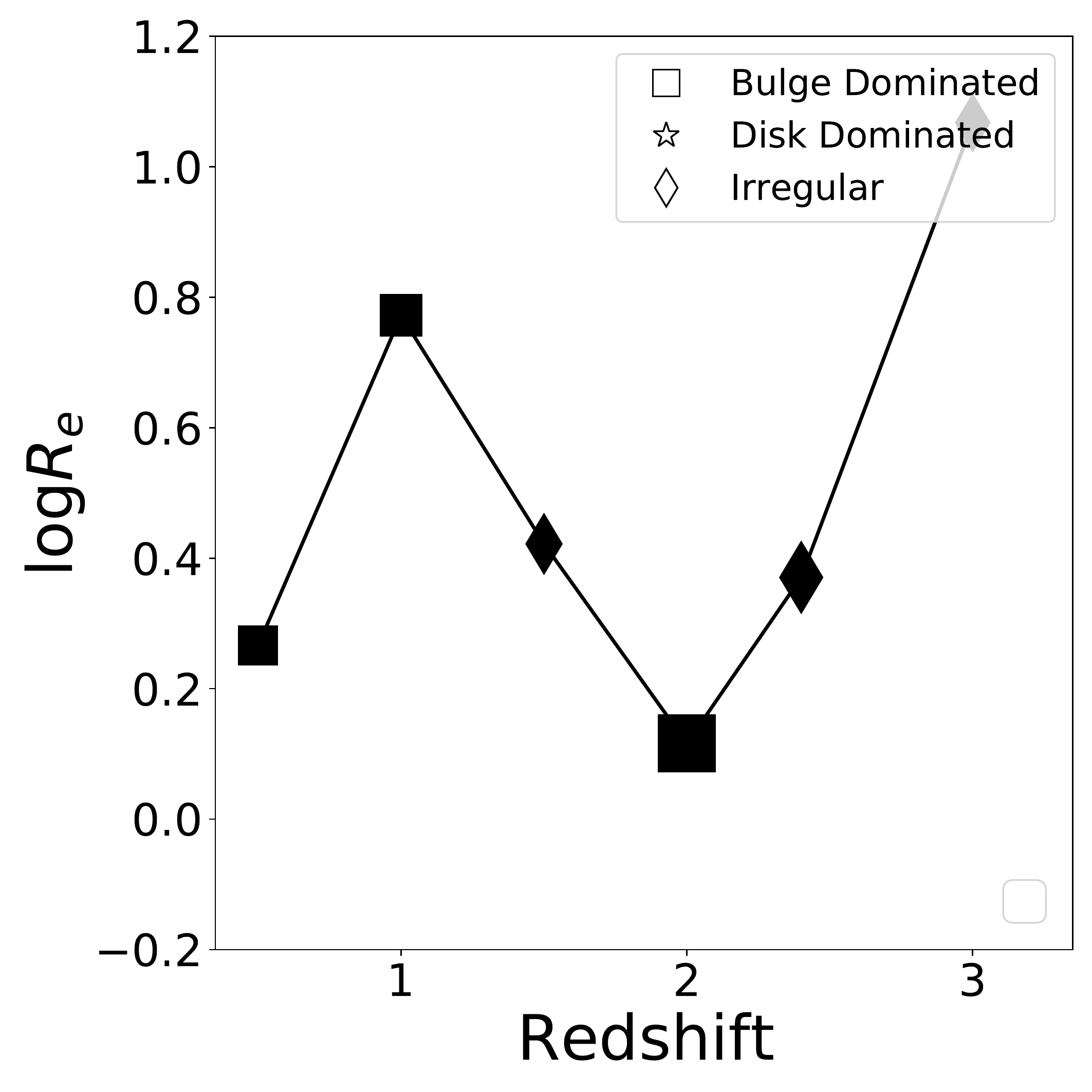}}%
     \qquad
      \subcaptionbox{}{\includegraphics[width=0.6\columnwidth]{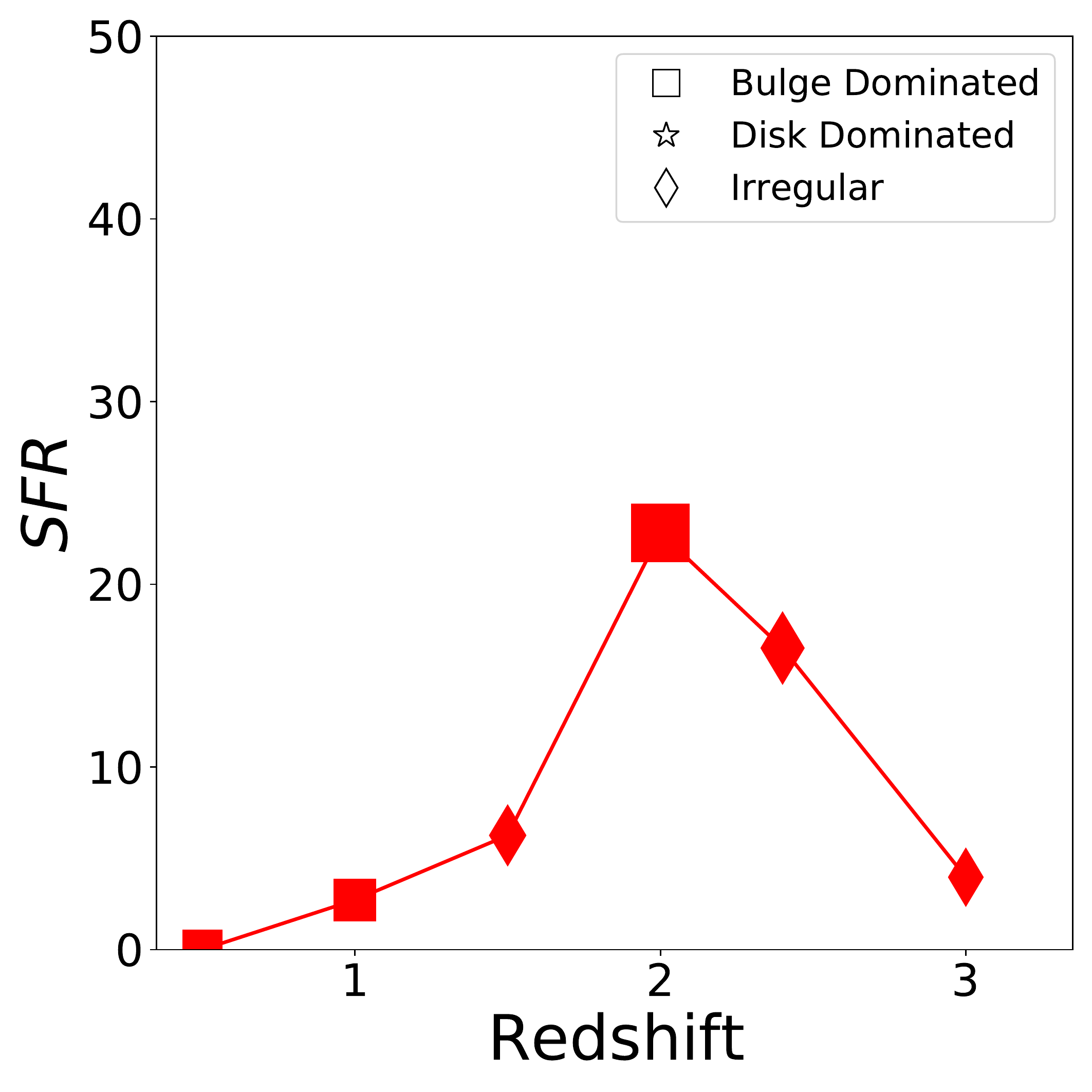}}%
          \qquad
     \subcaptionbox{}{\includegraphics[width=0.6\columnwidth]{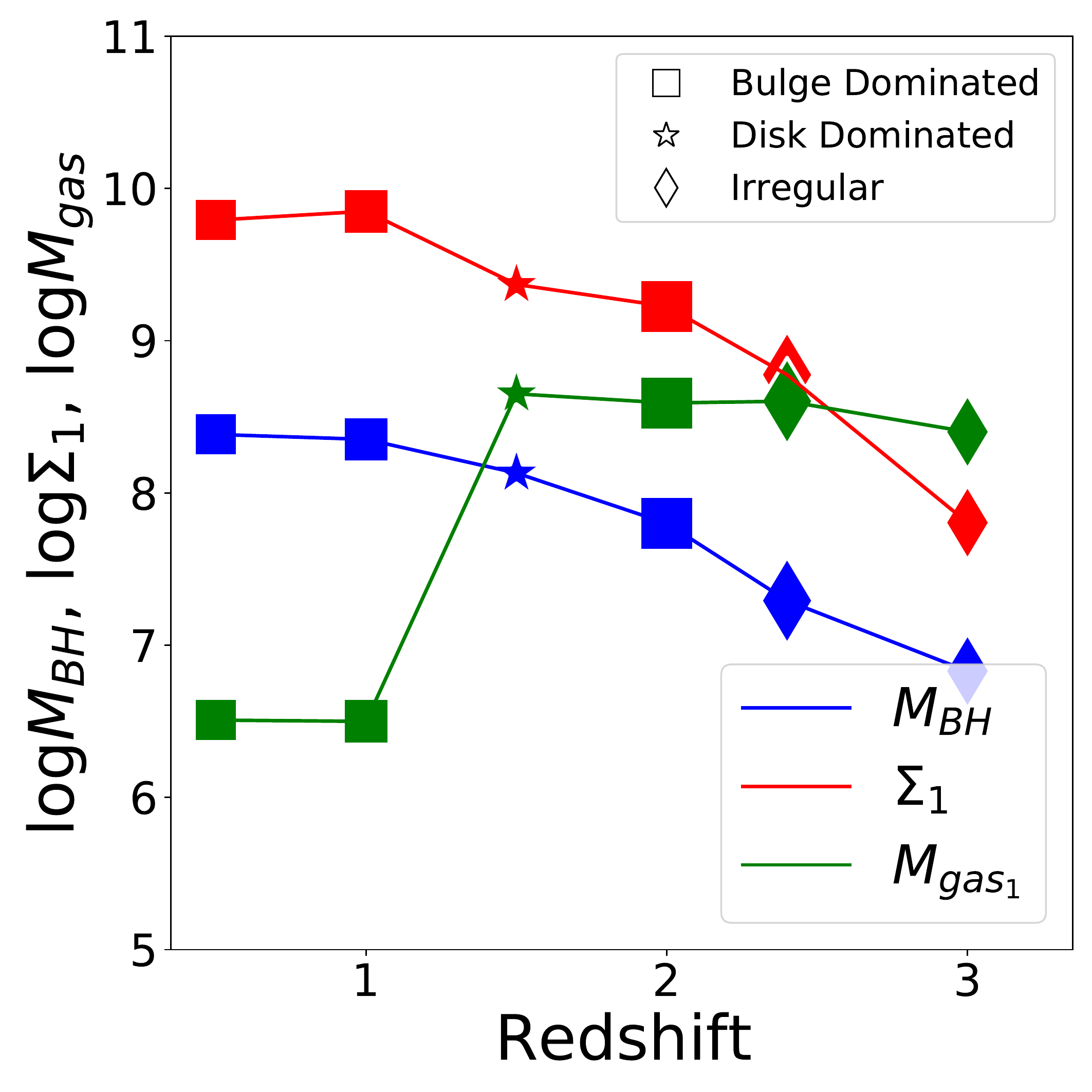}}%
     \qquad
      \subcaptionbox{}{\includegraphics[width=0.6\columnwidth]{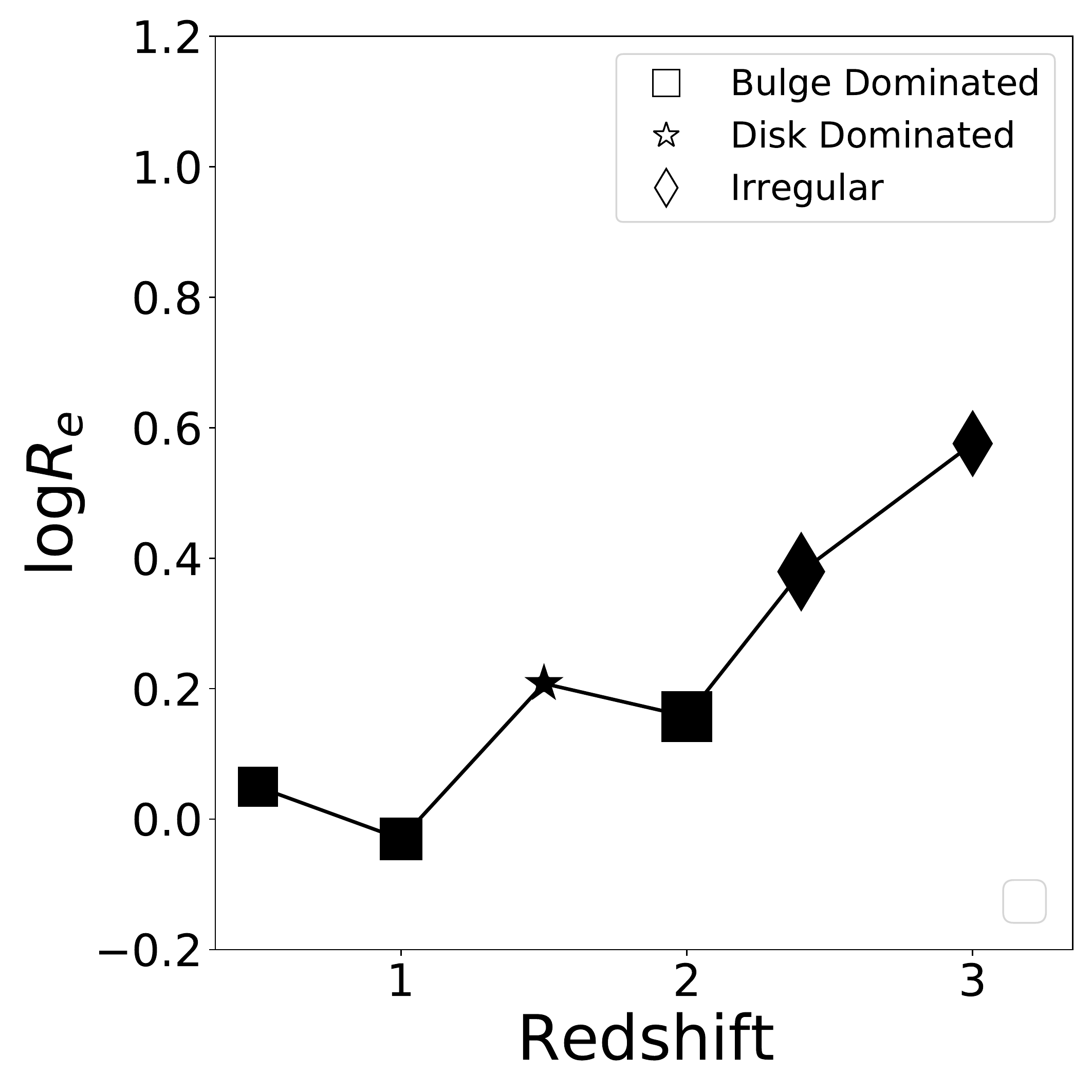}}%
     \qquad
      \subcaptionbox{}{\includegraphics[width=0.6\columnwidth]{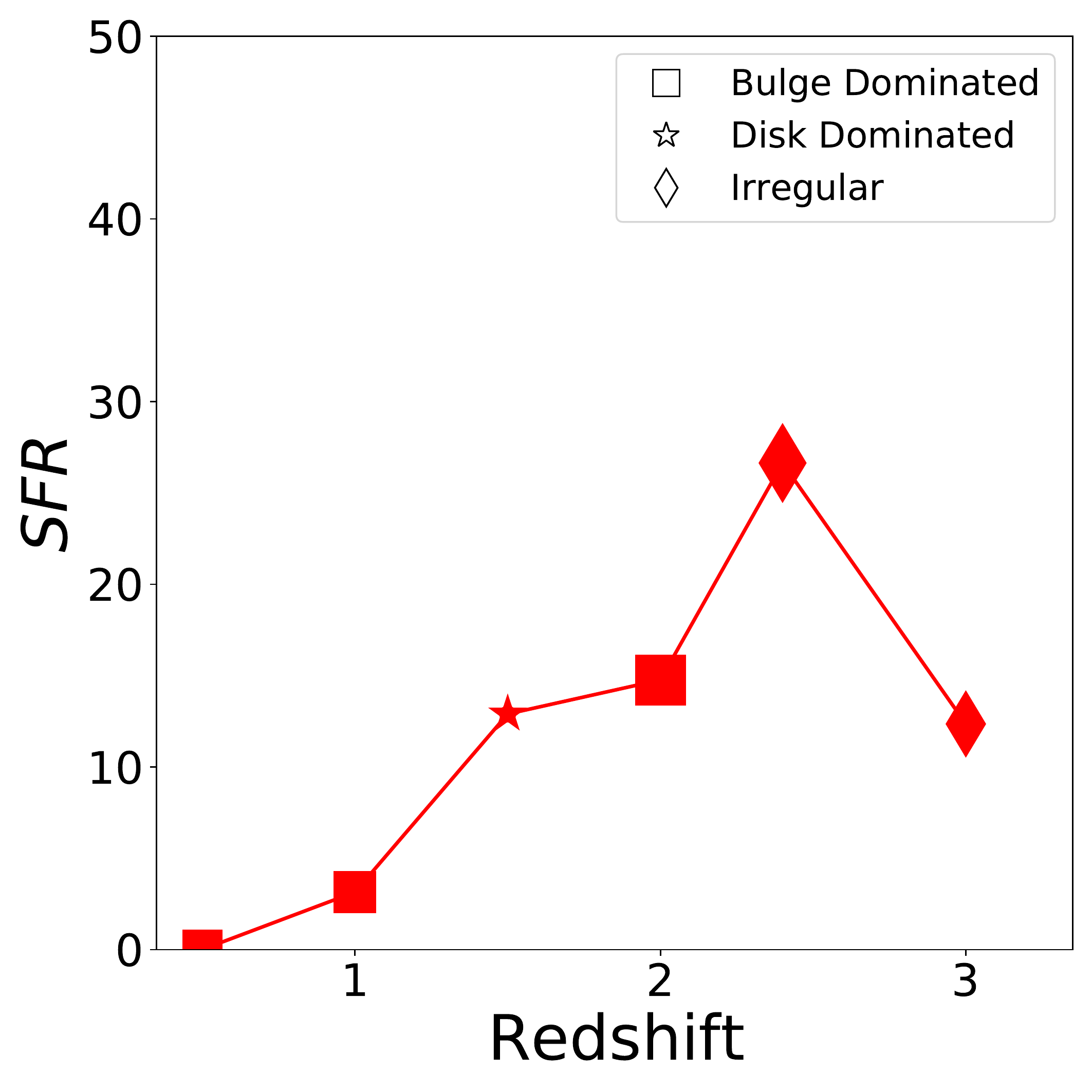}}%
          \qquad
     \subcaptionbox{}{\includegraphics[width=0.6\columnwidth]{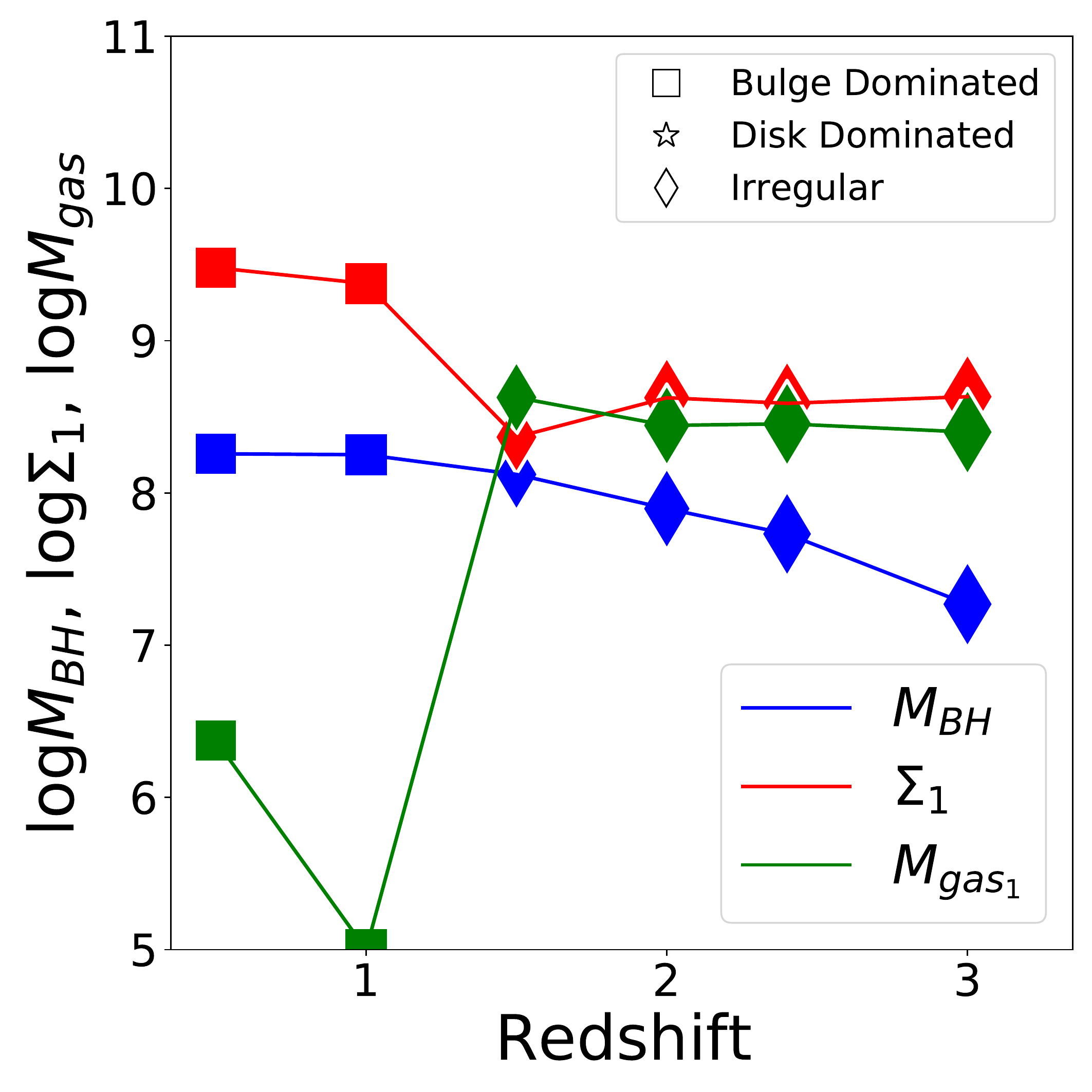}}%
     \qquad
      \subcaptionbox{}{\includegraphics[width=0.6\columnwidth]{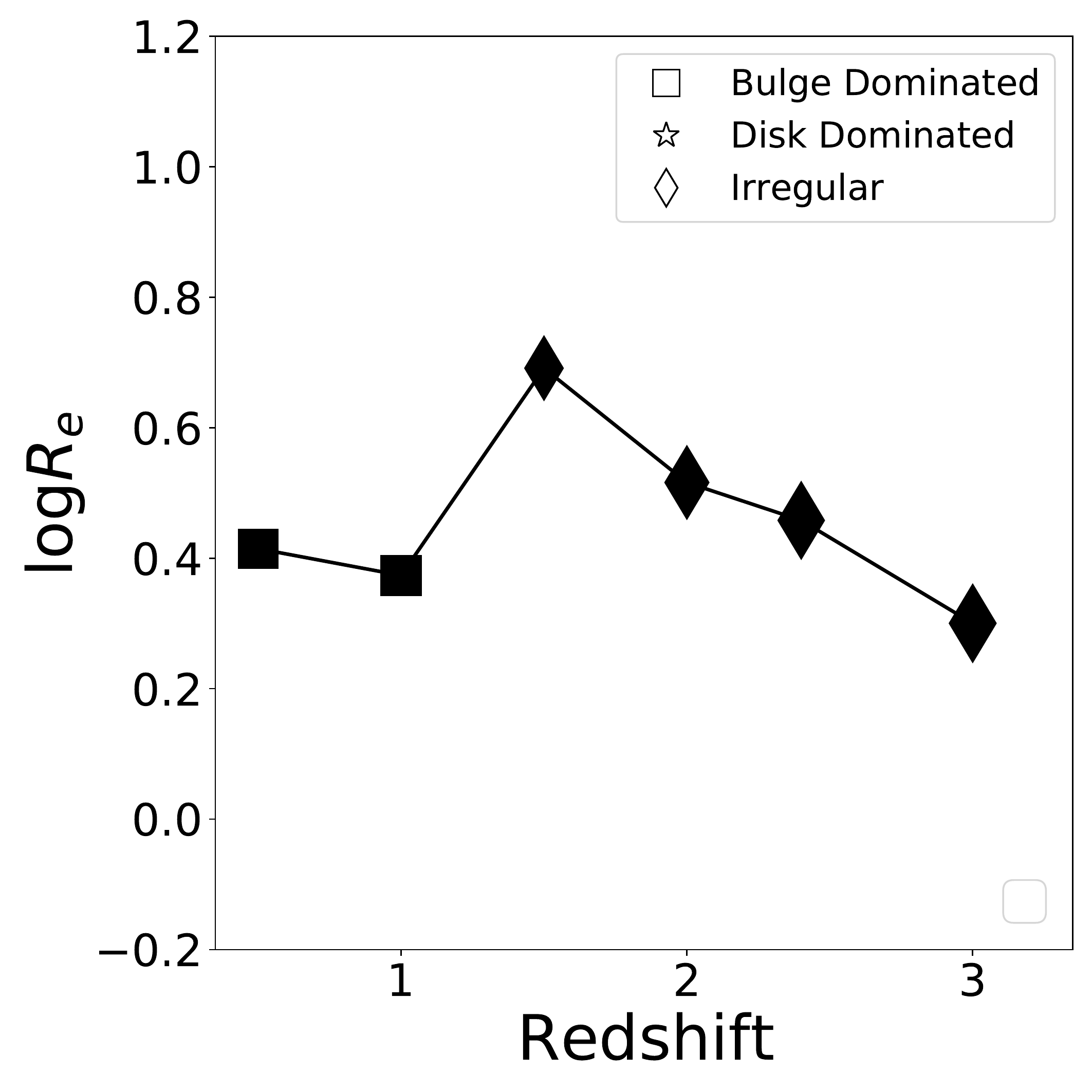}}%
     \qquad
      \subcaptionbox{}{\includegraphics[width=0.6\columnwidth]{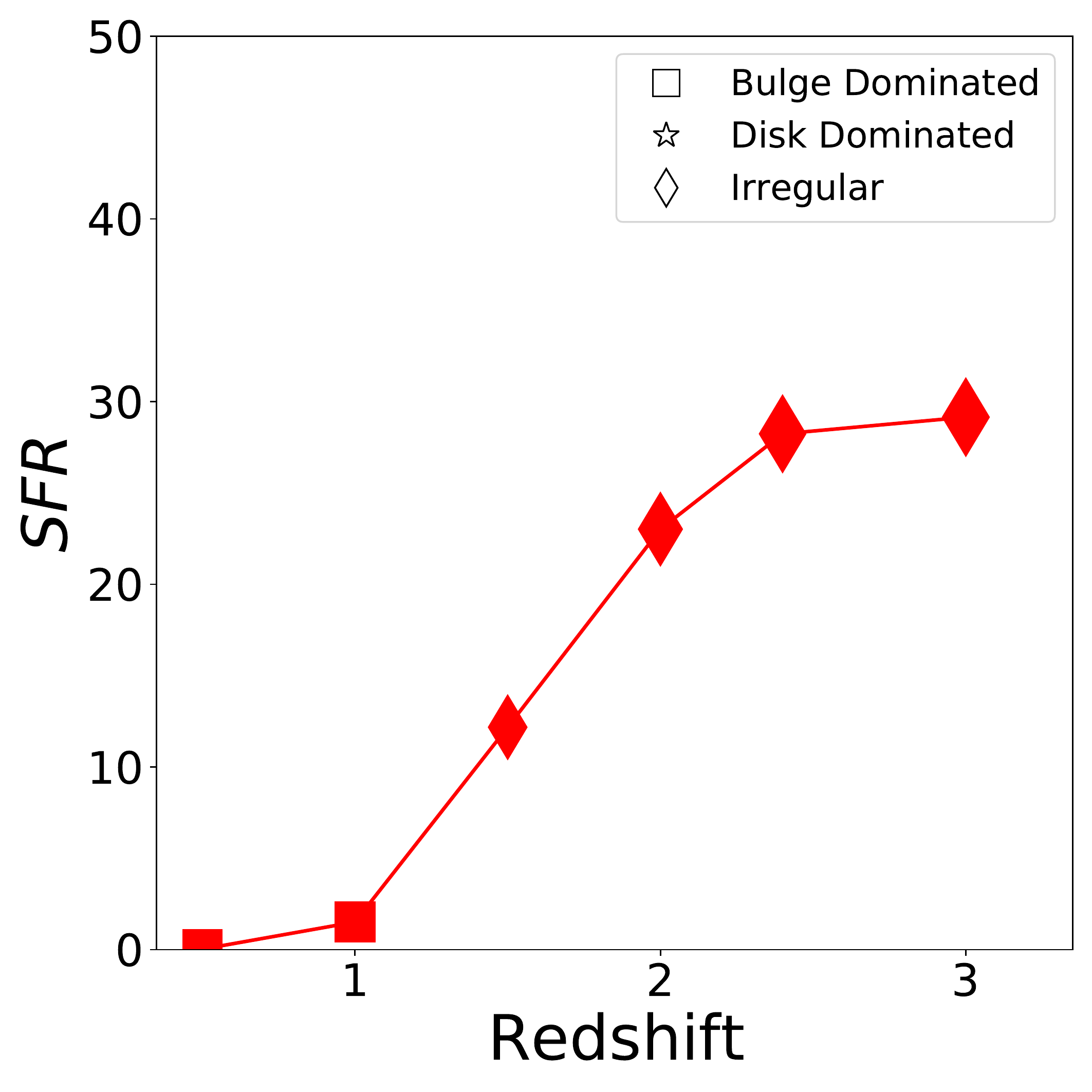}}%

      \vspace{-5 pt}
    \caption{Example tracks of $\Sigma_1$ (red), Gas mass in the central kpc (green), Black Hole Mass (blue), effective radius (black) and SFR in the central 5kpc (red) as a function of redshift for 8 massive quiescent ($10.5<\log M*/M_\odot<11$) galaxies at $z=0.5$. Each pair of plots corresponds to one galaxy. }%
    \label{fig:individual_tracks}%
\end{figure*}

Figure~\ref{fig:individual_tracks} shows the evolution of $\Sigma_1$, Gas mass, Black Hole mass and effective radius for 8 massive quiescent galaxies. Overall, we see the same trends highlighted for the average population. The central mass density $\Sigma_1$ tends to increase when the BH is in slow accretion mode at $z<1.5$. The gas mass in the central parts of galaxies also decreases as well as the effective radius,

\bsp	
\label{lastpage}
\end{document}